\documentclass[11pt]{article}

\usepackage[letterpaper, margin=1in]{geometry}
\usepackage[utf8]{inputenc} 
\usepackage[T1]{fontenc}    
\usepackage{lmodern}
\usepackage{amsmath}
\usepackage{amssymb}
\usepackage{amsthm}
\usepackage{mathtools}
\usepackage[round]{natbib}
\usepackage{graphicx}
\usepackage{url}
\usepackage{xcolor}
\usepackage{booktabs}       
\usepackage{nicefrac}       
\usepackage{microtype}      
\usepackage{tikz}
\usepackage{algorithm}
\floatstyle{plain}
\restylefloat{algorithm}
\usepackage{caption}
\usepackage[colorlinks, linkcolor=blue, citecolor=blue, urlcolor=magenta]{hyperref}

\theoremstyle{plain}
\newtheorem{theorem}{Theorem}[section]
\newtheorem{lemma}[theorem]{Lemma}
\newtheorem{proposition}[theorem]{Proposition}
\theoremstyle{definition}
\newtheorem{definition}[theorem]{Definition}
\newtheorem{example}{Example}[section]

\DeclareMathOperator{\poly}{poly}

\newenvironment{keywords}
  {\begin{quote}\small\noindent\textbf{Keywords:}\ \ignorespaces}
  {\end{quote}}
\newcommand{\acks}[1]{\section*{Acknowledgments}#1}
\makeatletter
\newcommand{\floatconts}[3]{%
  \def\@tempa{figure}%
  \ifx\@captype\@tempa
    {\centering #3\par}#2\label{#1}%
  \else
    #2\label{#1}{\centering #3\par}%
  \fi
}
\makeatother

\title{Stringological sequence prediction I: efficient algorithms for predicting highly repetitive sequences}

\author{%
  Vanessa Kosoy%
  \thanks{Faculty of Mathematics, Technion -- Israel Institute of Technology, Haifa, Israel 3200003.}%
  \hphantom{\textsuperscript{$*$}}\textsuperscript{,}%
  \thanks{Computational Rational Agents Laboratory (\href{https://coral-research.org}{coral-research.org}),
    2810 N Church St PMB 49028, Wilmington, DE 19802.}\\[0.5ex]
  \small \texttt{vanessa@alter.org}%
}
\date{}

\begin{document}

\maketitle


\begin{abstract}
  We propose novel algorithms for sequence prediction based on ideas from stringology. These algorithms are time and space efficient and satisfy mistake bounds related to particular stringological complexity measures of the sequence. In this work (the first in a series) we focus on two such measures: (i) the size of the smallest straight-line program that produces the sequence, and (ii) the number of states in the minimal automaton that can compute any symbol in the sequence when given its position in base $k$ as input. These measures are interesting because multiple rich classes of sequences studied in combinatorics of words (automatic sequences, morphic sequences, characteristic words) have low complexity and hence high predictability in this sense.
\end{abstract}


\section{Introduction}

Sequence (or ``time series'') prediction is a classical problem in artificial intelligence, machine learning and statistics \citep{Hutter2024, Cesa-Bianchi2006, Brockwell2002}. However, there is a dearth of examples of prediction algorithms which are simultaneously
\begin{itemize}
\item Computationally efficient.
\item Satisfy strong provable mistake bounds.
\item Guarantee asymptotically near-perfect prediction for natural rich classes of deterministic sequences.
\end{itemize}

Arguably, assuming no domain-specific knowledge, the ``gold standard'' for sequence prediction is Solomonoff induction \citep{Hutter2024}. On any sequence, it asymptotically performs as well as any computable predictor. It is also conceptually justifiable as a formalization of Occam's razor. However, Solomonoff induction is uncomputable, making it completely unsuitable for any practical implementation.

Much work on statistics focused on series of continuous variables \citep{Brockwell2002}. This led to algorithms that require assumptions such as particular form of probability distributions (e.g.\ normal). Most of these are inapplicable to the categorical (discrete) setting which is our own focus, and they don't yield interesting classes of deterministic sequences.

In practice, methods based on deep learning are incredibly successful in next-token prediction \citep{Brown2020}. However, the theoretical understanding of generalization bounds for deep learning is in its infancy \citep{Fan2020}. In particular, we don't have many examples where strong rigorous mistake bounds can be proved.

There are computationally efficient algorithms with strong provable generalization bounds for some interesting classes of stochastic processes, for example context-tree weighting methods \citep{Kontoyiannis2020}. However, such classes often degenerate in the deterministic case (e.g.\ admitting only periodic sequences).

One notable known example which does come close to our desiderata is using the perceptron algorithm for online linear classification \citep{Cesa-Bianchi2006} applied to some features computed from the sequence. However, the corresponding class of predictable deterministic sequences is still quite limited. Hence, finding conceptually different approaches appears to be a valuable goal.

Formally, we work in the following setting. There is a fixed finite alphabet $\Sigma$ and we are interested in predictors of the form $P:\Sigma^*\to\Sigma$ (here, $P(u)$ is the symbol predicted to follow the prefix $u$). The assumptions about the data are expressed as a \emph{complexity measure} $C:\Sigma^*\to\mathbb{N}$. We are then interested in bounding the number of mistakes $P$ makes on a sequence $u \in \Sigma^*$ in terms of $m \coloneqq C(u)$ and $n \coloneqq |u|$. At the same time, we require that $P$ is computable in polynomial time. Moreover, we wish to simultaneously bound the size of the \emph{internal state} of the predictor in terms of $m$ and $n$ (this can be interpreted as the predictor \emph{compressing} the sequence). The latter is interesting because it leads to predictors that are space efficient and in some cases run in quasilinear\footnote{I.e.\ linear up to logarithmic factors.} time.

We study multiple natural complexity measures $C$ with strong mistake and compression bounds. (Two such measures are addressed in this work, further examples will be studied in sequels.) Conceptually, such complexity measures can be viewed as candidate tractable analogues of Kolmogorov complexity. Our algorithms simultaneously get a mistake bound of the form $O(m \poly(\log m,\log n))$ and a compression bound of the form $\poly(m,\log n)$. In particular, when operating on an infinite sequence for which $m$ grows polylogarithmically, such a predictor runs in quasilinear time and polylog space while making only polylog mistakes (see Definition~\ref{def:comp_eff}).

Any prediction algorithm that assumes no domain-specific knowledge is forced to rely on properties of data that are ubiquitous across many domains in the real world. One candidate for such a property is \emph{hierarchical} structure \citep{Simon1962, Nevill-Manning1997}. A natural formalism for describing sequences with hierarchical structure is straight-line programs (SLP) \citep{Burgisser1997}. The size of the smallest SLP producing a sequence is therefore one natural complexity measure for us to consider.

Luckily, algorithms on sequences with small SLP were widely studied in stringology, where the size of the smallest SLP is one of several standard measures of ``compressibility'' that are equivalent up to factors of $\log n$, where $n$ is the length of the sequence \citep{Kociumaka2022}. Indeed, one of these measures is the size of the LZ77 compression, which was originally proposed as a computationally tractable analogue of Kolmogorov complexity \citep{Lempel1976}. It is therefore somewhat remarkable that there is, as far as we know, no prior work that uses LZ77 for sequence prediction.

As a ``warmup'', we start from a closely related but easier complexity measure, namely the size of the smallest automaton that can compute any symbol in the sequence when given the \emph{time index in base $k$} as input, for some fixed integer $k\geq 2$, a concept known as ``$k$-automaticity''. Notably, \emph{infinite} sequences for which this measure is finite were widely studied in combinatorics on words: they are called \emph{automatic sequences} and have their own rich theory \citep{Allouche2003}. Hence, even this simple complexity measure captures a rich family of interesting examples. More precisely, this defines different complexity measures if the automaton is assumed to read digits in left-to-right (LTR) order vs.\ right-to-left (RTL). For LTR, we find an algorithm with $O(k m \log(m+1) (\log n)^2)$ mistakes and $O(k m \log(m+1) \log n)$ compression. We leave the treatment of RTL for a sequel.

We call \emph{straight-line complexity} (SLC) the size of the smallest SLP. It's straightforward to show that SLC is dominated by LTR $k$-automaticity for all values of $k$ (see Proposition~\ref{prp:slc_vs_aut}). Hence, a prediction algorithm that's effective w.r.t.\ SLC is automatically effective w.r.t.\ LTR $k$-automaticity for all $k$. Moreover, SLC is closely related to another well-studied class of sequences, namely the \emph{morphic sequences} \citep{Allouche2003}\footnote{Specifically, ``typical'' morphic sequences have SLC that grows as $O(\log n)$, although for some it grows as fast as $O(\sqrt{n})$, see Appendix~\ref{apx:seqs}.}. We successfully find a prediction algorithm for SLC with $O(m \log n)$ mistakes and $O(m \log n)$ compression. This algorithm, which directly uses LZ77 compression, is based on the \emph{$\delta$-SA compressed index} \citep{Kempa2023}.

\subsection{Related Work}

\citet{Porat1991} devised an algorithm for learning an automaton from its behavior on inputs given in strict lexicographic order. They did not acknowledge the connection, but this setting is equivalent to predicting an automatic sequence. However, they only showed that their algorithm produces a minimal automaton and don't have an explicit mistake bound.

More generally, the inference of automata and related classes has received much attention (e.g.\ \citealt{Higuera2010, Attias2026}), however most existing work is inapplicable to our setting.

Morphic sequences are closely related to L-systems (see e.g.\ \citealt{Rozenberg2001}). An L-system is defined by a starting word $u_0 \in \Gamma^*$ and a homomorphism $h:\Gamma^*\to\Gamma^*$, in which case the language it produces is the set of words $\{h^k(u_0) \mid k \in \mathbb{N}\}$. Inference of L-systems received some attention in the literature, see \citep{BenNaoum2009} for a survey. However, the problem of inferring an L-systems from examples of its language is quite different from the problem of morphic sequence prediction. Moreover, many of the techniques are heuristic \citep{Guo2020, Bernard2021, Lee2024} and the exact algorithms have exponential complexity (see e.g.\ \citealt{McQuillan2018}). In fact, \citet{Duffy2025} recently proved NP-hardness results about the problem. Remarkably, the sequence they construct in the proof is periodic, which makes it trivially easy in the prediction setting we study.

Given that our LZP algorithm is based on LZ77 compression, it is relevant to point out that the connection of compression to prediction and learning was the subject of considerable study, see e.g.\ \citep{Ryabko2016, David2016}. However, much of that literature isn't concerned with computational efficiency.

Also, some prediction algorithms are known that are based on LZ78 compression\footnote{LZ78 compression was introduced in \citep{Ziv1978}.}, e.g.\ \citep{Feder1992, Gopalratnam2004, Sagan2024}. However, none of that work explores a connection to automatic or morphic sequences. Notably, LZ78 compression is less suited to our purposes since it can never compress a word of length $n$ to a word of length less than $\Omega(\sqrt{n})$. By contrast, LZ77 compression can compress to a length of $\Theta(\log n)$.

In particular, comparing our results to \citealt{Feder1992}:
\begin{itemize}
  \item We bound the absolute number of mistakes, they only bound the ``regret" w.r.t. the optimal k-th order Markov predictor.
  \item On highly compressible sequences (such as automatic), our mistake bound is $O((\log n)^2)$. Their regret bound (Theorem 4) is $O(\frac{n}{\sqrt{\log n}})$ -- barely sublinear.
  \item For highly compressible sequences, our algorithm runs in quasilinear time, while this doesn't hold for the Feder et al predictor, at least not for a naive implementation.
\end{itemize}

Finally, there was ample research on other (not prediction) algorithmic problems involving automatic and morphic sequences \citep{Shallit2022}.

\section{Setting}
\label{sec:setting}

In this section, we formalize the framework of stringological sequence prediction. We define the online prediction protocol in terms of state-based algorithms, introduce the notion of stringological complexity measures, and establish rigorous criteria for statistical and computational efficiency. Finally, we detail the \emph{counting criterion}, which relates the learnability of a sequence class to the growth rate of the number of low-complexity words.

\subsection{Preliminaries and Notation}
\label{sec:notation}

Let $\Sigma$ be a fixed finite\footnote{We assume $\Sigma$ is finite purely for ease of presentation. We could instead assume that $\Sigma=\mathbb{N}$, in which case the factors of $\log|\Sigma|$ in the bounds would be replaced by $\log M$, where $M$ is the highest number that actually appeared in the sequence so far.} alphabet. We denote the set of finite words over $\Sigma$ by $\Sigma^*$ and the set of right-infinite sequences by $\Sigma^\omega$. For a word $u \in \Sigma^*$, we denote its length by $|u|$. The $i$-th symbol in a word $u$ is denoted $u[i]$. We use 0-based indexing, such that
\begin{equation}
u = u[0]\, u[1] \dots u[|u|-1]
\label{eq:indexing}
\end{equation}

The notation $u[i:j]$ refers to the factor (subword) $u[i] \dots u[j-1]$. The notation $u[:j]$ is the same as $u[0:j]$. A word $v$ is a prefix of $u$, denoted $v \sqsubseteq u$, if $u = vw$ for some $w \in \Sigma^*$. Logarithms are taken to base 2 unless otherwise specified. When taking logarithms of non-negative integers we use the convention $\log 0 \coloneqq 0$. The notation $O(\cdot)$ hides an additive constant as well as a multiplicative one, i.e.\ $g = O(f)$ means that $g \leq Cf + C$ for some constant $C>0$.

\subsection{The Prediction Protocol}

We operate in the standard deterministic online prediction setting. To discuss memory and time constraints rigorously, we model the predictor not merely as a function of the infinite history, but as a state-based machine.

\begin{definition}
\label{def:predictor}
A \emph{predictor} is a tuple $\Pi = (\mathcal{S}, s_{\mathrm{init}}, \mathcal{U}, \mathcal{P})$, where:
\begin{itemize}
\item $\mathcal{S} \subseteq \{0, 1\}^*$ is the set of possible internal states (represented as binary strings).
\item $s_{\mathrm{init}} \in \mathcal{S}$ is the initial state.
\item $\mathcal{U}: \mathcal{S} \times \Sigma \to \mathcal{S}$ is the \textbf{state-update function}. It takes the current state and the most recent observation to produce the next state.
\item $\mathcal{P}: \mathcal{S} \to \Sigma$ is the \textbf{state-prediction function}. It maps the current state to a predicted next symbol.
\end{itemize}
\end{definition}

The prediction process proceeds in rounds $t = 0, 1, \dots$ for a target sequence $x \in \Sigma^\omega$. At step $t$:
\begin{enumerate}
\item The predictor outputs a hypothesis $\hat{x}_t = \mathcal{P}(s_t)$.
\item The true symbol $x[t]$ is revealed.
\item The predictor updates its internal state: $s_{t+1} = \mathcal{U}(s_t, x[t])$.
\end{enumerate}

We assume $s_0 = s_{\mathrm{init}}$. The total number of mistakes made by $\Pi$ on a finite word $u \in \Sigma^T$ is denoted by:
\begin{equation}
M_\Pi(u) \coloneqq |\{t \in \{0, \dots, T-1\} \mid \hat{x}_t \neq x[t]\}|
\label{eq:mistakes}
\end{equation}

We also introduce a notation for the maximal size of the state during the processing of a word:
\begin{equation}
S_\Pi(u) \coloneqq \max_{0 \leq t < T} |s_t|
\label{eq:state_size}
\end{equation}

\subsection{Efficiency Criteria}

We evaluate performance against the inherent structural complexity of the individual sequence, rather than a probabilistic prior.

\begin{definition}
\label{def:word_complexity}
A \emph{word complexity measure} is a function $C: \Sigma^* \to \mathbb{N}$ that satisfies the following conditions:
\begin{itemize}
\item \textbf{(Polynomial bound.)} There exists a polynomial $p \in \mathbb{N}[x]$ s.t.\ for all $u \in \Sigma^*$,
\begin{equation}
C(u) \leq p(|u|)
\label{eq:word_com_bound}
\end{equation}
\item \textbf{(Approximate monotonicity.)} There exists a polynomial $q \in \mathbb{N}[x, y]$ s.t.\ for all $u \in \Sigma^*$ and $k \leq |u|$,
\begin{equation}
C(u[:k]) \leq q(\log|u|, \log C(u)) \cdot (C(u)+1)
\label{eq:mono}
\end{equation}
\end{itemize}
\end{definition}

We seek predictors that are statistically efficient, space efficient and time efficient. We bound these resources in terms of the sequence's complexity $C(x)$ and its logarithmic length $\log|x|$.

\begin{definition}[Statistical Efficiency]
\label{def:stat_eff}
The predictor $\Pi$ is \emph{statistically efficient} with respect to $C$ if the number of mistakes is quasilinear in the complexity. That is, for all $u \in \Sigma^*$
\begin{equation}
M_\Pi(u) \leq (C(u)+1) \cdot \poly(\log C(u), \log|u|)
\label{eq:stat_eff}
\end{equation}
\end{definition}

\begin{definition}[Computational Efficiency]
\label{def:comp_eff}
The predictor $\Pi$ is \emph{computationally efficient} with respect to $C$ if
\begin{itemize}
\item The time to compute $\hat{x}_t = \mathcal{P}(s_t)$ is bounded by $\poly(|s_t|)$.
\item The time to compute $s_{t+1} = \mathcal{U}(s_t, x_t)$ is bounded by $\poly(|s_t|)$.
\item For all $u \in \Sigma^*$:
\begin{equation}
S_\Pi(u) \leq \poly(C(u), \log|u|)
\label{eq:gen_com_bound}
\end{equation}
\end{itemize}
\end{definition}

We can thus think of $s_t$ as a compressed representation of the history $x[:t]$, and we refer to inequalities of the form Equation~\eqref{eq:gen_com_bound} as ``compression bounds''. (In many examples, this compression is lossless, but it doesn't have to be.) In particular, this bound implies that if the sequence complexity grows polylogarithmically, then the space complexity and the per-round processing time are also polylogarithmic\footnote{Since per-round processing time is polynomial in $|s_t|$ and $|s_t|$ is polylogarithmic in $t$ due to inequality~\eqref{eq:gen_com_bound}.}.

\subsection{The Counting Criterion}

To characterize which complexity measures admit statistically efficient predictors, we utilize a counting argument. This connects the ``volume'' of the concept class (words of low complexity) to the hardness of learning.

\begin{definition}
\label{def:counting_complexity}
Given a word complexity measure $C$, the counting complexity $\mathcal{N}_C: \mathbb{N} \times \mathbb{N} \to \mathbb{R}$ is defined as the logarithm of the number of words of length $n$ with complexity at most $m$:
\begin{equation}
\mathcal{N}_C(n, m) \coloneqq \log|\{u \in \Sigma^n \mid C(u) \leq m\}|.
\label{eq:counting_complexity}
\end{equation}
\end{definition}

The following fact serves as the fundamental condition for learnability in this framework.

\begin{theorem}
\label{thm:counting_criterion}
Let $C$ be a word complexity measure. Then, there exists a statistically efficient predictor for $C$ if and only if the counting complexity satisfies:
\begin{equation}
\mathcal{N}_C(n, m) \leq (m+1) \cdot \poly(\log m, \log n)
\label{eq:counting_criterion}
\end{equation}
\end{theorem}

See Appendix~\ref{apx:counting} for the (fairly straightforward) proof.

\section{Automaticity}
\label{sec:automaticity}

In this section, we investigate our simplest example of a word complexity measure: automaticity. It is defined as the minimal number of states in an automaton that computes any symbol in the word from a base $k$ representation of its position. Details follow.

\subsection{Definitions}

Fix an integer base $k \geq 2$. For any length $m \in \mathbb{N}$ and integer $t < k^m$, let $\langle t \rangle_k^m \in [k]^m$ denote the standard base-$k$ representation of $t$, padded with leading zeros to length $m$. That is, $\langle t \rangle_k^m = w_0 w_1 \dots w_{m-1}$ such that $t = \sum_{j=0}^{m-1} w_j k^{m-1-j}$.

We model the structure of a sequence $x$ via deterministic finite automata (DFAs) that compute $x[t]$ from this padded representation.

\begin{definition}
\label{def:automaticity}
Let $x \in \Sigma^*$ be a finite word. The (LTR) \emph{$k$-automaticity} of $x$, denoted $\mathrm{AC}_k(x)$, is the minimal number of states in a DFA $M$ (with input alphabet $[k]$) such that there exists an integer $m$ satisfying $|x| \leq k^m$ and:
\begin{equation}
\forall t < |x| : x[t] = M(\langle t \rangle_k^m)
\label{eq:automaticity}
\end{equation}
\end{definition}

Note that we are processing the time index in Left-to-Right (most significant digit first) order. It is also possible to define RTL automaticity, where the order of processing is least significant digit first. For our purposes the resulting complexity measures are very different and require different algorithms. In this work, we will focus solely on LTR automaticity, leaving RTL automaticity for a sequel.

For any $x \sqsubseteq y$ it's obvious that $\mathrm{AC}_k(x) \leq \mathrm{AC}_k(y)$. Hence, Equation~\eqref{eq:mono} is satisfied for this measure. It is also easy to see that $\mathrm{AC}_k(x) \leq k^2 |x|$, and therefore $\mathrm{AC}_k$ is a word complexity measure\footnote{This is demonstrated by an automaton that has a state for every element of $[k]^{\leq m}$ with $m = \lceil \log_k |x| \rceil$, s.t.\ $M(\langle t \rangle_k^m)$ detects the exact $t$ and outputs $x[t]$ accordingly.}.

\begin{example}
\label{ex:thue_morse}
Let $\Sigma = \{0, 1\}$. We define the \emph{Thue-Morse sequence} recursively by a family of prefixes $u_n$ of length $2^n$:
\begin{equation}
u_0 = 0, \quad u_{n+1} = u_n \overline{u}_n
\end{equation}
where $\overline{w}$ denotes the bitwise negation of $w$. Note that for all $n$, $u_n \sqsubseteq u_{n+1}$ and hence there is a unique $u_\infty \in \Sigma^\omega$ s.t.\ $u_n = u_\infty[:2^n]$. The beginning of this sequence is
\begin{equation}
u_\infty = 01101001100101101001011001101001 \dots
\end{equation}

It's possible to show that for all $n$ and $i < 2^n$
\begin{equation}
u_n[i] \equiv \sum_j \langle i \rangle_2^n[j] \pmod{2}
\end{equation}

From this it is easy to see that $u_n[i]$ can be computed by an automaton with two states for any reading direction. Hence,
\begin{equation}
\mathrm{AC}_2(u_n) = 2
\end{equation}
\end{example}

The automaticity of a sequence is highly sensitive to the choice of base $k$. (See Appendix~\ref{apx:base} for an example.) This is a serious issue with automaticity-based predictors as candidate algorithms for practical applications, because in most cases there is no way to single out a preferred value of $k$. In Section~\ref{sec:slc} we will see how we can overcome the dependence on $k$ by using the word complexity measure $\operatorname{SLC}$.

\subsection{An Efficient Predictor for Automaticity}

We propose the \textbf{Hierarchical Dictionary Plurality (HDP)} algorithm. The HDP algorithm maintains a compressed representation of the sequence history using a hierarchy of dictionaries, denoted $D^{(0)}, D^{(1)}, \dots$. The dictionary at level $j$, $D^{(j)}$, stores unique blocks of length $k^j$ encountered so far. To ensure space efficiency, these blocks are not stored as raw strings but as tuples of indices pointing to entries in the dictionary of the level below, $D^{(j-1)}$. See Appendix~\ref{apx:hdp_examples} for worked examples.

\subsubsection{State}

The internal state $s_t$ of the predictor at time $t$ consists of:
\begin{enumerate}
\item A list of dictionaries $\{D^{(j)}\}_{j \geq 0}$.
  \begin{enumerate}
  \item $D^{(0)}$ is fixed as the alphabet $\Sigma$ (conceptually mapping indices $0 \dots \sigma - 1$ to symbols).
  \item For $j > 0$, $D^{(j)}$ is a dynamic list of $k$-tuples. Each entry $w \in D^{(j)}$ is a tuple $(i_0, i_1, \dots, i_{k-1}) \in \mathbb{N}^k$, where $i_r$ is an index into $D^{(j-1)}$. This represents a word of length $k^j$ formed by concatenating the words at indices $i_0, \dots, i_{k-1}$ from level $j-1$.
  \end{enumerate}
\item A list of active buffers $\{u^{(j)}\}_{j \geq 0}$.
  \begin{enumerate}
  \item $u^{(j)} \in \mathbb{N}^{<k}$ stores the sequence of indices from level $j$ observed so far in the current, incomplete block of level $j+1$.
  \end{enumerate}
\end{enumerate}

Initially, $D^{(j)} = \emptyset$ for $j > 0$ and $u^{(j)}$ is the empty word for all $j \in \mathbb{N}$.

\subsubsection{Update Rule}

The update function processes a new symbol $x_t \in \Sigma$ by propagating it up the hierarchy. At level 0, the symbol is appended to the buffer $u^{(0)}$. When a buffer at level $j$ reaches size $k$, it forms a complete block. The algorithm checks if this block (represented as a tuple of indices) exists in $D^{(j+1)}$. If not, it is added. The index of this block is then passed as an input to the buffer at level $j+1$. This process continues recursively until a buffer does not overflow (see Algorithm~\ref{alg:hdp_update}).

\begin{algorithm}
\centering
\renewcommand{\arraystretch}{1.15}
\begin{tabular}{@{}r@{\quad}l@{}}
\toprule
1 & \textbf{Input:} base $k$, state $\{D^{(j)}, u^{(j)}\}_{j \geq 0}$, new symbol $a \in \Sigma$ \\
2 & let $\mathrm{idx} \leftarrow$ index of $a$ in $\Sigma$ \\
3 & \textbf{for} $j = 0, 1, \dots$ \\
4 & \quad append $\mathrm{idx}$ to $u^{(j)}$ \\
5 & \quad \textbf{if} $|u^{(j)}| < k$ \\
6 & \quad\quad \textbf{return} \\
7 & \quad \textbf{else} \\
8 & \quad\quad let $\mathrm{tuple} \leftarrow u^{(j)}$ \\
9 & \quad\quad $u^{(j)} \leftarrow$ empty \\
10 & \quad\quad \textbf{if} can select $i$ s.t.\ $D^{(j+1)}[i] = \mathrm{tuple}$ \\
11 & \quad\quad\quad $\mathrm{idx} \leftarrow i$ \\
12 & \quad\quad \textbf{else} \\
13 & \quad\quad\quad append $\mathrm{tuple}$ to $D^{(j+1)}$ \\
14 & \quad\quad\quad $\mathrm{idx} \leftarrow |D^{(j+1)}| - 1$ \\
\bottomrule
\end{tabular}
\caption{HDP update}
\label{alg:hdp_update}
\end{algorithm}

\subsubsection{Prediction Rule}

To predict the next symbol, HDP employs a hierarchical plurality vote. It identifies a set of ``valid versions'' $V$. A version is a pair $(q, b)$, where $q$ represents a candidate index for the next block at some level $j$, and $b \in \Sigma$ is the implied next symbol at level 0.

The algorithm initializes $V$ with all possible symbols. It then iteratively filters $V$ by checking consistency with the active buffers $u^{(j)}$ against the known transitions in the dictionaries $D^{(j+1)}$. Specifically, if the current buffer $u^{(j)}$ combined with a candidate next index $c$ forms a tuple present in $D^{(j+1)}$, the version survives.

If filtering at level $j$ eliminates all candidates, the algorithm halts and outputs the plurality vote (most frequent $b$) among the survivors from level $j$. Otherwise, it proceeds to $j+1$. This logic can be regarded as running the computation of the underlying automaton in reverse (see Algorithm~\ref{alg:hdp_predict}).

\begin{algorithm}
\centering
\renewcommand{\arraystretch}{1.15}
\begin{tabular}{@{}r@{\quad}l@{}}
\toprule
1 & \textbf{Input:} base $k$, state $\{D^{(j)}, u^{(j)}\}_{j \geq 0}$ \\
2 & $V \leftarrow \{(a, a) \mid a \in \Sigma\}$ \\
3 & \textbf{for} $j = 0, 1, \dots$ \\
4 & \quad $V_{\mathrm{next}} \leftarrow \emptyset$ \\
5 & \quad \textbf{for} $(q, b) \in V$ \\
6 & \quad\quad \textbf{for} $\mathrm{idx} < |D^{(j+1)}|$ such that $u^{(j)} q \sqsubseteq D^{(j+1)}[\mathrm{idx}]$ \\
7 & \quad\quad\quad add $(\mathrm{idx}, b)$ to $V_{\mathrm{next}}$ \\
8 & \quad \textbf{if} $V_{\mathrm{next}} = \emptyset$ \\
9 & \quad\quad \textbf{return} $\arg\max_{a \in \Sigma} |\{(q, a) \in V\}|$ \\
10 & \quad \textbf{else} \\
11 & \quad\quad $V \leftarrow V_{\mathrm{next}}$ \\
\bottomrule
\end{tabular}
\caption{HDP predict}
\label{alg:hdp_predict}
\end{algorithm}

\subsection{Bounds}

\begin{theorem}
\label{thm:hdp_bounds}
HDP is statistically and computationally efficient with respect to $\mathrm{AC}_k$. Specifically, for any $n \in \mathbb{N}$ and $x \in \Sigma^n$, letting $m \coloneqq \mathrm{AC}_k(x)$ and $\sigma \coloneqq |\Sigma|$, we have:
\begin{enumerate}
\item \textbf{Statistical Efficiency:} The number of mistakes is bounded by
\begin{equation}
M_{\mathrm{HDP}}(x) = O(km \log(m+1) (\log n)^2)
\label{eq:hdp_mistakes}
\end{equation}
\item \textbf{Computational Efficiency:} The predictor satisfies the compression bound
\begin{equation}
S_{\mathrm{HDP}}(x) = O(km (\log(m+1) \log n + \log \sigma))
\label{eq:hdp_compression}
\end{equation}
\end{enumerate}
\end{theorem}

Here, the predictor receives $k$ as a parameter, and is polynomial time in this input as well. See Appendix~\ref{apx:hdp} for the proof.

\section{Straight-Line Complexity}
\label{sec:slc}

We now turn to a more general complexity measure: straight-line complexity (SLC). While $k$-automaticity characterizes sequences generated by finite state machines reading the time index, SLC characterizes sequences generated by straight-line programs: a type of context-free grammar whose language consists of a single word.

\subsection{Definitions}

We start from the definition of a straight-line program.

\begin{definition}
\label{def:slp}
A \emph{straight-line program} (SLP) over $\Sigma$ is $P = (Q, q_0, \delta)$, where $Q$ is a finite set, $q_0 \in Q$ and $\delta: Q \to \overline{Q}^*$, where $\overline{Q} \coloneqq Q \sqcup \Sigma$. We require that
\begin{itemize}
\item For all $q \in Q \setminus \{q_0\}$, $|\delta(q)| \geq 2$.
\item $\delta$ is acyclic, i.e.\ there are no $p_0, p_1 \dots p_{l-1} \in Q$ s.t for all $i < l$, $p_{i+1 \bmod l}$ appears in $\delta(p_i)$.
\item $q_0$ is the unique element of $Q$ that doesn't appear in $\delta(q)$ for any $q \in Q$.
\end{itemize}

We define $\operatorname{val}_P: \overline{Q} \to \Sigma^*$ recursively by
\begin{itemize}
\item For $a \in \Sigma$, $\operatorname{val}_P(a) \coloneqq a$.
\item For $q \in Q$, let $n \coloneqq |\delta(q)|$. Then, $\operatorname{val}_P(q) \coloneqq \operatorname{val}_P(\delta(q)[0]) \dots \operatorname{val}_P(\delta(q)[n-1])$.
\end{itemize}

We define the \emph{value} of $P$ as $\operatorname{val}(P) \coloneqq \operatorname{val}_P(q_0)$.
\end{definition}

Thus, an SLP is essentially an acyclic directed graph with vertices $\overline{Q}$ and edges $(q, q')$ where $q'$ appears in $\delta(q)$. The outgoing edges of every vertex are ordered (by the position inside $\delta(q)$), $q_0$ is the unique source in $Q$, and the elements of $\Sigma$ are sinks. See Appendix~\ref{apx:slp_examples} for examples.

The size of an SLP is defined to be the number of edges, i.e.
\begin{equation}
|P| \coloneqq \sum_{q \in Q} |\delta(q)|
\label{eq:slp_size}
\end{equation}

We can now define our next word complexity measure of interest.

\begin{definition}
\label{def:slc}
The \emph{straight-line complexity} of a word $x$, denoted $\operatorname{SLC}(x)$, is the minimum size of an SLP $P$ s.t.\ $x = \operatorname{val}(P)$.
\end{definition}

To see that $\operatorname{SLC}$ satisfies Equation~\eqref{eq:mono}, we observe the following

\begin{proposition}
\label{prp:slc_mon}
For any $x, y \in \Sigma^*$ s.t.\ $x \sqsubseteq y$, it holds that $\operatorname{SLC}(x) \leq 2\operatorname{SLC}(y)$.
\end{proposition}

See Appendix~\ref{apx:slp} for the proof.

$\operatorname{SLC}$ also satisfies Equation~\eqref{eq:word_com_bound}, because $\operatorname{SLC}(x) \leq |x|$: consider the SLP with $Q = \{q_0\}$ and $\delta(q_0) = x$.

The SLC measure is closely related to the Lempel-Ziv factorization, a foundational concept in lossless data compression. To explore this relationship, we formally define the LZ77 parsing size.

\begin{definition}
\label{def:lz77_main}
The \emph{LZ77 factorization} of a word $x$ is the unique decomposition $x = f_0 f_1 \dots f_{z-1}$ with the following property. For every $j < z$, denote $l_j \coloneqq |f_0 f_1 \dots f_{j-1}|$. Then, either $|f_j| = 1$ and $l_j$ is the first occurrence of the symbol $x[l_j]$ in $x$, or the following two conditions hold:
\begin{enumerate}
\item There exists $i < l_j$ s.t.
\begin{equation}
f_j = x[i : i + |f_j|]
\label{eq:lz77_copy}
\end{equation}
\item Either $j = z - 1$ or there is no $i < l_j$ s.t.
\begin{equation}
x[l_j : l_j + |f_j| + 1] = x[i : i + |f_j| + 1]
\label{eq:lz77_greedy}
\end{equation}
\end{enumerate}

That is, $f_j$ is the longest non-empty prefix of the remaining suffix that has occurred previously in $x$ (and the previous occurence is allowed to overlap with $f_j$). The \emph{LZ77 complexity}, denoted $\operatorname{LZC}(x)$, is defined as the number of factors $z$ in this decomposition.
\end{definition}

See Appendix~\ref{apx:lz77} for a recap of elementary properties and examples of LZ77 factorizations.

$\operatorname{LZC}$ satisfies Equation~\eqref{eq:word_com_bound} since $\operatorname{LZC}(x) \leq |x|$ and Equation~\eqref{eq:mono} since $x \sqsubseteq y$ implies $\operatorname{LZC}(x) \leq \operatorname{LZC}(y)$.

A well-known result in stringology \citep{Rytter2003} establishes that $\operatorname{SLC}(x)$ and $\operatorname{LZC}(x)$ are equivalent up to logarithmic factors:
\begin{equation}
\operatorname{LZC}(x) \leq \operatorname{SLC}(x) = O(\operatorname{LZC}(x) \log|x|)
\label{eq:slc_lzc}
\end{equation}

Hence, any predictor that is statistically (resp.\ computationally) efficient w.r.t.\ $\operatorname{SLC}$ is statistically (resp.\ computationally) efficient w.r.t.\ $\operatorname{LZC}$ and vice versa.

\subsection{Comparison with Automaticity}

Straight-line complexity is a strictly more powerful word complexity measure than LTR $k$-automaticity (i.e.\ lower up to log factors). Intuitively, the recursive $k$-section of the time interval inherent in processing the Most Significant Digit First can be directly mapped to the hierarchical concatenation structure of an SLP.

\begin{proposition}
\label{prp:slc_vs_aut}
For any base $k \geq 2$ and word $x$, the straight-line complexity is bounded by the LTR $k$-automaticity:
\begin{equation}
\operatorname{SLC}(x) = O(k \cdot \mathrm{AC}_k(x) \cdot \log|x|)
\label{eq:slc_vs_aut}
\end{equation}
\end{proposition}

See Appendix~\ref{apx:slp} for the proof.

As the following example shows, $\operatorname{SLC}$ also captures structures that don't seem to be captured by $\mathrm{AC}_k$ for any value of $k$.

\begin{example}
\label{ex:fibonacci}
The \emph{Fibonacci word} $u_\infty \in \{0, 1\}^\omega$ is defined as the limit of the recursively defined sequence of words
\begin{align}
u_0 &\coloneqq 0 \\
u_1 &\coloneqq 01 \\
u_{j+2} &\coloneqq u_{j+1} u_j
\end{align}

The beginning of the Fibonacci word is
\begin{equation}
u_\infty = 0100101001001010010100100101001001\dots
\end{equation}

The recurrence relation defining $u_j$ naturally describes an SLP of size $O(j)$ (see Example~\ref{ex:slp_fibonacci}). Since $|u_j|$ is the $j$-th Fibonacci number, Proposition~\ref{prp:slc_mon} implies
\begin{equation}
\operatorname{SLC}(u_\infty[:n]) = O(\log n)
\end{equation}
\end{example}

More generally, $\operatorname{SLC}$ has mild growth for several well-studied class of sequences (see Appendix~\ref{apx:seqs} for details).

\subsection{Efficient Prediction}

We propose the \textbf{Lempel-Ziv Plurality (LZP)} algorithm, which maintains the LZ77 factorization of the sequence so far and predicts by a plurality vote between the earlier occurrences of its last factor. We first describe the prediction and update rules as functions of the history, ignoring computational cost, and then explain how they are computed efficiently. See Appendix~\ref{apx:lzp} for the formal description.

\paragraph{Prediction rule.} Let $x[:t]$ be the history at time $t \geq 1$, and let $w$ be the last factor of the LZ77 factorization of $x[:t]$. We call $w$ the \emph{current factor}, since the upcoming symbols may extend it. The \emph{voters} are the starting positions of the earlier occurrences of $w$, i.e.\ the positions $j < t - |w|$ s.t.\ $x[j : j + |w|] = w$. A voter $j$ represents the hypothesis that the current factor is a copy of the substring starting at position $j$, and hence it votes for the symbol $x[j + |w|]$ that follows this occurrence of $w$ (this symbol is known, since $j + |w| < t$). LZP predicts the symbol that receives the most votes (if there are no voters, the prediction is arbitrary).

\paragraph{Update rule.} After the true symbol $a \coloneqq x[t]$ is revealed, the voters that voted for $a$ are exactly the positions $j < t - |w|$ s.t.\ $x[j : j + |w| + 1] = wa$. Hence, if at least one voter voted for $a$, then the LZ77 factorization of $x[:t+1]$ is obtained from that of $x[:t]$ by extending the current factor to $wa$, and the voters for $wa$ are the voters that voted for $a$. Otherwise, the LZ77 factorization of $x[:t+1]$ is obtained by appending $a$ as a new factor, whose voters are the earlier occurrences of the symbol $a$. In other words, the voters that predicted correctly survive, and a new factor begins when there are no survivors. Thus, within each factor, LZP runs the plurality algorithm (the multiclass version of the halving algorithm, see \citealt{Cesa-Bianchi2006}), where the hypotheses are the positions from which the factor may have been copied.

\paragraph{Efficient implementation.} The state of LZP is not the history itself but its LZ77 factorization, in which every factor of length 1 is stored as a symbol, and every longer factor as its length together with the starting position of an earlier occurrence. This takes $O(\operatorname{LZC}(x[:t])(\log t + \log|\Sigma|))$ bits, which is polylogarithmic in $t$ for sequences such as the Fibonacci word (Example~\ref{ex:fibonacci}). Since LZP must run in time polynomial in the size of its state (Definition~\ref{def:comp_eff}), it has to compute the votes without decompressing the history. For this, note that the number of votes for $a$ is the number of occurrences of $wa$ in $x[:t]$, which can be counted by binary search in the suffix array of $x[:t]$ (the list of starting positions of its suffixes, sorted lexicographically). LZP performs this search using the $\delta$-SA compressed index of \citet{Kempa2023} and two simpler data structures from the same work. It builds them directly from its state, in time polynomial in the size of the state, and they answer the queries of the binary search in polylogarithmic time. The update rule is implemented similarly.

\begin{theorem}
\label{thm:lzp_bounds}
There exists a predictor (LZP) which is statistically and computationally efficient w.r.t.\ $\operatorname{LZC}$. For any $n \in \mathbb{N}$ and $x \in \Sigma^n$, let $m = \operatorname{LZC}(x)$ and $\sigma \coloneqq |\Sigma|$.
\begin{enumerate}
\item \textbf{Statistical Efficiency:} The number of mistakes is bounded by
\begin{equation}
M_{\mathrm{LZP}}(x) = O(m \log n)
\label{eq:lzp_mistakes}
\end{equation}
\item \textbf{Computational Efficiency:} The predictor satisfies the compression bound
\begin{equation}
S_{\mathrm{LZP}}(x) = O(m(\log n + \log \sigma))
\label{eq:lzp_compression}
\end{equation}
\end{enumerate}
\end{theorem}

\paragraph{Proof sketch for the mistake bound.} The bound follows from the standard analysis of the plurality algorithm, applied to each factor separately. Let $x = f_0 f_1 \dots f_{m-1}$ be the LZ77 factorization of $x$, and let $l_i \coloneqq |f_0 f_1 \dots f_{i-1}|$ be the starting position of $f_i$. Whenever $l_i < t \leq l_i + |f_i|$, the LZ77 factorization of $x[:t]$ is $f_0 f_1 \dots f_{i-1} \cdot x[l_i : t]$. (Indeed, the conditions that Definition~\ref{def:lz77_main} imposes on $f_0, f_1, \dots, f_{i-1}$ only involve $x[:l_i + 1]$, and the condition on the last factor holds for every non-empty prefix of $f_i$.) Consequently, when LZP predicts $x[l_i + k]$ for some $1 \leq k < |f_i|$, its current factor is $f_i[:k]$, and the voters form the version space
\begin{equation}
V_k \coloneqq \{j < l_i \mid x[j : j + k] = f_i[:k]\}
\label{eq:lzp_version_space}
\end{equation}
The voters in $V_k$ that voted for the correct symbol $x[l_i + k]$ form $V_{k+1}$. If LZP makes a mistake, the correct symbol received at most as many votes as the predicted one, and hence at most half of all votes, so that $|V_{k+1}| \leq \frac{1}{2}|V_k|$. Moreover, $|V_1| \leq l_i < n$. On the other hand, if $|f_i| \geq 2$, then $f_i$ has an earlier occurrence $x[p : p + |f_i|]$ with $p < l_i$, and hence $p \in V_k$ for all $k \leq |f_i|$. It follows that LZP makes at most $\log n$ mistakes on the symbols $x[l_i + 1], \dots, x[l_i + |f_i| - 1]$. Adding at most one mistake on the first symbol $x[l_i]$ of $f_i$, LZP makes at most $1 + \log n$ mistakes on each factor, and therefore $M_{\mathrm{LZP}}(x) \leq m(1 + \log n)$.

See Appendix~\ref{apx:lzp} for the full proof.

\section*{AI Disclosure}

The core mathematical content and the text of the introduction were created by 
the author with only minimal AI assistance. The other sections of the paper 
were mostly written by AI based on the author's detailed research notes and guidance, 
with some back-and-forth and small manual corrections. The notes contained the algorithms and high-level 
proof sketches but not all the proof details. All AI-written material was subsequently reviewed by the author,
who takes full responsibility for its correctness. The AIs used were Google Gemini and Anthropic Claude.

\acks{This work was supported by the Advanced Research+Invention Agency (ARIA) of the United
Kingdom, the AI Security Institute (AISI) of the United Kingdom, Survival and Flourishing Corp,
and Coefficient Giving in San Francisco, California. The author wishes to thank Alexander Appel,
Matthias Georg Mayer, Vinayak Pathak, and her spouse Marcus Ogren for reviewing drafts,
locating errors and providing useful suggestions. She also thanks David Holmes for finding a typo.}

\bibliography{main}


\appendix

%
%

\section{Proof of the Counting Criterion}
\label{apx:counting}

In this appendix, we provide the detailed proof for Theorem~\ref{thm:counting_criterion}. The theorem establishes a necessary and sufficient condition for the existence of a statistically efficient predictor for a complexity measure $C$. Specifically, it states that such a predictor exists if and only if the counting complexity $\mathcal{N}_C(n, m)$ grows polynomially in $\log n$ and quasilinearly in $m$.

\subsection{Sufficiency}

We first prove that if the counting complexity satisfies inequality~\eqref{eq:counting_criterion}, then there exists a statistically efficient predictor. Explicitly, we assume that there is a polynomial $p$ s.t.\ for all $n, m \in \mathbb{N}$
\begin{equation}
\mathcal{N}_C(n, m) \leq B(n, m) \coloneqq (m+1) \cdot p(\log m, \log n)
\label{eq:suff_hyp}
\end{equation}
Replacing every coefficient of $p$ by its absolute value can only increase $p$ on non-negative arguments, so we may assume w.l.o.g.\ that all coefficients of $p$ are non-negative. Since $\log$ is non-negative and non-decreasing on $\mathbb{N}$ (recall that $\log 0 = 0$ by convention, see Section~\ref{sec:notation}), it follows that $B$ is non-decreasing in each of its arguments. This is not necessarily true for $\mathcal{N}_C$ itself, since the number of low-complexity words of a \emph{specific} length may fluctuate. For this reason, we will bound the sizes of sets of words using $B$ rather than $\mathcal{N}_C$.

We construct the predictor using the \textbf{Plurality Algorithm}, a standard approach in online learning \citep{Cesa-Bianchi2006}\footnote{Common names for this are ``majority algorithm'' and ``halving algorithm''. We prefer the word ``plurality'' because we work in the multiclass setting (i.e.\ $|\Sigma|$ can be $> 2$), in which ``plurality'' is more accurate.}. We adapt it to the anytime setting, in which the complexity $m \coloneqq C(x)$ and the length $n \coloneqq |x|$ of the target word $x$ are unknown to the predictor. To handle these unknown parameters, we employ a ``doubling trick'' that operates in phases. We stress that the resulting predictor is not computationally efficient: it enumerates a version space that can contain exponentially many words, and it evaluates $C$ on each of them. This is not an issue, since Theorem~\ref{thm:counting_criterion} is only concerned with statistical efficiency.

\subsubsection{The Predictor}

The predictor maintains the history $h \in \Sigma^*$ observed so far, together with a complexity bound $c$ and a length bound $l$, which serve as its current guesses for the complexity and the length of the target word. Initially, $h = \lambda$ (the empty word), $c = 1$ and $l = 1$.\footnote{In terms of Definition~\ref{def:predictor}, we can simply take the internal state to be the history $h$. Since we are not concerned with computational efficiency here, the bounds $c$ and $l$ can be recomputed from $h$ at every step, by running the algorithm on $h$ from the beginning.}

At time $t$ (when $|h| = t$), the predictor considers the \emph{version space}, i.e.\ the set of all words that extend the history and satisfy the current bounds:
\begin{equation}
V \coloneqq \{u \in \Sigma^* \mid h \sqsubseteq u,\; C(u) \leq c,\; |u| \leq l\}
\label{eq:version_space}
\end{equation}
For each symbol $a \in \Sigma$, it counts the words in $V$ that continue the history with $a$:
\begin{equation}
N_a \coloneqq |\{u \in V \mid |u| > t,\; u[t] = a\}|
\label{eq:symbol_count}
\end{equation}
It then predicts a symbol with the largest count, breaking ties according to some fixed ordering of $\Sigma$:
\begin{equation}
\hat{x}_t \in \arg\max_{a \in \Sigma} N_a
\label{eq:plurality_predict}
\end{equation}

Every word in $V$ has length at least $t$, and at most one word in $V$ has length exactly $t$: the history $h$ itself, which belongs to $V$ if and only if $C(h) \leq c$ and $t \leq l$. This word has no symbol at position $t$, so it is not counted in any $N_a$, i.e.\ it does not participate in the vote. Nevertheless, it plays a role in the analysis: it ensures that $V$ is non-empty whenever the history satisfies the current bounds, even if the history has no proper extension that satisfies them (see Lemma~\ref{lem:suff_history} and the discussion following it).

After receiving the true symbol $x[t]$, the predictor appends it to $h$ and checks whether the updated history has \emph{reached} one of the bounds. If $|h| \geq l$, it doubles $l$. If $C(h) \geq c$, it replaces $c$ by $\max(2c, C(h))$; the maximum is needed because the complexity of the history may jump above $2c$. (Both updates may happen at the same time step.) Each such update starts a new \emph{phase}. After these checks, the history always satisfies $C(h) \leq c$ and $|h| < l$ (see Lemma~\ref{lem:suff_history}), so that it belongs to the version space used for the next prediction. On the other hand, every update at least doubles the corresponding bound, so the number of phases is only logarithmic (see Lemma~\ref{lem:suff_phases}). The procedure is summarized in Algorithm~\ref{alg:doubling}.

\begin{algorithm}
\centering
\renewcommand{\arraystretch}{1.15}
\begin{tabular}{@{}r@{\quad}l@{}}
\toprule
\multicolumn{2}{@{}l@{}}{\textbf{Variables:} history $h \leftarrow \lambda$, complexity bound $c \leftarrow 1$, length bound $l \leftarrow 1$.} \\
\midrule
1 & \textbf{for} $t = 0, 1, 2, \dots$ \\
2 & \quad $V \leftarrow \{u \in \Sigma^* \mid h \sqsubseteq u,\; C(u) \leq c,\; |u| \leq l\}$ \\
3 & \quad \textbf{for} $a \in \Sigma$ \\
4 & \quad\quad $N_a \leftarrow |\{u \in V \mid |u| > t,\; u[t] = a\}|$ \\
5 & \quad $\hat{x}_t \leftarrow \arg\max_{a \in \Sigma} N_a$ \quad (ties broken by a fixed ordering of $\Sigma$) \\
6 & \quad receive $x[t]$ \\
7 & \quad $h \leftarrow h \cdot x[t]$ \\
8 & \quad \textbf{if} $|h| \geq l$ \textbf{then} $l \leftarrow 2l$ \\
9 & \quad \textbf{if} $C(h) \geq c$ \textbf{then} $c \leftarrow \max(2c, C(h))$ \\
\bottomrule
\end{tabular}
\caption{Plurality Algorithm}
\label{alg:doubling}
\end{algorithm}

\subsubsection{Mistake Analysis}

Let $\Pi$ denote the predictor defined by Algorithm~\ref{alg:doubling}. Fix a target word $x \in \Sigma^n$ with $n \geq 1$ (for $n = 0$ there is nothing to prove), and let $m \coloneqq C(x)$. We bound the number of mistakes $M_\Pi(x)$, i.e.\ the number of mistakes $\Pi$ makes when the observed symbols are $x[0], x[1], \dots, x[n-1]$. Note that $h = x[:t]$ at the beginning of iteration $t$. For every $t \in \{0, \dots, n-1\}$, we denote by $c_t$ and $l_t$ the values of $c$ and $l$ at the beginning of iteration $t$ (i.e.\ the bounds used for the prediction $\hat{x}_t$), and by $V_t$ and $N_a(t)$ the values of $V$ and $N_a$ computed in iteration $t$:
\begin{equation}
\begin{aligned}
V_t &= \{u \in \Sigma^* \mid x[:t] \sqsubseteq u,\; C(u) \leq c_t,\; |u| \leq l_t\} \\
N_a(t) &= |\{u \in V_t \mid |u| > t,\; u[t] = a\}|
\end{aligned}
\end{equation}
A \emph{phase} is a maximal interval of iterations $\{s, s+1, \dots, e\} \subseteq \{0, \dots, n-1\}$ on which the bounds $(c_t, l_t)$ remain constant. Since every update strictly increases the corresponding bound, a new phase begins at iteration $t \geq 1$ if and only if the condition on line~8 or on line~9 of Algorithm~\ref{alg:doubling} holds at the end of iteration $t-1$.

\begin{lemma}
\label{lem:suff_history}
For every $t \in \{0, \dots, n-1\}$:
\begin{enumerate}
\item $l_t$ is the smallest power of two greater than $t$. In particular, $t < l_t \leq \max(1, 2t)$.
\item If $t \geq 1$, then $C(x[:t]) \leq c_t$.
\end{enumerate}
Consequently, $x[:t] \in V_t$ (and in particular $V_t \neq \emptyset$) for every $t \in \{1, \dots, n-1\}$.
\end{lemma}

\begin{proof}
We prove the first claim by induction on $t$. For $t = 0$, we have $l_0 = 1$. Now, assume that the claim holds for some $t$, and consider line~8 at the end of iteration $t$, where $|h| = t+1$. Since $t < l_t$, either $t + 1 = l_t$ or $t + 1 < l_t$. In the former case, $l_{t+1} = 2l_t = 2(t+1)$, which is the smallest power of two greater than $t+1$, since $t+1$ is itself a power of two. In the latter case, $l_{t+1} = l_t$, which is the smallest power of two greater than $t$ and is also greater than $t+1$. Hence, it is the smallest power of two greater than $t+1$.

For the second claim, consider line~9 at the end of iteration $t-1$, where $h = x[:t]$. If $C(x[:t]) \geq c_{t-1}$, then $c_t = \max(2c_{t-1}, C(x[:t])) \geq C(x[:t])$. Otherwise, $c_t = c_{t-1} > C(x[:t])$.

Finally, let $t \geq 1$. The word $u \coloneqq x[:t]$ satisfies $x[:t] \sqsubseteq u$, $C(u) \leq c_t$ by the second claim, and $|u| = t < l_t$ by the first claim. Hence, $x[:t] \in V_t$.
\end{proof}

Let us comment on the role of the words of length exactly $t$ in $V_t$. As noted above, the only such word is the history $x[:t]$, and Lemma~\ref{lem:suff_history} shows that it indeed belongs to $V_t$ (for $t \geq 1$). Crucially, this does not require $x[:t]$ to have any \emph{proper} extension satisfying the current bounds. If there is no such extension, then $V_t = \{x[:t]\}$, all counts $N_a(t)$ vanish, and $\hat{x}_t$ is determined by the tie-breaking rule. In this case, the new history $x[:t+1]$ violates the current bounds, and since $|x[:t+1]| = t+1 \leq l_t$, this means that $C(x[:t+1]) > c_t$. Hence, the condition on line~9 holds at the end of iteration $t$, the phase ends, and $V_{t+1}$ is computed using updated bounds, which $x[:t+1]$ satisfies by Lemma~\ref{lem:suff_history}. In other words, whenever the history leaves the current bounds, the algorithm moves to a new phase before making the next prediction, so the version space never becomes empty. (The only version space that can be empty is $V_0$, which happens if all words of length at most $1$ have complexity greater than $1$. This is harmless: the condition on line~8 always holds at the end of iteration $0$, so iteration $0$ forms a phase by itself, in which at most one mistake is made.)

\begin{lemma}
\label{lem:suff_halving}
Let $t \in \{0, \dots, n-2\}$ be s.t.\ iterations $t$ and $t+1$ belong to the same phase. Then $V_{t+1} \subseteq V_t$ and $|V_{t+1}| = N_{x[t]}(t)$. Moreover, if $\Pi$ makes a mistake at iteration $t$ (i.e.\ $\hat{x}_t \neq x[t]$), then $|V_{t+1}| \leq \frac{1}{2}|V_t|$.
\end{lemma}

\begin{proof}
For any $u \in \Sigma^*$, we have $x[:t+1] \sqsubseteq u$ if and only if $x[:t] \sqsubseteq u$, $|u| > t$ and $u[t] = x[t]$. Since $c_{t+1} = c_t$ and $l_{t+1} = l_t$, it follows that
\begin{equation}
V_{t+1} = \{u \in V_t \mid |u| > t,\; u[t] = x[t]\}
\end{equation}
which proves the first two claims. For each $a \in \Sigma$, let $V_t^a \coloneqq \{u \in V_t \mid |u| > t,\; u[t] = a\}$, so that $|V_t^a| = N_a(t)$. The sets $V_t^a$ for different $a$ are pairwise disjoint subsets of $V_t$ (and the history $x[:t]$ belongs to none of them). If $\hat{x}_t \neq x[t]$, then $N_{\hat{x}_t}(t) \geq N_{x[t]}(t)$ by the choice of $\hat{x}_t$, and therefore $|V_t| \geq N_{\hat{x}_t}(t) + N_{x[t]}(t) \geq 2N_{x[t]}(t) = 2|V_{t+1}|$.
\end{proof}

\begin{lemma}
\label{lem:suff_phase}
Let $\{s, \dots, e\}$ be a phase, and let $c$ and $l$ be its bounds (i.e.\ $c_t = c$ and $l_t = l$ for all $t \in \{s, \dots, e\}$). Then the number of mistakes made in iterations $s, \dots, e$ is at most $B(l, c) + \log(l+1) + 1$.
\end{lemma}

\begin{proof}
First, we bound the size of $V_s$. All words in $V_s$ have length at most $l$ and complexity at most $c$. Since $\mathcal{N}_C(j, c)$ counts words of length \emph{exactly} $j$, we split them according to their length. For each $j \in \{0, \dots, l\}$, the number of words of length $j$ and complexity at most $c$ is at most $2^{\mathcal{N}_C(j, c)} \leq 2^{B(j, c)} \leq 2^{B(l, c)}$, where the second inequality uses the monotonicity of $B$. (The first quantity equals $2^{\mathcal{N}_C(j, c)}$ unless there are no such words, in which case $\mathcal{N}_C(j, c) = \log 0 = 0$.) Summing over the $l+1$ possible lengths, we get
\begin{equation}
|V_s| \leq \sum_{j=0}^{l} 2^{\mathcal{N}_C(j, c)} \leq (l+1) \cdot 2^{B(l, c)}
\label{eq:suff_vs_bound}
\end{equation}

Next, let $M'$ be the number of mistakes made in iterations $s, \dots, e-1$, i.e.\ in all iterations of the phase except the last one. If $e = s$, then $M' = 0$. Otherwise, $e \geq 1$, and hence $V_e \neq \emptyset$ by Lemma~\ref{lem:suff_history}. Applying Lemma~\ref{lem:suff_halving} to every $t \in \{s, \dots, e-1\}$, we get $1 \leq |V_e| \leq 2^{-M'}|V_s|$, and hence $M' \leq \log|V_s|$. In both cases, $M' \leq B(l, c) + \log(l+1)$: in the first case because the right-hand side is non-negative, and in the second case by inequality~\eqref{eq:suff_vs_bound}.

Finally, at most one mistake is made in the last iteration $e$. (Lemma~\ref{lem:suff_halving} does not apply to this iteration, since the next iteration, if there is one, belongs to a different phase, whose version space is defined using different bounds.)
\end{proof}

\begin{lemma}
\label{lem:suff_phases}
Let $q$ be the polynomial from Definition~\ref{def:word_complexity}, where we assume w.l.o.g.\ that $q \geq 1$ (otherwise, we can replace $q$ by $q+1$), and let
\begin{equation}
m^* \coloneqq (m+1) \cdot q(\log n, \log m)
\label{eq:mstar}
\end{equation}
Then $l_t \leq 2n - 1$ and $c_t \leq 2m^*$ for all $t \in \{0, \dots, n-1\}$. Moreover, the number of phases is at most $1 + \log(2n) + \log(2m^*) = \log(8nm^*)$.
\end{lemma}

\begin{proof}
By approximate monotonicity (inequality~\eqref{eq:mono}), every prefix of $x$ has complexity at most $m^*$. Moreover, $m^* \geq 1$ since $q \geq 1$.

By Lemma~\ref{lem:suff_history}, $l_t \leq \max(1, 2t) \leq \max(1, 2n-2) \leq 2n-1$. Moreover, for $t \geq 1$, we have $l_t \neq l_{t-1}$ if and only if $t$ is a power of two (in which case $l_{t-1} = t$ and $l_t = 2t$). The powers of two in $\{1, \dots, n-1\}$ are the numbers $2^j$ for integers $0 \leq j < \log n$, and there are at most $\lceil \log n \rceil \leq \log(2n)$ of them. Hence, there are at most $\log(2n)$ iterations $t \in \{1, \dots, n-1\}$ with $l_t \neq l_{t-1}$.

Now consider an update of $c$ on line~9 at the end of some iteration $t-1$, where $1 \leq t \leq n-1$. This update happens only if $C(x[:t]) \geq c_{t-1}$, and it sets $c_t = \max(2c_{t-1}, C(x[:t]))$, so that
\begin{equation}
2c_{t-1} \leq c_t \leq 2C(x[:t]) \leq 2m^*
\end{equation}
Since $c_0 = 1 \leq 2m^*$ and $c$ only changes through such updates, it follows that $c_t \leq 2m^*$ for all $t$. Moreover, every update at least doubles $c$. Hence, if $r$ is the number of iterations $t \in \{1, \dots, n-1\}$ with $c_t \neq c_{t-1}$, then $2^r \leq c_{n-1} \leq 2m^*$, i.e.\ $r \leq \log(2m^*)$.

Finally, a new phase begins at iteration $t \geq 1$ only if $l_t \neq l_{t-1}$ or $c_t \neq c_{t-1}$. Therefore, the number of phases is at most $1 + \log(2n) + \log(2m^*)$.
\end{proof}

\paragraph{Total mistake bound.}
By Lemma~\ref{lem:suff_phases}, the bounds $c$ and $l$ of every phase satisfy $c \leq 2m^*$ and $l + 1 \leq 2n$. Hence, by Lemma~\ref{lem:suff_phase} and the monotonicity of $B$, the number of mistakes in every phase is at most $B(2n, 2m^*) + \log(2n) + 1$. Multiplying by the number of phases (Lemma~\ref{lem:suff_phases}), we get
\begin{equation}
M_\Pi(x) \leq \log(8nm^*) \cdot \left(B(2n, 2m^*) + \log(4n)\right)
\label{eq:total_mistakes_bound}
\end{equation}

It remains to express this bound in terms of $m$ and $n$. By definition, $B(2n, 2m^*) = (2m^*+1) \cdot p(\log(2m^*), \log(2n))$, and since $m^* \geq 1$,
\begin{equation}
2m^* + 1 \leq 3m^* = 3(m+1) \cdot q(\log n, \log m)
\end{equation}
Moreover, since $\log y \leq y$ for $y \geq 1$, we have
\begin{equation}
\log m^* \leq \log(m+1) + q(\log n, \log m) \leq 1 + \log m + q(\log n, \log m)
\end{equation}
where we used $\log(m+1) \leq 1 + \log m$, which holds for all $m \in \mathbb{N}$ (for $m = 0$ by the convention $\log 0 = 0$). Hence, $\log(2m^*)$, $\log(2n)$, $\log(4n)$ and $\log(8nm^*)$ are all bounded by polynomials in $\log m$ and $\log n$. Since $p$ and $q$ have non-negative coefficients, substituting these bounds into inequality~\eqref{eq:total_mistakes_bound} yields
\begin{equation}
M_\Pi(x) \leq (m+1) \cdot \poly(\log m, \log n) = (C(x)+1) \cdot \poly(\log C(x), \log|x|)
\label{eq:suff_final}
\end{equation}
where the polynomial depends only on $p$ and $q$, and not on $x$. This is exactly the statistical efficiency condition~\eqref{eq:stat_eff}.

\subsection{Necessity}

We now prove the converse: if there exists a statistically efficient predictor $\Pi$, then the counting complexity must satisfy the growth condition.

Assume $\Pi$ is a statistically efficient predictor. By definition, there exists a polynomial $p$ such that for any sequence $u$, the number of mistakes $M_\Pi(u)$ is bounded by:
\begin{equation}
k \coloneqq M_\Pi(u) \leq (C(u)+1) \cdot p(\log C(u), \log|u|)
\label{eq:nec_mistake_bound}
\end{equation}

We rely on a compression argument (information theoretic lower bound). We show that if the predictor is efficient, we can construct a compressed representation for any word of low complexity.

Let $S_{n,m} = \{u \in \Sigma^n \mid C(u) \leq m\}$. We want to bound $\log|S_{n,m}|$. Any word $u \in S_{n,m}$ can be uniquely reconstructed if we know:
\begin{enumerate}
\item The deterministic prediction algorithm $\Pi$.
\item The specific time steps where $\Pi$ made a mistake.
\item The correct symbol at those specific time steps.
\end{enumerate}

For a word of length $n$ with at most $k$ mistakes, we can encode the locations of the mistakes using $(k+1)\log n$ bits: $\log n$ for the number of mistakes and $k \log n$ for the location of each. The correct symbols at these steps require $k \log|\Sigma|$ bits. Thus, the description length $L(u)$ in bits is:
\begin{equation}
L(u) \leq (k+1)(\log n + \log|\Sigma|)
\label{eq:description_length}
\end{equation}

Since this encoding must distinguish every word in $S_{n,m}$, the number of such words cannot exceed $2^{\max L(u)}$. Taking the logarithm gives the counting complexity:
\begin{equation}
\mathcal{N}_C(n, m) = \log|S_{n,m}| \leq \max_{u \in S_{n,m}} L(u)
\label{eq:counting_upper}
\end{equation}

Substituting the mistake bound $k \leq (m+1) \cdot p(\log m, \log n)$:
\begin{equation}
\mathcal{N}_C(n, m) \leq ((m+1) \cdot p(\log m, \log n) + 1) \cdot (\log n + \log|\Sigma|)
\label{eq:nec_expand}
\end{equation}
\begin{equation}
\mathcal{N}_C(n, m) \leq (m+1) \cdot \poly(\log m, \log n)
\label{eq:nec_final}
\end{equation}

Thus, the existence of an efficient predictor implies the required bound on the counting complexity.

\section{Properties of Lempel-Ziv 77}
\label{apx:lz77}

In this appendix, we provide background on LZ77-type factorizations, both in the standard setting and in a $k$-aligned variant used in the analysis of the HDP algorithm (Appendix~\ref{apx:hdp}). We first define general (non-greedy) factorizations, then show that the greedy variants used in the main text are uniquely determined special cases that minimize the number of factors.

\subsection{Standard LZ77-type Factorizations}

We begin with a general notion of LZ77-type factorization, which relaxes the maximality (greediness) requirement of the standard LZ77 factorization given in the main text.

\begin{definition}
\label{def:lz77_type}
An \emph{LZ77-type factorization} of a word $x \in \Sigma^n$ is a decomposition $x = f_0 f_1 \dots f_{z-1}$ into non-empty factors with the following property. For every $j < z$, denote $l_j \coloneqq |f_0 f_1 \dots f_{j-1}|$. Then, either
\begin{itemize}
\item $|f_j| = 1$ and the symbol $x[l_j]$ does not appear in $x[:l_j]$, or
\item there exists $i < l_j$ s.t.\ $f_j = x[i : i + |f_j|]$.
\end{itemize}

In the first case, we call $f_j$ a \emph{literal factor}. In the second case, we call $f_j$ a \emph{copy factor} with \emph{source position} $i$. The number of factors $z$ is called the \emph{size} of the factorization.
\end{definition}

Note that a copy factor is allowed to \emph{overlap} with its source: it is possible that $i + |f_j| > l_j$. Intuitively, this corresponds to a byte-by-byte copy where previously copied symbols become available as source material for the remainder of the factor. Also note that when $|f_j| = 1$ and the symbol $x[l_j]$ \emph{does} appear in $x[:l_j]$, the factor falls under the second case as a copy factor of length 1.

\begin{example}
\label{ex:lz77_type}
Let $\Sigma = \{a, b\}$ and $x = abababab$ (length 8). The following are all valid LZ77-type factorizations of $x$:
\begin{enumerate}
\item $a \cdot b \cdot ab \cdot ab \cdot ab$ \hspace{1em} (5 factors). Here, $f_2$, $f_3$ and $f_4$ are each a copy of $x[0:2] = ab$.
\item $a \cdot b \cdot ab \cdot abab$ \hspace{1em} (4 factors). Here, $f_3 = abab$ is a copy from source position $i = 0$; the source $x[0:4]$ and the destination $x[4:8]$ are adjacent but do not overlap.
\item $a \cdot b \cdot ababab$ \hspace{1em} (3 factors). Here, $f_2 = ababab$ is a copy from source position $i = 0$, with $i + |f_2| = 6 > l_2 = 2$: the source $x[0:6]$ overlaps with the destination $x[2:8]$, illustrating the overlap phenomenon.
\end{enumerate}
\end{example}

The greedy LZ77 factorization (Definition~\ref{def:lz77_main}), is the special case where each factor is chosen to be as long as possible. We restate the definition here for clarity.

\begin{definition}
\label{def:lz77_greedy}
The \emph{(greedy) LZ77 factorization} of a word $x \in \Sigma^n$ is the unique LZ77-type factorization $x = f_0 f_1 \dots f_{z-1}$ that additionally satisfies the following maximality condition: for every $j < z - 1$, there is no $i < l_j$ s.t.
\begin{equation}
x[l_j : l_j + |f_j| + 1] = x[i : i + |f_j| + 1]
\label{eq:lz77_greedy_cond}
\end{equation}

That is, each non-final factor is as long as possible: if the factor is a copy, it is the longest prefix of the remaining suffix that has an earlier occurrence in $x$ (and if the next symbol is new, it necessarily has length 1). The \emph{LZ77 complexity} $\operatorname{LZC}(x)$ is the number of factors $z$ in this factorization.
\end{definition}

The greedy LZ77 factorization is uniquely determined: starting from position $l_0 = 0$, each factor $f_j$ is fixed by the requirement that it be the longest possible match (or a literal if the symbol is new). Since $l_{j+1} = l_j + |f_j|$, the entire factorization is determined inductively.

\begin{example}
\label{ex:lz77_greedy}
Continuing Example~\ref{ex:lz77_type}, the greedy LZ77 factorization of $x = abababab$ is
\begin{equation}
a \cdot b \cdot ababab
\end{equation}
with 3 factors (factorization 3 from Example~\ref{ex:lz77_type}). This is the unique factorization satisfying the maximality condition. For instance, factorization 1 from Example~\ref{ex:lz77_type} is not greedy because $f_2 = ab$ could be extended: $x[2:5] = aba = x[0:3]$, violating maximality.
\end{example}

A fundamental property of the greedy LZ77 factorization is that it minimizes the number of factors among all LZ77-type factorizations (see Theorem 10 in \citealt{Storer1982}).

\begin{theorem}[Storer-Szymanski]
\label{prp:lz77_greedy_optimal}
For any word $x \in \Sigma^*$, the greedy LZ77 factorization of $x$ has the minimum number of factors among all LZ77-type factorizations of $x$.
\end{theorem}

\subsection{\texorpdfstring{$k$}{k}-aligned LZ77-type Factorizations}
\label{sec:k_lz77}

For the analysis of the HDP algorithm, we require a variety of LZ77-type factorizations where factors are constrained to be aligned with blocks of size $k^m$. This ensures compatibility with the hierarchical dictionary structure maintained by HDP.

\begin{definition}
\label{def:k_lz77_type}
Fix $k \geq 2$. A \emph{$k$-LZ77-type factorization} of a word $x \in \Sigma^n$ is a decomposition $x = f_0 f_1 \dots f_{z-1}$ into non-empty factors with the following properties. For every $j < z$, denote $L_j \coloneqq |f_0 f_1 \dots f_{j-1}|$. Then, there exist $m_j, l_j \in \mathbb{N}$ such that:
\begin{itemize}
\item $L_j = k^{m_j} l_j$ \hspace{1em} (the factor starts at an aligned position).
\item Either $|f_j| = k^{m_j}$ or ($j = z - 1$ and $|f_j| < k^{m_j}$).
\end{itemize}

Moreover, either $m_j = 0$ and $x[L_j]$ does not appear in $x[:L_j]$ (a literal factor), or there exists $i < l_j$ s.t.
\begin{equation}
f_j \sqsubseteq x[k^{m_j} i : k^{m_j}(i+1)]
\label{eq:k_lz77_copy}
\end{equation}
(a copy factor at \emph{level} $m_j$, copying a previously seen aligned block).

We call $m_j$ the \emph{level} of factor $f_j$, and the number of factors $z$ the \emph{size} of the factorization.
\end{definition}

In other words, each factor occupies a complete aligned block of size $k^{m_j}$ (except possibly the last factor, which may be shorter), and either introduces a new symbol at level 0 or matches a prefix of an aligned block of the same size that appeared earlier in $x$. Note that, unlike the standard case, overlapping copies cannot occur here: since both the source and destination blocks are aligned to multiples of $k^{m_j}$ and $i < l_j$ implies $i + 1 \leq l_j$, the source block $x[k^{m_j} i : k^{m_j}(i+1)]$ ends at or before the start of $f_j$.

Any $k$-LZ77-type factorization is in particular an LZ77-type factorization in the sense of Definition~\ref{def:lz77_type}.

Also note that the choice of level $m_j$ is constrained by the starting position $L_j$: we must have $k^{m_j} \mid L_j$. For example, a factor starting at a position that is an odd multiple of $k$ cannot have level greater than 1.

\begin{example}
\label{ex:k_lz77_type}
Let $\Sigma = \{0, 1\}$, $k = 2$, and $x = 01010101$ (length 8). The aligned blocks are:
\begin{itemize}
\item Level 0 (size 1): $0, 1, 0, 1, 0, 1, 0, 1$
\item Level 1 (size 2): $01, 01, 01, 01$
\item Level 2 (size 4): $0101, 0101$
\item Level 3 (size 8): $01010101$
\end{itemize}

The following are both valid $2$-LZ77-type factorizations:
\begin{enumerate}
\item $0 \cdot 1 \cdot 0 \cdot 1 \cdot 0 \cdot 1 \cdot 01$ \hspace{1em} (7 factors), all at level 0 except $f_6$ at level 1.
\item $0 \cdot 1 \cdot 01 \cdot 0101$ \hspace{1em} (4 factors), with levels $m_0 = m_1 = 0$, $m_2 = 1$, $m_3 = 2$.
\end{enumerate}

In factorization 2, $f_2 = 01$ at level 1 copies the aligned block $x[0:2] = 01$, and $f_3 = 0101$ at level 2 copies the aligned block $x[0:4] = 0101$.
\end{example}

The greedy $k$-LZ77 factorization is the variant that always promotes factors to the highest possible level.

\begin{definition}
\label{def:k_lz77_greedy}
Fix $k \geq 2$. The \emph{(greedy) $k$-LZ77 factorization} of a word $x \in \Sigma^n$ is the unique $k$-LZ77-type factorization $x = f_0 f_1 \dots f_{z-1}$ that additionally satisfies the following maximality condition: for every $j < z$, if $l_j$ is a multiple of $k$, then there is no $i < l_j / k$ s.t.
\begin{equation}
x[L_j : \min(L_j + k^{m_j + 1}, n)] \sqsubseteq x[k^{m_j + 1} i : k^{m_j + 1}(i+1)]
\label{eq:k_lz77_greedy_cond}
\end{equation}
\end{definition}

The maximality condition states that no factor can be ``promoted'' to the next level: if the current factor and its neighbors could be merged into a larger aligned block that has appeared previously, the greedy factorization would have already used that larger block. This determines the factorization uniquely, by the following inductive argument. At each position $L_j$, the greedy algorithm considers all levels $m$ such that $k^m \mid L_j$ and $L_j + k^m \leq n$ (plus possibly $m$ with $L_j + k^m > n$ for the last factor), in decreasing order. It selects the highest level at which the aligned block has a previous occurrence (or falls back to a level-0 literal if the symbol is new). Hence, the factor $f_j$ is uniquely determined, and $L_{j+1} = L_j + |f_j|$ determines the start of the next factor.

\begin{example}
\label{ex:k_lz77_greedy}
Continuing Example~\ref{ex:k_lz77_type}, the greedy $2$-LZ77 factorization of $x = 01010101$ is
\begin{equation}
0 \cdot 1 \cdot 01 \cdot 0101
\end{equation}
with 4 factors (factorization 2 from Example~\ref{ex:k_lz77_type}). To see that factorization 1 from Example~\ref{ex:k_lz77_type} is not greedy, observe that at position $L_2 = 2$, the level-0 factor $f_2 = 0$ can be promoted: $l_2 = 2$ is a multiple of $k = 2$, and the level-1 block $x[2:4] = 01$ matches the earlier block $x[0:2] = 01$, violating the maximality condition.
\end{example}

\section{Analysis of HDP}
\label{apx:hdp}




We now prove Theorem~\ref{thm:hdp_bounds}, establishing the compression and mistake bounds for HDP with respect to the LTR $k$-automaticity measure $\mathrm{AC}_k$, as well as its polynomial running time.

First, some notation. We will regard a DFA $M$ with input alphabet $\Gamma$ and output alphabet $\Sigma$ as a tuple $(Q, q_0, \delta, \tau)$, where:
\begin{itemize}
\item $Q$ is the set of states.
\item $q_0$ is the initial state.
\item $\delta: Q \times \Gamma \to Q$ is the transition function.
\item $\tau: Q \to \Sigma$ is the output function.
\end{itemize}

We will use the notation $\delta^*: Q \times \Gamma^* \to Q$ to denote the standard recursive extension of the transition function to input words, and we write $M(w) \coloneqq \tau(\delta^*(q_0, w))$ for the output of $M$ on an input word $w \in \Gamma^*$.

Now, let's connect the content of the dictionaries $D^{(j)}$ to the states of the minimal DFA. For this purpose, given $x \in \Sigma^n$ and $h, i \in \mathbb{N}$ s.t.\ $k^h i < n$, we call the factor $x[k^h i : \min(k^h(i+1), n)]$ the \emph{$i$-th aligned block of size $k^h$} of $x$. We call this block \emph{complete} when $k^h(i+1) \leq n$, i.e.\ when its length is exactly $k^h$. For each $h$, all aligned blocks of size $k^h$ are complete, except possibly the last one, which is cut off by the end of $x$.

\begin{lemma}
\label{lem:block}
Consider any $n, L \in \mathbb{N}$ s.t.\ $n \leq k^L$, $x \in \Sigma^n$ and a DFA $M$ with input alphabet $[k]$ and output alphabet $\Sigma$ s.t.\ $\forall t < n : x[t] = M(\langle t \rangle_k^L)$. Consider also some $h \leq L$ and $i, j \in \mathbb{N}$ s.t.\ $j \leq i$ and $k^h i < n$ (and hence $j \leq i < k^{L-h}$). Assume that
\begin{equation}
\delta^*(q_0, \langle i \rangle_k^{L-h}) = \delta^*(q_0, \langle j \rangle_k^{L-h})
\end{equation}

Then,
\begin{equation}
x[k^h i : \min(k^h(i+1), n)] \sqsubseteq x[k^h j : \min(k^h(j+1), n)]
\end{equation}
That is, the $i$-th aligned block of size $k^h$ is a prefix of the $j$-th one. In particular, if the $i$-th block is complete, then the two blocks are equal.
\end{lemma}

Intuitively, this lemma says that the content of an aligned block in $x$ is completely determined by the DFA state reached after reading the ``address'' of that block, up to truncation at the end of $x$. Recall that each position $t$ in $x$ is computed by feeding the base-$k$ digits of $t$ into the automaton $M$, most significant digit first. We can split these digits into two parts: a \emph{high-order prefix} of length $L - h$ (the address, which selects which aligned block of size $k^h$ we are in) and a \emph{low-order suffix} of length $h$ (which selects the position within that block). The high-order prefix steers the automaton from $q_0$ to some intermediate state $q$; the low-order suffix then determines, starting from $q$, the actual output symbol. If two different block indices $i$ and $j$ happen to steer the automaton to the \emph{same} intermediate state, then positions at the same offset inside the two blocks see the same low-order computation, and hence carry the same symbol. In other words, the block's content depends only on which state the automaton is in when it ``enters'' the block, not on \emph{how} it got there. The only caveat is the end of $x$: the later block might be the last aligned block of size $k^h$ and be incomplete, in which case it is a proper prefix of the earlier block rather than equal to it.

\begin{proof}
Let $q_{\mathrm{target}}$ denote the state reached by the automaton after processing either of the addresses $\langle i \rangle_k^{L-h}$ and $\langle j \rangle_k^{L-h}$. By the hypothesis:
\begin{equation}
q_{\mathrm{target}} = \delta^*(q_0, \langle i \rangle_k^{L-h}) = \delta^*(q_0, \langle j \rangle_k^{L-h})
\end{equation}

The two blocks start at the positions $T_i \coloneqq k^h i$ and $T_j \coloneqq k^h j$, and their lengths are $d_i \coloneqq \min(k^h, n - T_i)$ and $d_j \coloneqq \min(k^h, n - T_j)$ respectively. Since $j \leq i$, we have $T_j \leq T_i$ and hence $d_i \leq d_j$. Therefore, it is enough to show that the two blocks agree on their first $d_i$ symbols. Let $p$ be an integer offset representing the relative position within these blocks, such that $0 \leq p < d_i$.

The absolute positions in the sequence $x$ are $t_i(p) = T_i + p$ and $t_j(p) = T_j + p$. Both are positions in $x$, since $t_j(p) \leq t_i(p) < T_i + d_i \leq n$, and hence the symbols at these positions are computed by $M$. Since $p < d_i \leq k^h$, the base-$k$ representation of the full time index $t_i(p)$ is the concatenation of the representation of $i$ (padded to length $L - h$) and the representation of $p$ (padded to length $h$), and similarly for $t_j(p)$. That is:
\begin{align}
\langle t_i(p) \rangle_k^L &= \langle i \rangle_k^{L-h} \cdot \langle p \rangle_k^h \\
\langle t_j(p) \rangle_k^L &= \langle j \rangle_k^{L-h} \cdot \langle p \rangle_k^h
\end{align}

Using the definition of $M(\cdot)$ and the recursive property of the transition function $\delta^*(q, uv) = \delta^*(\delta^*(q, u), v)$, we have:
\begin{align}
x[t_i(p)] &= M(\langle i \rangle_k^{L-h} \cdot \langle p \rangle_k^h) = \tau(\delta^*(\delta^*(q_0, \langle i \rangle_k^{L-h}), \langle p \rangle_k^h)) \nonumber \\
           &= \tau(\delta^*(q_{\mathrm{target}}, \langle p \rangle_k^h))
\end{align}

Similarly for $t_j(p)$:
\begin{align}
x[t_j(p)] &= M(\langle j \rangle_k^{L-h} \cdot \langle p \rangle_k^h) = \tau(\delta^*(\delta^*(q_0, \langle j \rangle_k^{L-h}), \langle p \rangle_k^h)) \nonumber \\
           &= \tau(\delta^*(q_{\mathrm{target}}, \langle p \rangle_k^h))
\end{align}

Since the right-hand sides are identical, $x[t_i(p)] = x[t_j(p)]$ for all $0 \leq p < d_i$, i.e.\ $x[T_i : T_i + d_i] = x[T_j : T_j + d_i]$. Since $d_i \leq d_j$, it follows that the $i$-th block $x[T_i : T_i + d_i]$ is a prefix of the $j$-th block $x[T_j : T_j + d_j]$.

Finally, if the $i$-th block is complete, then $d_i = k^h \geq d_j$. Hence $d_i = d_j$, and the two blocks are equal.
\end{proof}

In particular, for any $h$, the complete aligned blocks of size $k^h$ of $x$ take at most $|Q|$ distinct values, since any two of them whose addresses lead $M$ to the same state are equal. Since for $h \geq 1$ the entries of $D^{(h)}$ represent distinct complete aligned blocks of size $k^h$, this shows that $|D^{(h)}| \leq |Q|$ for all $h \geq 1$.

\subsection{Compression Bound}

\paragraph{Theorem~\ref{thm:hdp_bounds} (Part 2):} For any $x \in \Sigma^n$, let $m = \mathrm{AC}_k(x)$ and $\sigma = |\Sigma|$. The state size of HDP is bounded by:
\begin{equation}
S_{\mathrm{HDP}}(x) = O(km(\log(m+1) \log n + \log \sigma))
\end{equation}

\begin{proof}
Let $M$ be a DFA with $m$ states that generates $x$ in the sense of Definition~\ref{def:automaticity}. That is, there exists some $L \geq \log_k n$ s.t.\ for any position $t < n$, $x[t] = M(\langle t \rangle_k^L)$.

For $j \geq 1$, every entry of the dictionary $D^{(j)}$ represents a complete aligned block $x[ik^j : (i+1)k^j]$ for some $i$, and distinct entries represent distinct blocks, since a block is added to $D^{(j)}$ only if it is not already there (see Algorithm~\ref{alg:hdp_update}). By Lemma~\ref{lem:block} (with $h = j$), the content of such a complete block is entirely determined by the state $\delta^*(q_0, \langle i \rangle_k^{L-j})$ of the DFA $M$ after processing the $L - j$ most significant digits of the positions in the block. Since there are only $m$ states in $M$, there are at most $m$ distinct complete aligned blocks of length $k^j$ in $x$. Consequently, $|D^{(j)}| \leq m$ for every $j \geq 1$, at every time step.

Now, fix some time $t < n$. Let $J$ be the highest level $j$ s.t.\ $D^{(j)}$ is non-empty at time $t$, or $J = 0$ if all the dictionaries are empty. The dictionary $D^{(j)}$ is non-empty only if a complete aligned block of size $k^j$ has already been processed, i.e.\ only if $k^j \leq t < n$. Hence, $J \leq \log_k n$. Moreover, for $j \geq 1$, the entries of the buffer $u^{(j)}$ are indices into $D^{(j)}$, and hence $u^{(j)}$ is empty for $j > J$. Therefore, the state $s_t$ consists of the dictionaries $D^{(1)}, \dots, D^{(J)}$ and the buffers $u^{(0)}, \dots, u^{(J)}$. (The dictionary $D^{(0)} = \Sigma$ is fixed, so it does not need to be stored.)

We encode $s_t$ as a binary string by writing $J$, then $|D^{(j)}|$ and the entries of $D^{(j)}$ for each $1 \leq j \leq J$, and then $|u^{(j)}|$ and the entries of $u^{(j)}$ for each $0 \leq j \leq J$. The integers $J$, $|D^{(j)}|$ and $|u^{(j)}|$ are written using a prefix-free code that uses $2\lfloor \log(a+1) \rfloor + 1$ bits for $a \in \mathbb{N}$ (namely, $\lfloor \log(a+1) \rfloor$ zeros followed by the binary representation of $a + 1$). An entry of $D^{(j)}$ is a $k$-tuple of indices into $D^{(j-1)}$, and an entry of $u^{(j)}$ is an index into $D^{(j)}$, where $|D^{(0)}| = \sigma$. We write every index into $D^{(j')}$ using $\lceil \log |D^{(j')}| \rceil$ bits (the decoder knows $|D^{(j')}|$ by the time it reads such an index).

Since $|D^{(j)}| \leq m$ for $j \geq 1$ and $|u^{(j)}| < k$, the integers take $O((J+1)(\log(m+1) + \log k))$ bits in total. The entries of the dictionaries take at most $km \lceil \log \sigma \rceil + Jkm \lceil \log m \rceil$ bits, and the entries of the buffers take at most $k \lceil \log \sigma \rceil + Jk \lceil \log m \rceil$ bits. Hence,
\begin{equation}
|s_t| \leq O\big((J+1)(\log(m+1) + \log k)\big) + 2km\big(\lceil \log \sigma \rceil + J \lceil \log m \rceil\big)
\end{equation}

If $n = 1$, the only relevant state is the initial state $s_0$, whose size is constant. Otherwise, $\log n \geq 1$, and we also have $\log(m+1) \geq 1$. Moreover, $J \leq \log_k n \leq \log n$, $(J+1) \log k \leq \log n + \log k$, and $\lceil \log a \rceil \leq 2 \log a$ for every integer $a \geq 1$. It follows that every term on the right-hand side is $O(km(\log(m+1) \log n + \log \sigma))$.
\end{proof}

\subsection{Mistake Bound}

The proof of the mistake bound applies the analysis of the plurality (halving) algorithm separately to every factor of the $k$-LZ77 factorization of $x$ (Definition~\ref{def:k_lz77_greedy}), and within every factor, separately to every level of the dictionary hierarchy. It proceeds in three steps. First, we bound the number of factors in terms of automaticity (Lemma~\ref{lem:k_lz77}). Second, we show that every prediction of HDP is a plurality vote over a \emph{version space}: the set of complete aligned blocks of size $k^j$ that occurred earlier in $x$ and are consistent with the observed part of the current aligned block of size $k^j$, where $j$ is the highest level at which this set is non-empty (Lemmas~\ref{lem:hdp_state} and~\ref{lem:hdp_predict}). Third, we show that during a factor of level $h$, the vote always takes place at some level $j \geq h$, and that the version space of every such level can only shrink during the factor, and shrinks at least by half whenever a mistake is made by a vote at this level (Lemma~\ref{lem:hdp_factor}). Since the version spaces of the levels $j \geq 1$ have at most $m$ elements, it follows that during every factor, at most $1 + \log m$ mistakes are made at every level.

\begin{lemma}
\label{lem:k_lz77}
For any $x \in \Sigma^n$, the number of factors in the $k$-LZ77 factorization of $x$ satisfies
\begin{equation}
z \leq k \cdot \mathrm{AC}_k(x) \cdot (\lceil \log_k n \rceil + 1)
\end{equation}
\end{lemma}

\begin{proof}
Let $m = \mathrm{AC}_k(x)$, and let $M$ be a DFA with $m$ states and $L \in \mathbb{N}$ s.t.\ $n \leq k^L$ and $x[t] = M(\langle t \rangle_k^L)$ for all $t < n$. We analyze the $k$-LZ77 factorization by mapping the factors to nodes in a $k$-ary tree representing all aligned blocks of $x$. A node at height $h$ represents an aligned block of size $k^h$, and its children are the aligned blocks of size $k^{h-1}$ that it contains.

We classify an aligned block of size $k^h$ as \textbf{Old} if it is a prefix of some aligned block of the same size that starts earlier in $x$, and as \textbf{New} otherwise. For a complete block, this means that it is New if and only if it is the first occurrence of its content among the aligned blocks of its size (ordered by start index). Now, consider two aligned blocks of size $k^h$, for some $h \leq L$, with indices $i_1 < i_2$ s.t.\ $\delta^*(q_0, \langle i_1 \rangle_k^{L-h}) = \delta^*(q_0, \langle i_2 \rangle_k^{L-h})$. By Lemma~\ref{lem:block}, the $i_2$-th block is a prefix of the $i_1$-th block, and hence it is Old. Therefore, the addresses of distinct New blocks of size $k^h$ lead $M$ to distinct states, and since $M$ has $m$ states, there are at most $m$ ``New'' blocks at any level $h \leq L$.

Consider a factor $f_j$ in the $k$-LZ77 factorization. By Definition~\ref{def:k_lz77_type}, $f_j$ is an aligned block of size $k^{m_j}$ (which is complete, unless $j = z - 1$ and $|f_j| < k^{m_j}$), i.e.\ a node at height $m_j$. Let $B$ be its parent, i.e.\ the aligned block of size $k^{m_j+1}$ that contains $f_j$, and let $P$ be the position where $B$ starts. Since $P$ is a multiple of $k^{m_j+1}$, any aligned block of size at most $k^{m_j}$ that contains position $P$ starts at $P$, whereas any aligned block of size at least $k^{m_j+1}$ that contains position $P$ also contains $f_j$. Since the factors are disjoint aligned blocks, it follows that the factor $f_{j'}$ that contains position $P$ starts at $P$ and has level $m_{j'} \leq m_j$.

Now, suppose that $B$ is Old, i.e.\ a prefix of some aligned block $B'$ of size $k^{m_j+1}$ that starts before $P$ (so that $B'$ is complete). Then, the aligned block of size $k^{m_{j'}+1}$ that starts at $P$ is a prefix of $B$ of length at most $k^{m_{j'}+1}$, and hence it is also a prefix of the first $k^{m_{j'}+1}$ symbols of $B'$. These symbols form an aligned block of size $k^{m_{j'}+1}$ that starts before $P$. Since $P$ is a multiple of $k^{m_{j'}+1}$ (i.e.\ $l_{j'} = P / k^{m_{j'}}$ is a multiple of $k$), this contradicts the maximality condition of Definition~\ref{def:k_lz77_greedy} for the factor $f_{j'}$. Thus, $B$ is New, i.e.\ every factor $f_j$ is a child of a ``New'' block.

The total number of ``New'' blocks across all levels $h$ (from $0$ to $H = \lceil \log_k n \rceil$, where $H \leq L$ since $n \leq k^L$) is:
\begin{equation}
N_{\mathrm{new}} = \sum_{h=0}^{H} (\text{New blocks at level } h) \leq \sum_{h=0}^{H} m = m(\lceil \log_k n \rceil + 1)
\end{equation}

Since every factor is a child of a ``New'' block, and each ``New'' block has at most $k$ children (the sub-blocks of size $k^{h-1}$), the total number of factors $z$ is bounded by:
\begin{equation}
z \leq k \cdot N_{\mathrm{new}} = km(\lceil \log_k n \rceil + 1)
\end{equation}
\end{proof}

\paragraph{Version spaces.}
Fix $x \in \Sigma^n$. Recall that $s_t$ denotes the state of HDP at the beginning of round $t$, i.e.\ after the updates with the symbols $x[0], \dots, x[t-1]$, and that $\hat{x}_t$ is computed from $s_t$ by Algorithm~\ref{alg:hdp_predict}. When we refer to the dictionaries $D^{(j)}$ and the buffers $u^{(j)}$ at round $t$, we mean their values in $s_t$.

Algorithm~\ref{alg:hdp_update} never modifies or removes a dictionary entry: it only appends new entries (line~13). Hence, an index $q$ of $D^{(j)}$ refers to the same entry at all times after its creation, and we can recursively define the word $\mathrm{str}_j(q)$ that it represents (see the description of the state of HDP in Section~\ref{sec:automaticity}). For $j = 0$, we identify the indices of $D^{(0)} = \Sigma$ with the symbols, and set $\mathrm{str}_0(a) \coloneqq a$ for every $a \in \Sigma$. For $j \geq 1$, if $D^{(j)}[q] = (c_0, \dots, c_{k-1})$, then
\begin{equation}
\mathrm{str}_j(q) \coloneqq \mathrm{str}_{j-1}(c_0)\, \mathrm{str}_{j-1}(c_1) \cdots \mathrm{str}_{j-1}(c_{k-1})
\end{equation}
By induction on $j$, this is a word of length $k^j$. Similarly, for a word $u = c_0 c_1 \dots c_{r-1}$ of indices of $D^{(j)}$, such as the buffer $u^{(j)}$, we write $\mathrm{str}_j(u) \coloneqq \mathrm{str}_j(c_0) \cdots \mathrm{str}_j(c_{r-1})$, which is a word of length $k^j r$.

For $j, t \in \mathbb{N}$, we denote by $\lfloor t \rfloor_j \coloneqq k^j \lfloor t / k^j \rfloor$ the number $t$ rounded down to a multiple of $k^j$. For $t < n$, this is the starting position of the aligned block of size $k^j$ that contains position $t$, and $x[\lfloor t \rfloor_j : t]$ is the part of this block observed before round $t$. Note that $0 \leq t - \lfloor t \rfloor_j < k^j$. Moreover, since $\lfloor t / k^{j+1} \rfloor = \lfloor \lfloor t / k^j \rfloor / k \rfloor$, we have
\begin{equation}
\lfloor t \rfloor_j - \lfloor t \rfloor_{j+1} = k^j r_j(t), \quad \text{where} \quad r_j(t) \coloneqq \lfloor t / k^j \rfloor \bmod k \in \{0, \dots, k-1\}
\label{eq:hdp_offsets}
\end{equation}
That is, inside the aligned block of size $k^{j+1}$ that contains position $t$, the aligned block of size $k^j$ that contains position $t$ is preceded by $r_j(t)$ complete aligned blocks of size $k^j$.

For $t \in \{0, \dots, n\}$ and $j \geq 1$, we denote by $\mathcal{B}_j(t)$ the set of complete aligned blocks of size $k^j$ of $x[:t]$, i.e.\ the set of words that occurred as complete aligned blocks of size $k^j$ before round $t$:
\begin{equation}
\mathcal{B}_j(t) \coloneqq \{x[k^j i : k^j(i+1)] \mid i \in \mathbb{N},\; k^j(i+1) \leq t\}
\label{eq:hdp_blocks}
\end{equation}
In accordance with $D^{(0)} = \Sigma$, we also set $\mathcal{B}_0(t) \coloneqq \Sigma$. For $t \in \{0, \dots, n-1\}$ and $j \in \mathbb{N}$, the \emph{version space} of level $j$ at round $t$ consists of the words in $\mathcal{B}_j(t)$ that are consistent with the observed part of the current aligned block of size $k^j$:
\begin{equation}
\mathcal{W}_j(t) \coloneqq \{w \in \mathcal{B}_j(t) \mid x[\lfloor t \rfloor_j : t] \sqsubseteq w\}
\label{eq:hdp_version_space}
\end{equation}
Every $w \in \mathcal{W}_j(t)$ \emph{votes} for its symbol at the offset of position $t$, namely $w[t - \lfloor t \rfloor_j]$. For $a \in \Sigma$, we denote the number of votes for $a$ by
\begin{equation}
N_j(t, a) \coloneqq |\{w \in \mathcal{W}_j(t) \mid w[t - \lfloor t \rfloor_j] = a\}|
\label{eq:hdp_votes}
\end{equation}
Note that $\mathcal{W}_0(t) = \Sigma$ and $N_0(t, a) = 1$ for all $t$ and $a$: the version space of level $0$ carries no information about the history.

The following lemma describes the state of HDP in these terms: for $j \geq 1$, the entries of $D^{(j)}$ represent the distinct complete aligned blocks of size $k^j$ that occurred so far, each of them exactly once, and the buffer $u^{(j)}$ lists the complete aligned blocks of size $k^j$ that precede the current one inside the current aligned block of size $k^{j+1}$.

\begin{lemma}
\label{lem:hdp_state}
Let $t \in \{0, \dots, n\}$. Then the state $s_t$ satisfies the following for every $j \in \mathbb{N}$:
\begin{enumerate}
\item The map $q \mapsto \mathrm{str}_j(q)$ is a bijection from the set of indices of $D^{(j)}$ onto $\mathcal{B}_j(t)$.
\item $\mathrm{str}_j(u^{(j)}) = x[\lfloor t \rfloor_{j+1} : \lfloor t \rfloor_j]$. In particular, $|u^{(j)}| = r_j(t) < k$.
\end{enumerate}
\end{lemma}

\begin{proof}
For $j = 0$, the first claim holds trivially, since $\mathrm{str}_0$ is the identity map of $\Sigma$. The second part of the second claim follows from the first part and equation~\eqref{eq:hdp_offsets}, since $|\mathrm{str}_j(u^{(j)})| = k^j |u^{(j)}|$. We prove the rest by induction on $t$. For $t = 0$, all the dictionaries $D^{(j)}$ with $j \geq 1$ and all the buffers are empty, and so are the sets $\mathcal{B}_j(0)$ with $j \geq 1$ and the words $x[\lfloor 0 \rfloor_{j+1} : \lfloor 0 \rfloor_j] = x[0 : 0]$.

Now, assume that the claims hold for some $t < n$, and consider the update of $s_t$ with the symbol $x[t]$ (Algorithm~\ref{alg:hdp_update}). Let $g$ be the largest integer s.t.\ $k^g$ divides $t + 1$. Then, position $t$ is the last position of an aligned block of size $k^j$ if and only if $j \leq g$, and in this case this block is $x[t+1-k^j : t+1]$. We claim that for every $j \leq g$, the loop on line~3 reaches iteration $j$, and at the beginning of this iteration, $\mathrm{str}_j(\mathrm{idx}) = x[t+1-k^j : t+1]$, and the dictionaries $D^{(1)}, \dots, D^{(j)}$ satisfy the first claim of the lemma for $t + 1$ (i.e.\ with $\mathcal{B}_{j'}(t+1)$ in place of $\mathcal{B}_{j'}(t)$).

We prove this by induction on $j$. For $j = 0$, it holds by line~2. Assume that it holds for some $j \leq g$. Since $k^j$ divides $t+1$, we have $\lfloor t \rfloor_j = t + 1 - k^j$. Hence, by the second claim for $t$, after line~4 the buffer satisfies $\mathrm{str}_j(u^{(j)}) = x[\lfloor t \rfloor_{j+1} : t+1]$. If $j = g$, then $k^{j+1}$ does not divide $t+1$, so that $\lfloor t \rfloor_{j+1} > t + 1 - k^{j+1}$ and $|u^{(j)}| < k$. Hence, the update returns on line~6, and the loop does not reach iteration $g + 1$. If $j < g$, then $k^{j+1}$ divides $t+1$, so that $\lfloor t \rfloor_{j+1} = t + 1 - k^{j+1}$ and $|u^{(j)}| = k$. Hence, lines~8--9 empty the buffer and set $\mathrm{tuple}$ to a $k$-tuple of indices of $D^{(j)}$ s.t.\ $\mathrm{str}_j(\mathrm{tuple}) = \beta \coloneqq x[t+1-k^{j+1} : t+1]$. Since all the words $\mathrm{str}_j(c)$ have the same length $k^j$, and $\mathrm{str}_j$ is injective on the indices of $D^{(j)}$ (by the first claim for $D^{(j)}$), an entry $D^{(j+1)}[q]$ is equal to $\mathrm{tuple}$ if and only if $\mathrm{str}_{j+1}(q) = \beta$. Moreover, $\mathcal{B}_{j+1}(t+1) = \mathcal{B}_{j+1}(t) \cup \{\beta\}$. Therefore, by the first claim for $t$, either $\beta \in \mathcal{B}_{j+1}(t)$, in which case line~10 finds the unique index $q$ s.t.\ $\mathrm{str}_{j+1}(q) = \beta$ and $D^{(j+1)}$ does not change, or $\beta \notin \mathcal{B}_{j+1}(t)$, in which case line~13 appends to $D^{(j+1)}$ a new entry that represents $\beta$. In both cases, $D^{(j+1)}$ now satisfies the first claim for $t+1$, and the loop proceeds to iteration $j+1$ with $\mathrm{str}_{j+1}(\mathrm{idx}) = \beta$, as claimed.

It remains to verify the claims for $t+1$ for the dictionaries $D^{(j)}$ with $j > g$ and for the buffers. For $j > g$, the dictionary $D^{(j)}$ was not modified, and $\mathcal{B}_j(t+1) = \mathcal{B}_j(t)$, since no aligned block of size $k^j$ ends at position $t$. For $j < g$, the buffer $u^{(j)}$ is now empty, and indeed $\lfloor t+1 \rfloor_{j+1} = \lfloor t+1 \rfloor_j = t+1$. For $j = g$, we showed that $\mathrm{str}_g(u^{(g)}) = x[\lfloor t \rfloor_{g+1} : t+1]$, which is equal to $x[\lfloor t+1 \rfloor_{g+1} : \lfloor t+1 \rfloor_g]$, since $\lfloor t+1 \rfloor_g = t+1$, and since $k^{g+1}$ does not divide $t+1$, so that $\lfloor t+1 \rfloor_{g+1} = \lfloor t \rfloor_{g+1}$. Finally, for $j > g$, the buffer $u^{(j)}$ was not modified, and $\lfloor t+1 \rfloor_j = \lfloor t \rfloor_j$ and $\lfloor t+1 \rfloor_{j+1} = \lfloor t \rfloor_{j+1}$, since neither $k^j$ nor $k^{j+1}$ divides $t + 1$.
\end{proof}

Next, we show that Algorithm~\ref{alg:hdp_predict} computes a plurality vote over a version space. A version $(q, b)$ of level $j$ corresponds to the word $\mathrm{str}_j(q) \in \mathcal{W}_j(t)$, and $b$ is the vote of this word. The algorithm climbs the levels as long as the version space is non-empty, and it votes at the highest level whose version space is non-empty.

\begin{lemma}
\label{lem:hdp_predict}
Let $t \in \{0, \dots, n-1\}$. Then there exists $J_t \in \mathbb{N}$ s.t.\ for every $j \in \mathbb{N}$, $\mathcal{W}_j(t) \neq \emptyset$ if and only if $j \leq J_t$. Moreover, when Algorithm~\ref{alg:hdp_predict} computes $\hat{x}_t$, the loop on line~3 reaches iteration $j$ if and only if $j \leq J_t$, and at the beginning of every iteration $j \leq J_t$,
\begin{equation}
V = \{(q, \mathrm{str}_j(q)[t - \lfloor t \rfloor_j]) \mid q \text{ is an index of } D^{(j)} \text{ s.t.\ } \mathrm{str}_j(q) \in \mathcal{W}_j(t)\}
\label{eq:hdp_v_char}
\end{equation}
Consequently, the algorithm returns in iteration $J_t$, and its prediction is a plurality vote over the version space of level $J_t$, i.e.\ $N_{J_t}(t, \hat{x}_t) \geq N_{J_t}(t, a)$ for every $a \in \Sigma$.
\end{lemma}

\begin{proof}
First, we show that $\mathcal{W}_{j+1}(t) \neq \emptyset$ implies $\mathcal{W}_j(t) \neq \emptyset$. Let $w \in \mathcal{W}_{j+1}(t)$, and write $w = w_0 w_1 \cdots w_{k-1}$, where $|w_s| = k^j$ for all $s$. Since $w$ is a complete aligned block of size $k^{j+1}$ of $x[:t]$, every $w_s$ is a complete aligned block of size $k^j$ of $x[:t]$, and hence $w_s \in \mathcal{B}_j(t)$ (for $j = 0$, this holds trivially). Let $r \coloneqq r_j(t)$. By equation~\eqref{eq:hdp_offsets},
\begin{equation}
x[\lfloor t \rfloor_{j+1} : t] = x[\lfloor t \rfloor_{j+1} : \lfloor t \rfloor_j] \cdot x[\lfloor t \rfloor_j : t]
\label{eq:hdp_split}
\end{equation}
where the first factor on the right-hand side has length $k^j r$, and the second one has length less than $k^j$. Hence, $x[\lfloor t \rfloor_{j+1} : t] \sqsubseteq w$ implies $x[\lfloor t \rfloor_j : t] \sqsubseteq w_r$, and therefore $w_r \in \mathcal{W}_j(t)$. It follows that the set of $j \in \mathbb{N}$ s.t.\ $\mathcal{W}_j(t) \neq \emptyset$ is an initial segment of $\mathbb{N}$. This set contains $0$, since $\mathcal{W}_0(t) = \Sigma$, and it is finite, since for $j \geq 1$ we have $\mathcal{B}_j(t) = \emptyset$ whenever $k^j > t$. Hence, $J_t$ exists.

Next, we show by induction on $j$ that if the loop reaches iteration $j$, then equation~\eqref{eq:hdp_v_char} holds at the beginning of this iteration, and the loop proceeds to iteration $j + 1$ if and only if $\mathcal{W}_{j+1}(t) \neq \emptyset$. Since the loop reaches iteration $0$, this implies that it reaches iteration $j$ if and only if $\mathcal{W}_1(t), \dots, \mathcal{W}_j(t)$ are all non-empty, i.e.\ if and only if $j \leq J_t$. For $j = 0$, line~2 sets $V = \{(a, a) \mid a \in \Sigma\}$, which agrees with equation~\eqref{eq:hdp_v_char} since $\mathcal{W}_0(t) = \Sigma$, $\mathrm{str}_0(a) = a$ and $\lfloor t \rfloor_0 = t$. Now, assume that the loop reaches iteration $j$ and that equation~\eqref{eq:hdp_v_char} holds at its beginning, and consider the set $V_{\mathrm{next}}$ that is computed in this iteration (lines~4--7). By Lemma~\ref{lem:hdp_state}, $u^{(j)}$ is a word of $r \coloneqq r_j(t)$ indices s.t.\ $\mathrm{str}_j(u^{(j)}) = x[\lfloor t \rfloor_{j+1} : \lfloor t \rfloor_j]$. Let $q'$ be an index of $D^{(j+1)}$, let $D^{(j+1)}[q'] = (c_0, \dots, c_{k-1})$, and let $w \coloneqq \mathrm{str}_{j+1}(q') = \mathrm{str}_j(c_0) \cdots \mathrm{str}_j(c_{k-1})$. By equation~\eqref{eq:hdp_split}, and since all the words $\mathrm{str}_j(c_s)$ have length $k^j$, we have $x[\lfloor t \rfloor_{j+1} : t] \sqsubseteq w$ if and only if
\begin{equation}
\mathrm{str}_j(c_0 c_1 \cdots c_{r-1}) = \mathrm{str}_j(u^{(j)}) \quad \text{and} \quad x[\lfloor t \rfloor_j : t] \sqsubseteq \mathrm{str}_j(c_r)
\end{equation}
Since $\mathrm{str}_j$ is injective (Lemma~\ref{lem:hdp_state}), the first condition is equivalent to $c_0 c_1 \cdots c_{r-1} = u^{(j)}$. Since $\mathrm{str}_j(c_r) \in \mathcal{B}_j(t)$ (Lemma~\ref{lem:hdp_state}), the second condition is equivalent to $\mathrm{str}_j(c_r) \in \mathcal{W}_j(t)$, i.e.\ (by the induction hypothesis) to $c_r$ being the first component of an element of $V$, namely of $(c_r, \mathrm{str}_j(c_r)[t - \lfloor t \rfloor_j])$. Together, the two conditions state that $u^{(j)} c \sqsubseteq D^{(j+1)}[q']$ for some $(c, b) \in V$, i.e.\ that line~7 adds $(q', b)$ to $V_{\mathrm{next}}$. In this case $b$ is unique, and since $t - \lfloor t \rfloor_{j+1} = k^j r + (t - \lfloor t \rfloor_j)$, we have $b = \mathrm{str}_j(c_r)[t - \lfloor t \rfloor_j] = w[t - \lfloor t \rfloor_{j+1}]$. Since, moreover, $w \in \mathcal{B}_{j+1}(t)$ by Lemma~\ref{lem:hdp_state}, we conclude that $V_{\mathrm{next}}$ is equal to the right-hand side of equation~\eqref{eq:hdp_v_char} with $j + 1$ in place of $j$. Since $\mathrm{str}_{j+1}$ maps the indices of $D^{(j+1)}$ onto $\mathcal{B}_{j+1}(t) \supseteq \mathcal{W}_{j+1}(t)$, it follows that $V_{\mathrm{next}} \neq \emptyset$ if and only if $\mathcal{W}_{j+1}(t) \neq \emptyset$. By lines~8--11, the loop proceeds to iteration $j+1$ if and only if $V_{\mathrm{next}} \neq \emptyset$, and in this case $V = V_{\mathrm{next}}$ at the beginning of iteration $j + 1$. This completes the induction.

Finally, since the loop reaches iteration $J_t$ but not iteration $J_t + 1$, the algorithm returns on line~9 in iteration $J_t$. By equation~\eqref{eq:hdp_v_char}, and since $\mathrm{str}_{J_t}$ is injective, the number of elements of the form $(q, a)$ in $V$ is $N_{J_t}(t, a)$ for every $a \in \Sigma$. Hence, $\hat{x}_t$ maximizes $N_{J_t}(t, \cdot)$.
\end{proof}

\paragraph{Version spaces within a factor.}
Let $x = f_0 f_1 \dots f_{z-1}$ be the $k$-LZ77 factorization of $x$. For every $i < z$, let $L_i \coloneqq |f_0 \cdots f_{i-1}|$ be the starting position of $f_i$, and let
\begin{equation}
T_i \coloneqq \{L_i, L_i + 1, \dots, L_i + |f_i| - 1\}
\end{equation}
be the set of rounds in which the symbols of $f_i$ are predicted. The sets $T_0, \dots, T_{z-1}$ partition $\{0, \dots, n-1\}$.

Consider a factor $f_i$ of level $h \geq 1$. By Definition~\ref{def:k_lz77_type}, $f_i$ is a copy factor, i.e.\ a prefix of a complete aligned block of size $k^h$ that occurs earlier in $x$. We call such a block a \emph{source block} of $f_i$. A source block belongs to the version space of level $h$ in every round of $f_i$, and therefore the vote in these rounds always takes place at some level $j \geq h$. For every such $j$, all the rounds of $f_i$ lie in the same aligned block of size $k^j$. Hence, during $f_i$, the version space of level $j$ evolves as in the plurality algorithm: in every round, the words that disagree with the revealed symbol are removed, and no words are added. The following lemma makes this precise.

\begin{lemma}
\label{lem:hdp_factor}
Let $i < z$ be s.t.\ the level $h$ of $f_i$ is at least $1$. Then:
\begin{enumerate}
\item For every $t \in T_i$, we have $\mathcal{W}_h(t) \neq \emptyset$, and hence $J_t \geq h$.
\item For every $j \geq h$ and every $t$ s.t.\ $t, t + 1 \in T_i$,
\begin{equation}
\mathcal{W}_j(t+1) = \{w \in \mathcal{W}_j(t) \mid w[t - \lfloor t \rfloor_j] = x[t]\}
\label{eq:hdp_shrink}
\end{equation}
In particular, $\mathcal{W}_j(t+1) \subseteq \mathcal{W}_j(t)$ and $|\mathcal{W}_j(t+1)| = N_j(t, x[t])$. Moreover, if $J_t = j$ and $\hat{x}_t \neq x[t]$, then $|\mathcal{W}_j(t+1)| \leq \frac{1}{2}|\mathcal{W}_j(t)|$.
\end{enumerate}
\end{lemma}

\begin{proof}
Since $h \geq 1$, the factor $f_i$ is a copy factor. That is, $L_i = k^h l_i$ for some $l_i \in \mathbb{N}$, and there exists $p < l_i$ s.t.\ $f_i \sqsubseteq \beta \coloneqq x[k^h p : k^h(p+1)]$, i.e.\ $\beta$ is a source block of $f_i$. Moreover, $|f_i| \leq k^h$, so that all the positions in $T_i$ belong to the aligned block of size $k^h$ that starts at $L_i$.

Let $t \in T_i$. Since $k^h(p+1) \leq k^h l_i = L_i \leq t$, the source block $\beta$ is a complete aligned block of size $k^h$ of $x[:t]$, i.e.\ $\beta \in \mathcal{B}_h(t)$. Moreover, $\lfloor t \rfloor_h = L_i$, and $x[L_i : t]$ is a prefix of $f_i$, and hence of $\beta$. Therefore, $\beta \in \mathcal{W}_h(t)$, and $J_t \geq h$ by Lemma~\ref{lem:hdp_predict}. This proves the first claim.

Now, let $j \geq h$, and let $t$ be s.t.\ $t, t+1 \in T_i$. Then $L_i < t + 1 < L_i + k^h$, and since $L_i$ is a multiple of $k^h$, it follows that $t + 1$ is not a multiple of $k^h$, and hence not a multiple of $k^j$. Therefore, no aligned block of size $k^j$ ends at position $t$, so that $\mathcal{B}_j(t+1) = \mathcal{B}_j(t)$, and moreover $\lfloor t+1 \rfloor_j = \lfloor t \rfloor_j$. Denote $P \coloneqq \lfloor t \rfloor_j$. For every $w \in \mathcal{B}_j(t)$, since $|w| = k^j > t - P$, we have $x[P : t+1] \sqsubseteq w$ if and only if $x[P : t] \sqsubseteq w$ and $w[t - P] = x[t]$. This proves equation~\eqref{eq:hdp_shrink}, and the next two claims follow from it and from equation~\eqref{eq:hdp_votes}.

Finally, assume that $J_t = j$ and $\hat{x}_t \neq x[t]$. By Lemma~\ref{lem:hdp_predict}, $N_j(t, \hat{x}_t) \geq N_j(t, x[t])$. The words that vote for $\hat{x}_t$ and the words that vote for $x[t]$ form disjoint subsets of $\mathcal{W}_j(t)$, and therefore
\begin{equation}
|\mathcal{W}_j(t)| \geq N_j(t, \hat{x}_t) + N_j(t, x[t]) \geq 2 N_j(t, x[t]) = 2 |\mathcal{W}_j(t+1)|
\end{equation}
\end{proof}

\paragraph{Theorem~\ref{thm:hdp_bounds} (Part 1):} The number of mistakes is bounded by:
\begin{equation}
M_{\mathrm{HDP}}(x) = O(km \log(m+1) (\log n)^2)
\end{equation}

\begin{proof}
Let $x \in \Sigma^n$ and $m = \mathrm{AC}_k(x)$. If $n \leq 1$, then $M_{\mathrm{HDP}}(x) \leq 1$, and there is nothing to prove, since $O(\cdot)$ hides an additive constant (see Section~\ref{sec:notation}). Hence, we assume that $n \geq 2$. Note that $m \geq 1$, since a DFA has at least one state. As above, let $x = f_0 f_1 \dots f_{z-1}$ be the $k$-LZ77 factorization of $x$. For every $i < z$, let $\mu_i$ be the number of mistakes made in the rounds $t \in T_i$, so that $M_{\mathrm{HDP}}(x) = \sum_{i < z} \mu_i$. We will show that
\begin{equation}
\mu_i \leq \max\{1, \log_k n \cdot (1 + \log m)\}
\label{eq:hdp_factor_mistakes}
\end{equation}

If the level of $f_i$ is $0$, then $|f_i| = 1$, so that $T_i$ consists of a single round, and $\mu_i \leq 1$. (An earlier occurrence of the symbol $f_i$ does not help here, since the version space of level $0$ is always $\Sigma$.)

Now, assume that the level $h$ of $f_i$ is at least $1$. For every $j \in \mathbb{N}$, let $\mu_{i,j}$ be the number of mistakes made in the rounds $t \in T_i$ with $J_t = j$, so that $\mu_i = \sum_{j} \mu_{i,j}$. By the first claim of Lemma~\ref{lem:hdp_factor}, $\mu_{i,j} = 0$ for $j < h$. Moreover, if $J_t = j \geq 1$ for some round $t$, then $\mathcal{W}_j(t) \neq \emptyset$, and hence $x[:t]$ contains a complete aligned block of size $k^j$, i.e.\ $k^j \leq t \leq n - 1$. Therefore, $\mu_{i,j} = 0$ unless $h \leq j \leq \log_k(n-1)$, and since $h \geq 1$, there are at most $\log_k n$ such levels $j$.

We now show that $\mu_{i,j} \leq 1 + \log m$ for every $j \geq h$. Let $\mu \coloneqq \mu_{i,j}$, assume that $\mu \geq 1$, and let $t_1 < t_2 < \dots < t_\mu$ be the rounds $t \in T_i$ with $J_t = j$ in which HDP makes a mistake. For every $s < \mu$, all the rounds $t_s, t_s + 1, \dots, t_{s+1}$ belong to $T_i$, since $T_i$ is an interval. Hence, by the second claim of Lemma~\ref{lem:hdp_factor}, applied with $t = t_s + 1, \dots, t_{s+1} - 1$ for the first inequality and with $t = t_s$ for the second one, we have
\begin{equation}
|\mathcal{W}_j(t_{s+1})| \leq |\mathcal{W}_j(t_s + 1)| \leq \frac{1}{2}|\mathcal{W}_j(t_s)|
\end{equation}
It follows that $|\mathcal{W}_j(t_\mu)| \leq 2^{-(\mu - 1)}|\mathcal{W}_j(t_1)|$. On the one hand, $|\mathcal{W}_j(t_\mu)| \geq 1$, since $J_{t_\mu} = j$. On the other hand, $\mathcal{W}_j(t_1) \subseteq \mathcal{B}_j(t_1)$, which is a set of complete aligned blocks of size $k^j$ of $x$. By Lemma~\ref{lem:block}, applied to a DFA with $m$ states that generates $x$ (see the remark following Lemma~\ref{lem:block}), there are at most $m$ distinct such blocks, and hence $|\mathcal{W}_j(t_1)| \leq m$. Therefore, $2^{\mu - 1} \leq m$, i.e.\ $\mu_{i,j} \leq 1 + \log m$.

Summing over the at most $\log_k n$ levels $j$ s.t.\ $\mu_{i,j} \neq 0$, we get $\mu_i \leq \log_k n \cdot (1 + \log m)$, which completes the proof of inequality~\eqref{eq:hdp_factor_mistakes}. Summing this inequality over all factors and using Lemma~\ref{lem:k_lz77}, we get
\begin{equation}
\begin{aligned}
M_{\mathrm{HDP}}(x) &\leq z \cdot \max\{1, \log_k n \cdot (1 + \log m)\} \\
&\leq km (\lceil \log_k n \rceil + 1) \cdot \max\{1, \log_k n \cdot (1 + \log m)\}
\end{aligned}
\label{eq:hdp_explicit}
\end{equation}

Finally, we simplify this bound. Since $n \geq 2$, $m \geq 1$ and $k \geq 2$, we have $\log n \geq 1$, $\log(m+1) \geq 1$ and $\log_k n \leq \log n$. Moreover, $1 + \log m = \log(2m) \leq \log((m+1)^2) = 2 \log(m+1)$. Hence, $\lceil \log_k n \rceil + 1 \leq \log n + 2 \leq 3 \log n$ and $\max\{1, \log_k n \cdot (1 + \log m)\} \leq 2 \log n \log(m+1)$. Substituting these bounds into inequality~\eqref{eq:hdp_explicit}, we conclude that
\begin{equation}
M_{\mathrm{HDP}}(x) \leq 6km \log(m+1) (\log n)^2
\end{equation}
\end{proof}

\subsection{Running Time}
\label{sec:hdp_time}

Finally, we bound the time that HDP takes in every round. In particular, Proposition~\ref{prp:hdp_time} below verifies the first two conditions of Definition~\ref{def:comp_eff}, namely that the prediction and the update can be computed in time polynomial in $|s_t|$ (and in $k$, which HDP receives as a parameter). Together with the compression bound, this shows that HDP is computationally efficient.

\begin{proposition}
\label{prp:hdp_time}
Let $x \in \Sigma^n$ with $n \geq 2$, and let $m \coloneqq \mathrm{AC}_k(x)$ and $\sigma \coloneqq |\Sigma|$. Then, in every round $t < n$, HDP computes the prediction $\hat{x}_t$ (Algorithm~\ref{alg:hdp_predict}) in $O(\sigma k m^2 \log n)$ time, and the update of $s_t$ with $x[t]$ (Algorithm~\ref{alg:hdp_update}) in $O(km \log n)$ time. Moreover, these computations take $O(\sigma k |s_t|^3)$ and $O(k |s_t|^2)$ time, respectively. Here, time is measured in the word RAM model with words of $\Theta(\log n + \log \sigma)$ bits.
\end{proposition}

\begin{proof}
Fix a round $t < n$. As in the proof of the compression bound, let $J$ be the highest level $j$ s.t.\ $D^{(j)}$ is non-empty in $s_t$, or $J = 0$ if all the dictionaries are empty. Moreover, let $d \geq 1$ be s.t.\ $|D^{(j)}| \leq d$ for every $j \geq 1$. We first show that the prediction takes $O(\sigma k d^2 (J+1))$ time, and the update takes $O(kd(J+1))$ time.

Algorithm~\ref{alg:hdp_predict} returns at the latest in iteration $J$ of the loop on line~3, since $D^{(J+1)}$ is empty, and hence $V_{\mathrm{next}} = \emptyset$ in this iteration. In iteration $0$, the set $V$ has $\sigma$ elements, and in every later iteration $j$, its elements are pairs $(q, b)$ s.t.\ $q$ is an index of $D^{(j)}$ and $b \in \Sigma$, so that $|V| \leq \sigma d$. For every element of $V$, line~6 loops over the at most $d$ indices of $D^{(j+1)}$, and for each of them, the test $u^{(j)} q \sqsubseteq D^{(j+1)}[\mathrm{idx}]$ takes $O(k)$ time, since $|u^{(j)}| < k$. Hence, every iteration takes $O(\sigma k d^2)$ time, including the plurality vote on line~9, which takes $O(|V| + \sigma)$ time.

In $s_t$, the buffer $u^{(J+1)}$ is empty, since its entries are indices into the empty dictionary $D^{(J+1)}$. Hence, Algorithm~\ref{alg:hdp_update} returns at the latest in iteration $J + 1$ of the loop on line~3, and executes lines~8--14 in at most $J + 1$ iterations. In each of them, line~10 compares $\mathrm{tuple}$ with each of the at most $d$ entries of $D^{(j+1)}$, which takes $O(kd)$ time, and the other lines take $O(k)$ time.

By the proof of the compression bound, we can take $d = m$, and moreover $J \leq \log_k n \leq \log n$. Since $n \geq 2$, we have $J + 1 \leq 2 \log n$, which proves the first claim.

For the second claim, recall the encoding of $s_t$ from the proof of the compression bound. It writes the $J + 1$ integers $J, |D^{(1)}|, \dots, |D^{(J)}|$ using at least one bit each, and hence $J + 1 \leq |s_t|$. Moreover, if $|D^{(j)}| \geq 2$ for some $j \geq 1$, then $|D^{(j-1)}| \geq 2$, since the entries of $D^{(j)}$ are distinct $k$-tuples of indices into $D^{(j-1)}$. In this case, the encoding writes each of these indices using at least one bit, and hence $|D^{(j)}| \leq |s_t|$. Since $|s_t| \geq 1$, we can therefore take $d = |s_t|$, and the second claim follows.
\end{proof}

\section{Lempel-Ziv Plurality}
\label{apx:lzp}

In this appendix, we define the Lempel-Ziv Plurality (LZP) algorithm and provide the proof for Theorem~\ref{thm:lzp_bounds}.

\subsection{Algorithm Description}

\subsubsection{State}

The state of LZP is a representation of the past sequence $x$ via its LZ77 factorization. Let the LZ77 factorization be $x = f_0 f_1 \dots f_{z-1}$. The representation is a list $R \in (\Sigma \sqcup (\mathbb{N} \times \mathbb{N}))^*$ with an entry for every $i < z$. Let $k_i \coloneqq |f_i|$. Then,
\begin{itemize}
\item If $k_i = 1$, we just store the symbol $R[i] \coloneqq f_i[0]$.
\item If $k_i > 1$ and $l_i \coloneqq |f_0 f_1 \dots f_{i-1}|$, we store the pair $R[i] \coloneqq (j, k_i)$, where $j < l_i$ is any s.t.
\begin{equation}
x[j : j + k_i] = f_i
\end{equation}
\end{itemize}

Initially, $R$ is the empty list.

\subsubsection{The \texorpdfstring{$\delta$}{δ}-SA Compressed Index}

To efficiently implement the algorithm while satisfying the space and time constraints defined in Section~\ref{sec:setting}, we utilize the $\delta$-SA data structure proposed by \citet{Kempa2023}, together with two simpler data structures from the same work. The $\delta$-SA is a compressed representation of the so-called suffix array and its inverse, designed specifically for highly repetitive sequences.

All three data structures can be constructed from the LZ77 factorization of $x$, represented as above, in time polynomial in $\operatorname{LZC}(x)$ and $\log|x|$ (see Proposition~\ref{prp:delta_sa} below). This property is crucial, as it allows us to build them directly from the compressed state without ever expanding the sequence to its full length.

To rigorously define the queries supported by the index, we first define the \emph{lexicographical rank}. We assume a fixed total order on the alphabet $\Sigma$. The \emph{standard lexicographical order} on $\Sigma^*$, denoted $\leq_{\mathrm{lex}}$, is defined as follows: for any two distinct strings $u, v \in \Sigma^*$, we say $u <_{\mathrm{lex}} v$ if and only if either $u$ is a proper prefix of $v$ (i.e.\ $u \sqsubset v$), or there exists an index $k < \min(|u|, |v|)$ such that $u[:k] = v[:k]$ and $u[k] < v[k]$.

Fix some $x \in \Sigma^n$. We will use the shorthand notation $x[i:] \coloneqq x[i:n]$ for suffixes. The lexicographical rank of a suffix $x[i:]$, denoted $\mathrm{rank}(i)$, is its position among all suffixes of $x$ sorted by $\leq_{\mathrm{lex}}$. Formally:
\begin{equation}
\mathrm{rank}(i) = |\{j \in \{0, \dots, n-1\} \mid x[j:] <_{\mathrm{lex}} x[i:]\}|
\end{equation}
Thus, the ranks form a permutation of $\{0, \dots, n-1\}$. Together, these data structures support the following stringological queries:

\begin{itemize}
\item \textbf{Random Access (RA):} Given $i \in \{0, \dots, n-1\}$, the query $\mathrm{RA}[i]$ returns $x[i]$.

\item \textbf{Inverse Suffix Array (ISA):} Given a position $i \in \{0, \dots, n-1\}$, the query $\mathrm{ISA}[i]$ returns the lexicographical rank of the suffix starting at $i$. That is, $\mathrm{ISA}[i] \coloneqq \mathrm{rank}(i)$. This allows the predictor to determine the lexicographical order of different histories efficiently.

\item \textbf{Suffix Array (SA):} Given a rank $r \in \{0, \dots, n-1\}$, the query $\mathrm{SA}[r]$ returns the starting position $i$ of the suffix whose lexicographical rank is $r$. Formally, $\mathrm{SA}[r] = i \Leftrightarrow \mathrm{rank}(i) = r$. This serves as the inverse operation to the ISA query.

\item \textbf{Longest Common Extension (LCE):} Given two positions $i, j \in \{0, \dots, n-1\}$, the query $\mathrm{LCE}(i, j)$ returns the length of the longest common prefix of the suffixes $x[i:n]$ and $x[j:n]$. Formally:
\begin{equation}
\mathrm{LCE}(i, j) \coloneqq \max\{0 \leq k \leq \min(n-i, n-j) \mid x[i:i+k] = x[j:j+k]\}
\end{equation}
\end{itemize}

\begin{proposition}
\label{prp:delta_sa}
Fix a constant $\epsilon \in (0, 1)$. Let $x \in \Sigma^n$ and $z \coloneqq \operatorname{LZC}(x)$. Given the representation $R$ of the LZ77 factorization of $x$, it is possible to construct, in $O(z (\log n)^7)$ time, data structures that answer RA and LCE queries in $O(\log n)$ time, and SA and ISA queries in $O((\log n)^{4+\epsilon})$ time. Here, time is measured in the word RAM model with words of $\Theta(\log n)$ bits, and the alphabet size $\sigma$ is regarded as a constant.
\end{proposition}

\begin{proof}
We use three data structures of \citet{Kempa2023}, citing the full version of their paper \citep{Kempa2023full} for the precise statements: the $\delta$-SA itself, which answers SA and ISA queries in $O((\log n)^{4+\epsilon})$ time (their Theorem~1.1), and the data structures of their Theorems~5.24 and~5.25, which answer RA and LCE queries, respectively, in $O(\log n)$ time. \citet{Kempa2023full} use the same model of computation, and define the LZ77 factorization in the same way as Definition~\ref{def:lz77_main}. Given this factorization, each of the three data structures can be constructed in $O(\delta (\log n)^7)$ time, where $\delta \coloneqq \max_{1 \leq \ell \leq n} d_\ell / \ell$ is the \emph{substring complexity} of $x$ \citep{Kociumaka2022}, and $d_\ell$ is the number of distinct substrings of $x$ of length $\ell$. It is known that $\delta \leq z$ (see e.g.\ \citealt[Section~3]{Navarro2022}). Indeed, the leftmost occurrence of any substring of $x$ contains the last position of some LZ77 factor (otherwise, it would lie inside a copy factor without containing its last position, and hence it would have an earlier occurrence inside the source of this factor). Since each of these $z$ positions is contained in at most $\ell$ intervals of length $\ell$, we get $d_\ell \leq \ell z$.

Finally, the input format of \citet{Kempa2023full} differs from $R$ only in that it also specifies an earlier occurrence for every copy factor of length $1$. Since the first occurrence of every symbol is a literal factor, these occurrences can be filled in by a left-to-right scan over $R$ in $O(z)$ time.
\end{proof}

For our purposes, the utility of the RA, ISA, SA and LCE queries is that they enable the following two queries.

\begin{definition}[Internal Pattern Count]
\label{def:ipc}
For a fixed $x \in \Sigma^n$, given any $i < n$ and $a \in \Sigma$, we define
\begin{equation}
\mathrm{IPC}[i, a] \coloneqq |\{j < n \mid x[i:]\, a \sqsubseteq x[j:]\}|
\end{equation}
\end{definition}

\begin{definition}[Internal Pattern Match]
\label{def:ipm}
For a fixed $x \in \Sigma^n$, given any $i < n$ and $a \in \Sigma$, we define $\mathrm{IPM}[i, a]$ to be any $j$ s.t.\ $x[i:]\, a \sqsubseteq x[j:]$, or $\bot$ if there is no such $j$.
\end{definition}

\begin{lemma}
\label{lem:ipc_ipm}
Given oracle access to RA, ISA, SA and LCE, $\mathrm{IPC}[i, a]$ and $\mathrm{IPM}[i, a]$ can be computed using $O(\log n)$ oracle queries and $O(\log n)$ additional time.
\end{lemma}

\begin{proof}
To compute the count ($\mathrm{IPC}$) or the location ($\mathrm{IPM}$) of the pattern $x[i:]\, a$, we identify the contiguous range of suffixes in the Suffix Array that begin with this pattern. The pattern consists of the suffix $x[i:]$ followed by the symbol $a$. Since $x[i:]$ is itself a suffix of $x$, its position in the Suffix Array is given exactly by $\mathrm{ISA}[i]$.

Crucially, because $x[i:]$ is a proper prefix of any other suffix $x[j:]$ that matches the pattern (i.e., $x[i:] \sqsubset x[j:]$), $x[i:]$ is lexicographically strictly smaller than any such $x[j:]$. Consequently, the index $\mathrm{ISA}[i]$ marks the inclusive lower bound of the range of suffixes starting with $x[i:]$.

The algorithm proceeds in two stages:
\begin{enumerate}
\item We determine the upper bound of the range of suffixes starting with $x[i:]$ by performing a binary search to the right of $\mathrm{ISA}[i]$. We use the LCE oracle to verify if a candidate suffix $x[j:]$ extends $x[i:]$.
\item Within this identified range, we perform two additional binary searches to isolate the sub-range where the character at offset $n - i$ (the position immediately following the prefix $x[i:]$) matches $a$.
\end{enumerate}

The range $[r_{\mathrm{start}}, r_{\mathrm{end}}]$ returned by this procedure corresponds to all occurrences of $x[i:]\, a$. For $\mathrm{IPC}$, we return the size of this range. For $\mathrm{IPM}$, we return $\mathrm{SA}[r_{\mathrm{start}}]$ (or $\bot$ if the range is empty). The core logic is detailed in Algorithm~\ref{alg:range_search}, and the wrapper queries in Algorithm~\ref{alg:ipc_ipm}. Since Algorithm~\ref{alg:range_search} consists of three binary searches over ranges of at most $n$ ranks, it is easy to see that it (and hence each procedure of Algorithm~\ref{alg:ipc_ipm}) makes $O(\log n)$ oracle queries and spends $O(\log n)$ additional time.
\end{proof}

\begin{algorithm}
\centering
\renewcommand{\arraystretch}{1.05} 
\begin{tabular}{@{}r@{\quad}l@{}}
\toprule
1 & \textbf{Input:} index $i < n$, symbol $a \in \Sigma$ \\
2 & $L \leftarrow n - i$ \\
3 & $R_{\mathrm{start}} \leftarrow \mathrm{ISA}[i]$ \\
4 & $\mathrm{left} \leftarrow R_{\mathrm{start}}$, $\mathrm{right} \leftarrow n - 1$ \\
5 & $R_{\mathrm{limit}} \leftarrow R_{\mathrm{start}}$ \\
6 & \textbf{while} $\mathrm{left} \leq \mathrm{right}$ \\
7 & \quad $\mathrm{mid} \leftarrow \lfloor(\mathrm{left} + \mathrm{right}) / 2\rfloor$ \\
8 & \quad $j \leftarrow \mathrm{SA}[\mathrm{mid}]$ \\
9 & \quad \textbf{if} $\mathrm{LCE}(i, j) \geq L$ \\
10 & \quad\quad $R_{\mathrm{limit}} \leftarrow \mathrm{mid}$ \\
11 & \quad\quad $\mathrm{left} \leftarrow \mathrm{mid} + 1$ \\
12 & \quad \textbf{else} \\
13 & \quad\quad $\mathrm{right} \leftarrow \mathrm{mid} - 1$ \\
14 & \textbf{if} $R_{\mathrm{limit}} = R_{\mathrm{start}}$ \\
15 & \quad \textbf{return} $\bot$ \\
16 & $r_{\mathrm{start}} \leftarrow \bot$, $r_{\mathrm{end}} \leftarrow \bot$ \\
17 & $\mathrm{left} \leftarrow R_{\mathrm{start}} + 1$, $\mathrm{right} \leftarrow R_{\mathrm{limit}}$ \\
18 & \textbf{while} $\mathrm{left} \leq \mathrm{right}$ \\
19 & \quad $\mathrm{mid} \leftarrow \lfloor(\mathrm{left} + \mathrm{right}) / 2\rfloor$ \\
20 & \quad $j \leftarrow \mathrm{SA}[\mathrm{mid}]$ \\
21 & \quad $c \leftarrow \mathrm{RA}[j + L]$ \\
22 & \quad \textbf{if} $c \geq a$ \\
23 & \quad\quad $r_{\mathrm{start}} \leftarrow \mathrm{mid}$ \\
24 & \quad\quad $\mathrm{right} \leftarrow \mathrm{mid} - 1$ \\
25 & \quad \textbf{else} \\
26 & \quad\quad $\mathrm{left} \leftarrow \mathrm{mid} + 1$ \\
27 & \textbf{if} $r_{\mathrm{start}} = \bot$ \textbf{or} $\mathrm{RA}[\mathrm{SA}[r_{\mathrm{start}}] + L] \neq a$ \\
28 & \quad \textbf{return} $\bot$ \\
29 & $\mathrm{left} \leftarrow r_{\mathrm{start}}$, $\mathrm{right} \leftarrow R_{\mathrm{limit}}$ \\
30 & \textbf{while} $\mathrm{left} \leq \mathrm{right}$ \\
31 & \quad $\mathrm{mid} \leftarrow \lfloor(\mathrm{left} + \mathrm{right}) / 2\rfloor$ \\
32 & \quad $j \leftarrow \mathrm{SA}[\mathrm{mid}]$ \\
33 & \quad $c \leftarrow \mathrm{RA}[j + L]$ \\
34 & \quad \textbf{if} $c \leq a$ \\
35 & \quad\quad $r_{\mathrm{end}} \leftarrow \mathrm{mid}$ \\
36 & \quad\quad $\mathrm{left} \leftarrow \mathrm{mid} + 1$ \\
37 & \quad \textbf{else} \\
38 & \quad\quad $\mathrm{right} \leftarrow \mathrm{mid} - 1$ \\
39 & \textbf{return} $(r_{\mathrm{start}}, r_{\mathrm{end}})$ \\
\bottomrule
\end{tabular}
\caption{Pattern range search via suffix array oracle}
\label{alg:range_search}
\end{algorithm}

\begin{algorithm}
\centering
\renewcommand{\arraystretch}{1.15}
\begin{tabular}{@{}r@{\quad}l@{}}
\toprule
1 & \textbf{Procedure} $\mathrm{IPC}(i, a)$ \\
2 & \quad $\mathrm{range} \leftarrow \mathrm{PatternRangeSearch}(i, a)$ \\
3 & \quad \textbf{if} $\mathrm{range} = \bot$ \\
4 & \quad\quad \textbf{return} $0$ \\
5 & \quad \textbf{else} \\
6 & \quad\quad $(r_{\mathrm{start}}, r_{\mathrm{end}}) \leftarrow \mathrm{range}$ \\
7 & \quad\quad \textbf{return} $r_{\mathrm{end}} - r_{\mathrm{start}} + 1$ \\
8 & \textbf{Procedure} $\mathrm{IPM}(i, a)$ \\
9 & \quad $\mathrm{range} \leftarrow \mathrm{PatternRangeSearch}(i, a)$ \\
10 & \quad \textbf{if} $\mathrm{range} = \bot$ \\
11 & \quad\quad \textbf{return} $\bot$ \\
12 & \quad \textbf{else} \\
13 & \quad\quad $(r_{\mathrm{start}}, r_{\mathrm{end}}) \leftarrow \mathrm{range}$ \\
14 & \quad\quad \textbf{return} $\mathrm{SA}[r_{\mathrm{start}}]$ \\
\bottomrule
\end{tabular}
\caption{IPC and IPM queries}
\label{alg:ipc_ipm}
\end{algorithm}

\subsubsection{Update Rule}

The Lempel-Ziv Plurality predictor maintains the LZ77 factorization of the sequence processed so far. When a new symbol $a$ is received, the algorithm must update this factorization efficiently.

Recall that the state is represented by the list $R$. Let the current sequence be $x$ with factorization $f_0 f_1 \dots f_{z-1}$. The new symbol $a$ effectively asks whether the last factor $f_{z-1}$ can be extended to include $a$ while satisfying the LZ77 constraints, or if $a$ must begin a new factor $f_z$.

Using the methods of the previous section, specifically the internal pattern match query IPM, we can resolve this decision efficiently. Indeed, extending the last factor $f_{z-1}$ by $a$ is valid if and only if the concatenated string $f_{z-1} a$ appears previously in $x$.

The update procedure is formally described in Algorithm~\ref{alg:lzp_update}.

\begin{algorithm}
\centering
\renewcommand{\arraystretch}{1.15}
\begin{tabular}{@{}r@{\quad}l@{}}
\toprule
1 & \textbf{Input:} LZP state list $R$, new symbol $a \in \Sigma$ \\
2 & \textbf{if} $R$ is empty \\
3 & \quad $R.\mathrm{append}(a)$ \\
4 & \quad \textbf{return} \\
5 & $z \leftarrow R.\mathrm{len}$ \\
6 & $\mathrm{entry} \leftarrow R[z-1]$ \\
7 & \textbf{if} ``entry'' is pair $(j, k)$ \\
8 & \quad $\mathrm{len} \leftarrow k$ \\
9 & \textbf{else} \\
10 & \quad $\mathrm{len} \leftarrow 1$ \\
11 & $n \leftarrow \sum_{e \in R} (\text{if } e \text{ is pair then } e.k \text{ else } 1)$ \\
12 & $\mathrm{pos} \leftarrow n - \mathrm{len}$ \\
13 & Build the data structures of Proposition~\ref{prp:delta_sa} from $R$ \\
14 & $\mathrm{match} \leftarrow \mathrm{IPM}(\mathrm{pos}, a)$ \\
15 & \textbf{if} $\mathrm{match} \neq \bot$ \\
16 & \quad $R[z-1] \leftarrow (\mathrm{match}, \mathrm{len} + 1)$ \\
17 & \textbf{else} \\
18 & \quad $R.\mathrm{append}(a)$ \\
\bottomrule
\end{tabular}
\caption{LZP update}
\label{alg:lzp_update}
\end{algorithm}

\subsubsection{Prediction Rule}

The prediction phase of the Lempel-Ziv Plurality algorithm relies on a ``plurality vote'' based on the history of the last LZ77 factor. Let $x$ be the sequence processed so far, and let $f_{z-1}$ be the most recent factor in its LZ77 factorization. To predict the next symbol, the algorithm considers all previous occurrences of $f_{z-1}$ in $x$ and examines the symbol immediately following each occurrence. The predicted symbol $\hat{x}_t$ is the one that appears most frequently as a continuation of $f_{z-1}$.

Formally, let $w = f_{z-1}$ and $n = |x|$. We consider the set of indices $V$ where the pattern $w$ appeared previously:
\begin{equation}
V \coloneqq \{j < n - |w| \mid x[j : j + |w|] = w\}
\end{equation}
For each symbol $a \in \Sigma$, we count the ``votes'' from these occurrences:
\begin{equation}
N_a \coloneqq |\{j \in V \mid x[j + |w|] = a\}|
\end{equation}
The algorithm outputs $u = \arg\max_{a \in \Sigma} N_a$.

Efficiently computing these counts is possible using the internal pattern count query defined in Algorithm~\ref{alg:ipc_ipm}. Observe that querying $\mathrm{IPC}[\mathrm{pos}, a]$, where ``pos'' is the starting index of the last factor $f_{z-1}$, returns exactly the number of times the suffix $x[\mathrm{pos}:]\, a$ (which is $wa$) occurs in $x$. Since the current sequence $x$ ends precisely after $w$, any occurrence of $wa$ must effectively start at some $j < n - |w|$, thereby strictly preceding the current factor. Thus, $\mathrm{IPC}[\mathrm{pos}, a]$ is equivalent to $N_a$.

The prediction procedure is detailed in Algorithm~\ref{alg:lzp_predict}.

\begin{algorithm}
\centering
\renewcommand{\arraystretch}{1.15}
\begin{tabular}{@{}r@{\quad}l@{}}
\toprule
1 & \textbf{Input:} LZP state list $R$, alphabet $\Sigma$ \\
2 & \textbf{if} $R$ is empty \\
3 & \quad \textbf{return} arbitrary symbol \\
4 & $z \leftarrow R.\mathrm{len}$ \\
5 & $\mathrm{entry} \leftarrow R[z-1]$ \\
6 & \textbf{if} ``entry'' is pair $(j, k)$ \\
7 & \quad $\mathrm{len} \leftarrow k$ \\
8 & \textbf{else} \\
9 & \quad $\mathrm{len} \leftarrow 1$ \\
10 & $n \leftarrow \sum_{e \in R} (\text{if } e \text{ is pair then } e.k \text{ else } 1)$ \\
11 & $\mathrm{pos} \leftarrow n - \mathrm{len}$ \\
12 & Build the data structures of Proposition~\ref{prp:delta_sa} from $R$ \\
13 & $\mathrm{best\_count} \leftarrow -1$ \\
14 & $\mathrm{prediction} \leftarrow \bot$ \\
15 & \textbf{for} $a$ \textbf{in} $\Sigma$ \\
16 & \quad $\mathrm{count} \leftarrow \mathrm{IPC}(\mathrm{pos}, a)$ \\
17 & \quad \textbf{if} $\mathrm{count} > \mathrm{best\_count}$ \\
18 & \quad\quad $\mathrm{best\_count} \leftarrow \mathrm{count}$ \\
19 & \quad\quad $\mathrm{prediction} \leftarrow a$ \\
20 & \textbf{return} $\mathrm{prediction}$ \\
\bottomrule
\end{tabular}
\caption{LZP predict}
\label{alg:lzp_predict}
\end{algorithm}

\subsection{Analysis}

We now prove Theorem~\ref{thm:lzp_bounds}, establishing the compression and mistake bounds for LZP with respect to the LZ77 complexity $\operatorname{LZC}$, as well as its polynomial running time.

First, we clarify the relationship between the LZP state size and the complexity measures. Recall that the LZ77 factorization of $x$ is denoted $f_0 f_1 \dots f_{z-1}$, where $z = \operatorname{LZC}(x)$.

\subsubsection{Compression Bound}
\label{sec:lzp_compression}

\paragraph{Theorem~\ref{thm:lzp_bounds} (Part 2):} The predictor satisfies the compression bound:
\begin{equation}
S_{\mathrm{LZP}}(x) = O(m(\log n + \log \sigma))
\end{equation}

\begin{proof}
The internal state of the LZP predictor at time $t$ (where $x$ has length $n$) consists solely of the list $R$ representing the LZ77 factorization of the history. Note that while the algorithm constructs the data structures of Proposition~\ref{prp:delta_sa} to perform queries efficiently, these are rebuilt transiently at every step (in $O(m (\log n)^7)$ time, see Section~\ref{sec:lzp_time}) and are not part of the persistent state $s_t$.

Let $m = \operatorname{LZC}(x)$. The list $R$ contains exactly $m$ entries. Each entry is either a literal symbol $a \in \Sigma$ or a pair $(j, k)$ representing a backward reference.
\begin{itemize}
\item A literal requires $O(\log \sigma)$ bits.
\item A reference pair $(j, k)$ requires storing a position $j < n$ and a length $k \leq n$. This requires $O(\log n)$ bits.
\end{itemize}

Therefore, the bit size of the state $R$ is bounded by:
\begin{equation}
S_{\mathrm{LZP}}(x) = |R|_{\mathrm{bits}} \leq m \cdot O(\log n + \log \sigma)
\end{equation}

This confirms the compression bound.
\end{proof}

\subsubsection{Mistake Bound}

\paragraph{Theorem~\ref{thm:lzp_bounds} (Part 1):} The number of mistakes is bounded by:
\begin{equation}
M_{\mathrm{LZP}}(x) = O(m \log n)
\end{equation}

\begin{proof}
Let $x \in \Sigma^n$ and let its LZ77 factorization be $f_0 f_1 \dots f_{m-1}$. We analyze the mistakes committed by the LZP predictor during the processing of a single factor $f_i$.

\paragraph{Case 1: Literal Factor.}
If $f_i$ is a single symbol that has not appeared previously in $x$, it contributes at most $1$ mistake.

\paragraph{Case 2: Copied Factor.}
Suppose $f_i$ is a factor of length $L_i$ that copies a previous substring. Let $w = f_i$. By definition, there exists a position $p$ in the history ($p < \mathrm{start}(f_i)$) such that $x[p : p + L_i] = w$.

We define a ``Version Space'' $V_k$ representing the set of valid history indices consistent with the prefix of $f_i$ observed so far. Let $u_k = f_i[:k]$ be the prefix of length $k$.
\begin{equation}
V_k \coloneqq \{j < \mathrm{start}(f_i) \mid x[j : j + k] = u_k\}
\end{equation}

Obviously, $|V_0| \leq n$. Because $f_i$ is a valid LZ77 copy, the ``true'' history index $p$ is guaranteed to satisfy $p \in V_k$ for all $0 \leq k < L_i$. This ensures the version space is never empty: $|V_k| \geq 1$.

The LZP algorithm predicts the next symbol by taking a plurality vote among the extensions of indices in $V_k$. Let $x_{\mathrm{true}} = f_i[k]$ be the correct next symbol. The algorithm calculates the counts for each symbol $a \in \Sigma$:
\begin{equation}
C_a = |\{j \in V_k \mid x[j+k] = a\}|
\end{equation}
The prediction is $\hat{x} = \arg\max_{a \in \Sigma} C_a$.

The set of surviving versions $V_{k+1}$ consists exactly of those indices in $V_k$ that correctly predict the observed symbol $x_{\mathrm{true}}$:
\begin{equation}
V_{k+1} = \{j \in V_k \mid x[j+k] = x_{\mathrm{true}}\}
\end{equation}
Thus, the size of the new version space is exactly the count of the correct symbol: $|V_{k+1}| = C_{x_{\mathrm{true}}}$.

A mistake occurs if $\hat{x} \neq x_{\mathrm{true}}$. If a mistake occurs, it implies that $x_{\mathrm{true}}$ did not win the plurality vote. In particular, $x_{\mathrm{true}}$ could not have held a strict majority of the votes in $V_k$. Therefore, whenever a mistake is made, the number of surviving versions must satisfy:
\begin{equation}
|V_{k+1}| = C_{x_{\mathrm{true}}} \leq \frac{1}{2} |V_k|
\end{equation}

Let $\mu_i$ be the number of mistakes made during the processing of factor $f_i$. Since each mistake reduces the version space cardinality by a factor of at least $2$, we have:
\begin{equation}
|V_{\mathrm{final}}| \leq |V_0| \cdot \left(\frac{1}{2}\right)^{\mu_i}
\end{equation}

(Here, $V_{\mathrm{final}}$ is the set of versions at the end of the factor.)

Since the true index $p$ remains in the set throughout, we must have $|V_{\mathrm{final}}| \geq 1$. Therefore:
\begin{equation}
1 \leq n \cdot \left(\frac{1}{2}\right)^{\mu_i}
\end{equation}
Taking logarithms:
\begin{equation}
0 \leq \log n - \mu_i
\end{equation}
\begin{equation}
\mu_i \leq \log n
\end{equation}

There might be one extra mistake in the beginning, when the $f_i$ factor doesn't exist yet.

\paragraph{Total Mistakes:}
The total number of mistakes is the sum of mistakes over all $m$ factors. Since each factor contributes at most $\log n + 1$, we get
\begin{equation}
M_{\mathrm{LZP}}(x) \leq m(\log n + 1)
\end{equation}
\end{proof}

\subsubsection{Running Time}
\label{sec:lzp_time}

Finally, we verify the first two conditions of Definition~\ref{def:comp_eff}, namely that the prediction $\mathcal{P}(s_t)$ and the update $\mathcal{U}(s_t, x[t])$ can be computed in time polynomial in $|s_t|$ (the third condition is the compression bound of Section~\ref{sec:lzp_compression}). We use the same model of computation and the same constant $\epsilon$ as in Proposition~\ref{prp:delta_sa}, and regard $\sigma$ as a constant. As in the description of the update and prediction rules, let $x$ be the sequence processed so far, $n \coloneqq |x| \geq 1$ (for $n = 0$, both algorithms return immediately) and $z \coloneqq \operatorname{LZC}(x)$, so that the state is the list $R$ with $z$ entries.

Both algorithms build the data structures of Proposition~\ref{prp:delta_sa} from $R$ (line~13 of Algorithm~\ref{alg:lzp_update} and line~12 of Algorithm~\ref{alg:lzp_predict}), which takes $O(z (\log n)^7)$ time. After that, each IPC or IPM query takes $O((\log n)^{5+\epsilon})$ time: by Lemma~\ref{lem:ipc_ipm}, it consists of $O(\log n)$ oracle queries, each answered in $O((\log n)^{4+\epsilon})$ time by Proposition~\ref{prp:delta_sa}, and $O(\log n)$ additional time. Algorithm~\ref{alg:lzp_update} makes one IPM query, Algorithm~\ref{alg:lzp_predict} makes $\sigma = O(1)$ IPC queries, and both spend $O(z)$ time on their remaining steps. Hence, both algorithms run in $O(z (\log n)^7)$ time.

It remains to bound this in terms of $|s_t|$. Each of the $z$ entries of $R$ occupies at least one bit, so $|s_t| \geq z$. Moreover, the longest LZ77 factor has some length $k \geq n/z$. If $k \geq 2$, then $k$ is stored in $R$, which requires at least $\log k \geq \log(n/z)$ bits, and otherwise $n = z$. In both cases, $|s_t| \geq \max(z, \log(n/z)) \geq \frac{1}{2}(z + \log n - \log z) \geq \frac{1}{2}\log n$. Therefore, both algorithms run in $O(z (\log n)^7) = O(|s_t|^8)$ time. Together with the compression bound, this shows that LZP is computationally efficient.

\section{Base-Sensitivity of Automaticity}
\label{apx:base}

In Section~\ref{sec:automaticity} we remarked that the automaticity of a sequence is highly sensitive to the choice of base $k$. Here we show a specific example.

\begin{example}
\label{ex:base}
Let $\Sigma = \{a, b\}$. For any $k \in \mathbb{N}$, let $n \coloneqq 2^k$ and
\begin{equation}
u_k \coloneqq (a^n b^n)^n
\end{equation}

Then, $\mathrm{AC}_2(u_k) = O(k)$ and $\mathrm{AC}_3(u_k) = \Omega(2^k)$.
\end{example}

Note that $|u_k| = 2n^2 = 2^{2k+1}$.

\subsection{\texorpdfstring{Upper Bound: $\mathrm{AC}_2(u_k) = O(k)$}{Upper Bound: AC\_2(u\_k) = O(k)}}

\begin{proof}
We construct a DFA $M$ over input alphabet $[2]$ with $O(k)$ states that computes $u_k[t]$ from $\langle t \rangle_2^{2k+1}$.

The symbol $u_k[t]$ depends only on $t \bmod 2n = t \bmod 2^{k+1}$:
\begin{equation}
u_k[t] = \begin{cases} a & \text{if } t \bmod 2^{k+1} < 2^k \\ b & \text{if } t \bmod 2^{k+1} \geq 2^k \end{cases}
\end{equation}

In the binary representation $\langle t \rangle_2^{2k+1} = w_0 w_1 \dots w_{2k}$, the residue $t \bmod 2^{k+1}$ is determined by the last $k+1$ bits $w_k w_{k+1} \dots w_{2k}$, and $t \bmod 2^{k+1} \geq 2^k$ if and only if $w_k = 1$. Hence,
\begin{equation}
u_k[t] = \begin{cases} a & \text{if } w_k = 0 \\ b & \text{if } w_k = 1 \end{cases}
\end{equation}

Define the DFA $M = (Q, s_0, \delta, \tau)$ with $Q = \{s_0, s_1, \dots, s_k, s_a, s_b\}$ ($k+3$ states), initial state $s_0$, and transitions:
\begin{itemize}
\item For $0 \leq i \leq k - 1$ and $d \in \{0, 1\}$: $\delta(s_i, d) = s_{i+1}$.
\item $\delta(s_k, 0) = s_a$ and $\delta(s_k, 1) = s_b$.
\item For $d \in \{0, 1\}$: $\delta(s_a, d) = s_a$ and $\delta(s_b, d) = s_b$.
\end{itemize}

The output function is $\tau(s_a) = a$ and $\tau(s_b) = b$ (the values of $\tau$ on the remaining states are irrelevant).

On input $\langle t \rangle_2^{2k+1} = w_0 w_1 \dots w_{2k}$, the automaton passes through $s_0, s_1, \dots, s_k$ after reading $w_0 \dots w_{k-1}$. It then reads $w_k$ and transitions to $s_a$ (if $w_k = 0$) or $s_b$ (if $w_k = 1$). The remaining bits $w_{k+1} \dots w_{2k}$ leave the state unchanged. Therefore, $M(\langle t \rangle_2^{2k+1}) = u_k[t]$ as required, and $\mathrm{AC}_2(u_k) \leq k + 3 = O(k)$.
\end{proof}

\subsection{\texorpdfstring{Lower Bound: $\mathrm{AC}_3(u_k) = \Omega(2^k)$}{Lower Bound: AC\_3(u\_k) = Omega(2\^k)}}

\begin{proof}
Let $M = (Q, q_0, \delta, \tau)$ be any DFA with input alphabet $[3]$ and output alphabet $\{a, b\}$ that computes $u_k$ in the sense of Definition~\ref{def:automaticity}, i.e., there exists $L \in \mathbb{N}$ with $|u_k| \leq 3^L$ such that $u_k[t] = M(\langle t \rangle_3^L)$ for all $t < |u_k|$. We show that $|Q| = \Omega(2^k)$.

Let $m \coloneqq \lfloor k \log_3 2 \rfloor$, so that $3^m \leq 2^k = n < 3^{m+1}$. In particular, $3^m > n/3$.

Consider the complete aligned blocks of size $3^m$ in $u_k$ (see Appendix~\ref{apx:hdp}), i.e., the subwords $u_k[i \cdot 3^m : (i+1) \cdot 3^m]$ for $i = 0, 1, \dots$ s.t.\ $(i+1) \cdot 3^m \leq |u_k|$. By Lemma~\ref{lem:block}, if two block indices $i$ and $j$ lead the DFA to the same state after reading their base-$3$ address (i.e., $\delta^*(q_0, \langle i \rangle_3^{L-m}) = \delta^*(q_0, \langle j \rangle_3^{L-m})$), then the corresponding complete blocks have identical content. Contrapositively,
\begin{equation}
|Q| \geq |\{\text{distinct complete aligned blocks of size } 3^m \text{ in } u_k\}|
\end{equation}

It remains to show that the right-hand side is $\Omega(2^k)$.

\paragraph{Structure of $u_k$.} The word $u_k = (a^n b^n)^n$ has period $2n$. The transitions from $a$ to $b$ (which we call \emph{$ab$-boundaries}) occur at positions $jn$ for odd $j$ with $1 \leq j \leq 2n - 1$. Since $3^m < n$, each aligned block of size $3^m$ contains at most one boundary.

\paragraph{Counting distinct blocks via boundary offsets.} For an $ab$-boundary at position $jn$ (with $j$ odd), the offset of the boundary within its enclosing aligned block is
\begin{equation}
r_j \coloneqq jn \bmod 3^m
\end{equation}

If $r_j \neq 0$, the block containing this boundary has the form $a^{r_j} b^{3^m - r_j}$. Two such blocks with different offsets are manifestly distinct.

We now show that all $3^m - 1$ nonzero offsets are achieved. Since $n = 2^k$ and $3^m$ is a power of $3$, we have $\gcd(n, 3^m) = 1$, so multiplication by $n$ is a bijection on $\mathbb{Z}/3^m\mathbb{Z}$. Moreover, since $3^m$ is odd, the $3^m$ odd integers $1, 3, 5, \dots, 2 \cdot 3^m - 1$ represent all distinct residue classes modulo $3^m$: indeed, if $j_1$ and $j_2$ are odd with $1 \leq j_1, j_2 \leq 2 \cdot 3^m - 1$ and $j_1 \equiv j_2 \pmod{3^m}$, then $j_1 - j_2$ is divisible by $3^m$ and even; since $|j_1 - j_2| < 2 \cdot 3^m$ and $3^m$ is odd, the only possibility is $j_1 = j_2$. Hence, the set $\{jn \bmod 3^m : j \text{ odd}, 1 \leq j \leq 2 \cdot 3^m - 1\}$ is a permutation of $\{0, 1, \dots, 3^m - 1\}$.

Since there are $n > 3^m$ odd values of $j$ in $\{1, 3, \dots, 2n - 1\}$ (there are exactly $n$ of them), and the first $3^m$ already exhaust all residues modulo $3^m$, every $r \in \{1, 2, \dots, 3^m - 1\}$ is realized. This yields $3^m - 1$ blocks of the form $a^r b^{3^m - r}$ with pairwise distinct content.

Therefore,
\begin{equation}
\mathrm{AC}_3(u_k) \geq 3^m - 1 > \frac{n}{3} - 1 = \frac{2^k - 3}{3} = \Omega(2^k)
\end{equation}
\end{proof}

\section{Properties of Straight-Line Programs}
\label{apx:slp}

In this appendix, we prove some basic properties of SLP, including Proposition~\ref{prp:slc_mon} and Proposition~\ref{prp:slc_vs_aut}.

\subsection{Relaxed Straight-Line Programs}
\label{sec:relaxed_slp}

Some of our constructions naturally produce objects that are not quite SLPs, because they contain unary rules (i.e.\ rules of length 1) for nonterminals other than the root, or nonterminals that are not needed to compute the value. We formalize such objects as follows, and show that they can always be converted into SLPs without increasing the size.

\begin{definition}
\label{def:relaxed_slp}
A \emph{relaxed SLP} over $\Sigma$ is $P = (Q, q_0, \delta)$, where $Q$ is a finite set, $q_0 \in Q$ and $\delta: Q \to \overline{Q}^*$ is acyclic (in the sense of Definition~\ref{def:slp}) and satisfies $|\delta(q)| \geq 1$ for all $q \in Q \setminus \{q_0\}$. Its value $\operatorname{val}(P)$ and size $|P|$ are defined as for SLPs.
\end{definition}

That is, compared to Definition~\ref{def:slp}, nonterminals other than $q_0$ may have unary rules, and $q_0$ is not required to be the unique element of $Q$ that doesn't appear in any rule. In particular, every SLP is a relaxed SLP.

Given a relaxed SLP $P = (Q, q_0, \delta)$, we say that $q \in Q$ is \emph{reachable} when there are $p_0, p_1 \dots p_l \in Q$ s.t.\ $p_0 = q_0$, $p_l = q$ and for all $i < l$, $p_{i+1}$ appears in $\delta(p_i)$. If all elements of $Q$ are reachable, then $q_0$ is the unique element of $Q$ that doesn't appear in any rule, as required by Definition~\ref{def:slp}. Indeed, every $q \in Q \setminus \{q_0\}$ appears in $\delta(p_{l-1})$, where $p_0, \dots, p_l$ are as above, whereas if $q_0$ appeared in $\delta(p)$ for some reachable $p$, then $\delta$ would not be acyclic.

\begin{lemma}
\label{lem:slp_normalize}
For any relaxed SLP $P$, there exists an SLP $P'$ s.t.\ $\operatorname{val}(P') = \operatorname{val}(P)$ and $|P'| \leq |P|$. In particular, $\operatorname{SLC}(\operatorname{val}(P)) \leq |P|$. Moreover, if $|\operatorname{val}(P)| \geq 2$, then $P'$ can be chosen s.t.\ all of its rules, including the rule of its root, have length at least 2.
\end{lemma}

\begin{proof}
Let $P = (Q, q_0, \delta)$, and let $R \subseteq Q$ be the set of reachable nonterminals. Every nonterminal that appears in the rule of an element of $R$ is also in $R$. Hence, $(R, q_0, \delta|_R)$ is a relaxed SLP with the same value as $P$, whose size is at most $|P|$ and all of whose nonterminals are reachable. Therefore, it is sufficient to prove the lemma for relaxed SLPs all of whose nonterminals are reachable. We do so by induction on $|Q|$.

Let us call a unary rule $\delta(q) = r$ \emph{removable} if $q \neq q_0$ or $r \in Q$. If $P$ has no removable unary rules, then the rules of all nonterminals other than $q_0$ have length at least 2, and hence $P$ is an SLP by the observation above. Moreover, the rule of $q_0$ then has length at least 2, unless it is empty or consists of a single terminal, i.e.\ unless $|\operatorname{val}(P)| \leq 1$. Hence, we can take $P' \coloneqq P$.

Otherwise, let $\delta(q) = r$ be a removable unary rule, and notice that $r \neq q$ by acyclicity. Let $P_1$ be the result of removing $q$ from $Q$, replacing every occurrence of $q$ in the rules of the other nonterminals by $r$ and, if $q = q_0$, making $r$ the new root (notice that $r \in Q$ in this case). Then:
\begin{itemize}
\item $P_1$ is acyclic, since every edge $(p, r)$ of $P_1$ that arises from the replacement corresponds to the path $p, q, r$ in $P$, and hence a cycle in $P_1$ would give rise to a cycle in $P$.
\item The rules of the remaining nonterminals keep their lengths. Hence, the rules of all nonterminals of $P_1$ other than its root are non-empty, and $|P_1| = |P| - 1$.
\item All nonterminals of $P_1$ are reachable: for any $p \in Q \setminus \{q\}$, removing $q$ from a sequence $p_0, \dots, p_l$ that witnesses the reachability of $p$ in $P$ yields such a sequence in $P_1$. (Notice that if $p_i = q$, then $i < l$ and $p_{i+1} = r$.)
\item Since $\operatorname{val}_P(q) = \operatorname{val}_P(r)$, replacing $q$ by $r$ doesn't change the value of any rule. Hence, by induction along the acyclic structure of $P_1$, $\operatorname{val}_{P_1}(p) = \operatorname{val}_P(p)$ for all $p \in \overline{Q} \setminus \{q\}$. In particular, $\operatorname{val}(P_1) = \operatorname{val}(P)$.
\end{itemize}
Thus, $P_1$ is a relaxed SLP with fewer nonterminals than $P$, all of which are reachable, and with the same value. Applying the induction hypothesis to $P_1$ yields an SLP $P'$ with the desired properties and $|P'| \leq |P_1| < |P|$.
\end{proof}

\subsection{Reduction to Binary Straight-Line Programs}

We can reduce any SLP to a ``binary'' SLP with only a mild increase in size.

\begin{proposition}
\label{prp:binary_slp}
For any SLP $P$ with $|\operatorname{val}(P)| \geq 2$, there exists an SLP $P' = (Q', q'_0, \delta')$ s.t.
\begin{itemize}
\item $\operatorname{val}(P') = \operatorname{val}(P)$
\item $|P'| \leq 2|P|$
\item For all $q \in Q'$, $|\delta'(q)| = 2$.
\end{itemize}
\end{proposition}

\begin{proof}
Let $P = (Q, q_0, \delta)$ be an SLP over $\Sigma$ with $|\operatorname{val}(P)| \geq 2$. By Lemma~\ref{lem:slp_normalize}, we can assume w.l.o.g.\ that all rules of $P$, including the rule of $q_0$, have length at least 2. We construct a binary SLP $P' = (Q', q'_0, \delta')$ with $\operatorname{val}(P') = \operatorname{val}(P)$ and $|P'| \leq 2|P|$.

We process each nonterminal $q \in Q$ in topological order (from sinks toward the source $q_0$). If $|\delta(q)| = 2$, the rule is already binary and we keep it unchanged. If $|\delta(q)| > 2$, we replace $q$ by a binary tree of fresh nonterminals.

Concretely, let $\delta(q) = r_0 r_1 \dots r_{l-1}$ where $l \geq 3$. We define a recursive binarization procedure $\mathrm{Bin}(r_0, \dots, r_{l-1})$ that returns a symbol whose value is $\operatorname{val}(r_0) \dots \operatorname{val}(r_{l-1})$:
\begin{itemize}
\item If $l = 1$, return $r_0$ itself, without creating a fresh nonterminal (whose rule would be unary). This case arises in recursive calls, e.g.\ $\mathrm{Bin}(r_0, r_1, r_2)$ calls $\mathrm{Bin}(r_0)$.
\item If $l = 2$, create a fresh nonterminal $q'$ with $\delta'(q') = r_0 r_1$ and return $q'$.
\item If $l \geq 3$, let $h = \lfloor l/2 \rfloor$. Recursively construct $q'_L = \mathrm{Bin}(r_0, \dots, r_{h-1})$ and $q'_R = \mathrm{Bin}(r_h, \dots, r_{l-1})$. Create a fresh nonterminal $q'$ with $\delta'(q') = q'_L q'_R$ and return $q'$.
\end{itemize}

We set $q_0$'s replacement in $Q'$ to be the result of this procedure applied to $\delta(q_0)$, and similarly for every nonterminal.

It remains to bound $|P'|$. The binarization of a single rule $\delta(q)$ of length $l$ produces a binary tree with $l$ leaves. Such a tree has exactly $l - 1$ internal nodes, each contributing 2 edges. Hence, the number of edges introduced is $2(l-1) \leq 2l$. Meanwhile, the original rule contributed $l$ edges. Therefore, binarizing one rule at most doubles its edge count.
\end{proof}

\subsection{Proof of Proposition~\ref{prp:slc_mon}}

\begin{proof}
Let $x, y \in \Sigma^*$ with $x \sqsubseteq y$, and let $P = (Q, q_0, \delta)$ be an optimal SLP for $y$, so $|P| = \operatorname{SLC}(y)$ and $\operatorname{val}(P) = y$. If $x = y$, there is nothing to prove. If $x$ is empty, then $\operatorname{SLC}(x) = 0$, as witnessed by the SLP with $Q = \{q_0\}$ and $\delta(q_0) = x$. Hence, we can assume that $1 \leq |x| < |y|$.

We construct a relaxed SLP $P'$ (Definition~\ref{def:relaxed_slp}) with $\operatorname{val}(P') = x$ and $|P'| \leq 2|P|$ by ``truncating'' $P$. By Lemma~\ref{lem:slp_normalize}, this implies $\operatorname{SLC}(x) \leq |P'| \leq 2\operatorname{SLC}(y)$.

For every $r \in \overline{Q}$ and $1 \leq n \leq |\operatorname{val}_P(r)|$, we recursively define a symbol $\mathrm{Trunc}(r, n)$ which evaluates to $\operatorname{val}_P(r)[:n]$:
\begin{itemize}
\item If $n = |\operatorname{val}_P(r)|$, then $\mathrm{Trunc}(r, n) \coloneqq r$ (no truncation is needed). In particular, this is always the case when $r \in \Sigma$.
\item Otherwise, $r \in Q$ and $\delta(r) = r_0 r_1 \dots r_{l-1}$. Find the unique index $j < l$ such that $\sum_{i<j} |\operatorname{val}_P(r_i)| < n \leq \sum_{i \leq j} |\operatorname{val}_P(r_i)|$, and let $n' \coloneqq n - \sum_{i<j} |\operatorname{val}_P(r_i)|$, so that $1 \leq n' \leq |\operatorname{val}_P(r_j)|$. Then, $\mathrm{Trunc}(r, n)$ is a fresh nonterminal with the rule $\delta'(\mathrm{Trunc}(r, n)) \coloneqq r_0 \dots r_{j-1}\, \mathrm{Trunc}(r_j, n')$. That is, we discard the entries $r_{j+1}, \dots, r_{l-1}$ and truncate the entry $r_j$.
\end{itemize}
Notice that when $j = 0$, the rule of $\mathrm{Trunc}(r, n)$ is unary (for example, this always happens when $|x| = 1$), which is why $P'$ is only a relaxed SLP.

The root of $P'$ is $\mathrm{Trunc}(q_0, |x|)$, which is a fresh nonterminal since $|x| < |y|$. The nonterminals of $P'$ are the elements of $Q$, with their original rules $\delta$, and the fresh nonterminals, with their rules $\delta'$. All of these rules are non-empty (in particular, the rule of $q_0$ is non-empty since $y$ is non-empty). Moreover, $P'$ is acyclic, since the rules of the elements of $Q$ only refer to elements of $\overline{Q}$, whereas the rule of each fresh nonterminal refers, besides elements of $\overline{Q}$, only to the fresh nonterminal created by the next recursive call (if any). Hence, $P'$ is indeed a relaxed SLP. By induction on the recursion, $\operatorname{val}_{P'}(\mathrm{Trunc}(r, n)) = \operatorname{val}_P(r)[:n]$, and in particular $\operatorname{val}(P') = y[:|x|] = x$.

It remains to bound the size of $P'$. The computation of $\mathrm{Trunc}(q_0, |x|)$ is a single chain of recursive calls, which visits a path in $P$ from the root toward the sinks. Since $P$ is acyclic, this path visits each nonterminal of $P$ at most once, and each visit creates at most one fresh nonterminal. The rule of the fresh nonterminal created while visiting $r \in Q$ has length $j + 1 \leq |\delta(r)|$, since we only removed entries from the right end of $\delta(r)$. Hence, the fresh rules contribute at most $\sum_{r \in Q} |\delta(r)| = |P|$ edges in total, whereas the original rules contribute $|P|$ edges. We conclude that $|P'| \leq 2|P|$.
\end{proof}

\subsection{Proof of Proposition~\ref{prp:slc_vs_aut}}

\begin{proof}
Let $x \in \Sigma^n$ and let $m \coloneqq \mathrm{AC}_k(x)$. If $n \leq 1$, then $\operatorname{SLC}(x) \leq 1$, and the claim holds since $O(\cdot)$ hides an additive constant (see Section~\ref{sec:notation}). Hence, we assume $n \geq 2$. By definition, there exist an integer $L$ with $n \leq k^L$ and a DFA $M = (Q, q_0, \delta, \tau)$ with $|Q| = m$ states, input alphabet $[k]$ and output alphabet $\Sigma$, such that $x[t] = M(\langle t \rangle_k^L)$ for all $t < n$.

Let $L' \coloneqq \lceil \log_k n \rceil$ (so that $n \leq k^{L'}$) and let $p \coloneqq L - L'$. Since every $t < n$ satisfies $t < k^{L'}$, the representation $\langle t \rangle_k^L$ begins with $p$ leading zeros. Therefore, for all $t < n$:
\begin{equation}
x[t] = \tau(\delta^*(q_0, 0^p \cdot \langle t \rangle_k^{L'})) = \tau(\delta^*(q^*, \langle t \rangle_k^{L'}))
\end{equation}

where $q^* \coloneqq \delta^*(q_0, 0^p)$. That is, $x$ is equally well generated by $M$ starting from state $q^*$ with input length $L'$.

We now construct a relaxed SLP $P$ (Definition~\ref{def:relaxed_slp}) of depth $L'$. For any state $q \in Q$ and level $0 \leq l \leq L'$, define the word
\begin{equation}
w_{q,l} \coloneqq \tau(\delta^*(q, \langle 0 \rangle_k^l)) \; \tau(\delta^*(q, \langle 1 \rangle_k^l)) \; \cdots \; \tau(\delta^*(q, \langle k^l - 1 \rangle_k^l))
\end{equation}
That is, $w_{q,l}$ is the word of length $k^l$ obtained by running $M$ from state $q$ on all $l$-digit base-$k$ inputs in order. In particular, $w_{q^*, L'}[:n] = x$.

Each word $w_{q,l}$ decomposes by the first input digit: reading digit $d \in [k]$ transitions from $q$ to $\delta(q, d)$, and the remaining $l - 1$ digits produce $w_{\delta(q,d), l-1}$. Therefore:
\begin{equation}
w_{q,l} = w_{\delta(q,0), l-1} \; w_{\delta(q,1), l-1} \; \cdots \; w_{\delta(q,k-1), l-1}
\end{equation}

We build $P$ bottom-up, introducing a symbol $A_{q,l}$ for each pair $(q, l)$ with $q \in Q$ and $0 \leq l \leq L'$, such that $\operatorname{val}_P(A_{q,l}) = w_{q,l}$.
\begin{itemize}
\item \emph{Base case ($l = 0$):} $w_{q,0} = \tau(q)$ is a single symbol, so $A_{q,0}$ is just the terminal $\tau(q) \in \Sigma$. No edges are needed.
\item \emph{Recursive case ($l \geq 1$):} We set $\delta_P(A_{q,l}) = A_{\delta(q,0), l-1} \; A_{\delta(q,1), l-1} \; \cdots \; A_{\delta(q,k-1), l-1}$. This rule has exactly $k$ edges.
\end{itemize}

The root of $P$ is $A_{q^*, L'}$, which is a nonterminal since $L' \geq 1$, and $\operatorname{val}(P) = w_{q^*, L'}$. Some of the nonterminals $A_{q,l}$ might be unreachable from the root, which is why $P$ is only a relaxed SLP.

The number of nonterminals with $l \geq 1$ is at most $m \cdot L'$, each contributing $k$ edges, so
\begin{equation}
|P| = m \cdot L' \cdot k = O(km \log_k n) = O(km \log n)
\end{equation}

By Lemma~\ref{lem:slp_normalize}, $\operatorname{SLC}(w_{q^*, L'}) \leq |P|$. Since $w_{q^*, L'}$ has $x$ as a prefix, Proposition~\ref{prp:slc_mon} gives $\operatorname{SLC}(x) \leq 2|P| = O(k \cdot \mathrm{AC}_k(x) \cdot \log|x|)$.
\end{proof}

\section{Some Classes of Sequences with Low SLC}
\label{apx:seqs}

In this appendix, we survey several well-studied classes of sequences from combinatorics on words and establish that they have slowly growing straight-line complexity. For each class, we bound $\operatorname{SLC}(x[:n])$ (or equivalently, $\operatorname{LZC}(x[:n])$ up to a factor of $\log n$) as a function of $n$, showing that the LZP algorithm achieves strong mistake bounds on these sequences.

\subsection{Automatic Sequences}

Automatic sequences are among the most extensively studied objects in combinatorics on words (see \citealt{Allouche2003} for a comprehensive treatment). A sequence $x \in \Sigma^\omega$ is called \emph{$k$-automatic} (for a fixed integer $k \geq 2$) if there exists a DFA $M$ with input alphabet $[k]$ and output alphabet $\Sigma$ such that for all $t \in \mathbb{N}$ and $m \geq \log_k t$,
\begin{equation}
x[t] = M(\langle t \rangle_k^m)
\end{equation}

By Definition~\ref{def:automaticity}, a $k$-automatic sequence $x$ satisfies $\mathrm{AC}_k(x[:n]) = O(1)$: the number of states in the minimal automaton is a constant (depending on $x$ but not on $n$). Proposition~\ref{prp:slc_vs_aut} then immediately gives:
\begin{equation}
\operatorname{SLC}(x[:n]) = O(k \cdot \mathrm{AC}_k(x) \cdot \log n) = O(\log n)
\end{equation}

Hence, by the equivalence of $\operatorname{SLC}$ and $\operatorname{LZC}$ up to logarithmic factors (Equation~\eqref{eq:slc_lzc}), we also have $\operatorname{LZC}(x[:n]) = O(\log n)$.

\subsection{Morphic Sequences}
\label{sec:morphic_examples}

Morphic sequences form one of the most important families of infinite words in combinatorics on words (see \citealt{Allouche2003}, Chapter 7). We begin by recalling the necessary definitions, then state the Lempel-Ziv complexity classification due to \citet{Constantinescu2006}.

\subsubsection{Background and Definitions}

Recall that a \emph{homomorphism} of free monoids $h: \Gamma^* \to \Gamma^*$ is a map satisfying $h(uv) = h(u)h(v)$ for all $u, v \in \Gamma^*$. Such a map is completely determined by its values on the individual symbols in $\Gamma$. We say that $h$ is \emph{non-erasing} if $|h(a)| \geq 1$ for every $a \in \Gamma$, i.e., no symbol is mapped to the empty word. A \emph{coding} is a homomorphism $\beta: \Gamma^* \to \Sigma^*$ that maps every symbol to a single letter (i.e., $|\beta(a)| = 1$ for all $a \in \Gamma$); it is equivalently just a function $\Gamma \to \Sigma$ applied symbol-by-symbol.

We say that $h$ is \emph{prolongable on a symbol $a \in \Gamma$} if $a$ is a proper prefix of $h(a)$, that is, $h(a) = aw$ for some non-empty $w \in \Gamma^*$. In this case, the iterates $a, h(a), h^2(a), \dots$ form an increasing chain of words (each is a prefix of the next), and their limit defines a unique infinite sequence $h^\omega(a) \in \Gamma^\omega$. Concretely, the $i$-th symbol of $h^\omega(a)$ is the $i$-th symbol of $h^n(a)$ for any sufficiently large $n$.

\begin{definition}
\label{def:morphic}
A sequence $y \in \Sigma^\omega$ is \emph{morphic} if there exist a finite alphabet $\Gamma$, a non-erasing homomorphism $h: \Gamma^* \to \Gamma^*$ prolongable on some $a \in \Gamma$, and a coding $\beta: \Gamma \to \Sigma$, such that
\begin{equation}
y[t] = \beta(h^\omega(a)[t]) \quad \text{for all } t \in \mathbb{N}
\end{equation}
If $\Gamma = \Sigma$ and $\beta$ is the identity, the sequence is called \emph{pure morphic}.
\end{definition}

The class of morphic sequences strictly contains the $k$-automatic sequences: a sequence is $k$-automatic if and only if it can be obtained from the above construction with $h$ being \emph{$k$-uniform}, meaning $|h(a)| = k$ for every $a \in \Gamma$ (see \citealt{Allouche2003}, Theorem 6.3.2).

\subsubsection{The Growth Function}

The asymptotic behavior of $\operatorname{SLC}$ and $\operatorname{LZC}$ for a morphic sequence is governed by the \emph{growth function} of the underlying morphism. For a letter $a \in \Gamma$ and a non-erasing homomorphism $h: \Gamma^* \to \Gamma^*$, the growth function is $h_a(n) \coloneqq |h^n(a)|$, measuring the length of the $n$-th iterate.

A classical result (see Lemma 12 in \citealt{Constantinescu2006} and citations therein) characterizes the possible asymptotic behaviors: there exist an integer $e_a \geq 0$ and an algebraic real $\rho_a \geq 1$ such that
\begin{equation}
h_a(n) = \Theta(n^{e_a} \rho_a^n)
\end{equation}
When $\rho_a > 1$, the growth is \emph{exponential} (this includes all $k$-uniform morphisms, which have $\rho_a = k$ and $e_a = 0$). When $\rho_a = 1$, the growth is \emph{polynomial} of degree $e_a$.

\subsubsection{The Constantinescu-Ilie Classification}

\citet{Constantinescu2006} gave a complete characterization of the Lempel-Ziv complexity classes for fixed points of non-erasing morphisms. We state their main result, translated into our notation.

\begin{theorem}[Constantinescu-Ilie]
\label{thm:ci_classification}
Let $w = h^\omega(a)$ be the fixed point of a non-erasing morphism $h$ prolongable on $a$, and let $h_a(n) = |h^n(a)|$ be the growth function.
\begin{enumerate}
\item If $w$ is ultimately periodic\footnote{A sequence $w \in \Sigma^\omega$ is \emph{ultimately periodic} if there exist $n_0, p \in \mathbb{N}$ with $p \geq 1$ such that $w[t+p] = w[t]$ for all $t \geq n_0$.}, then $\operatorname{LZC}(w[:n]) = \Theta(1)$.
\item If $w$ is not ultimately periodic and $h_a$ has exponential growth ($\rho_a > 1$), then $\operatorname{LZC}(w[:n]) = \Theta(\log n)$.
\item If $w$ is not ultimately periodic and $h_a$ has polynomial growth of degree $e_a \geq 2$ ($\rho_a = 1$), then $\operatorname{LZC}(w[:n]) = \Theta(n^{1/e_a})$.
\end{enumerate}
\end{theorem}

Note that the case of polynomial growth with $e_a = 1$ (linear growth, i.e., $|h^n(a)| = \Theta(n)$) is absent from case 3: Constantinescu and Ilie show that linear growth always forces the fixed point to be ultimately periodic, reducing it to case 1.

\subsection{Characteristic Words}

For $\theta \in (0, 1)$, the \emph{characteristic word} with slope $\theta$ is the binary sequence $y \in \{0, 1\}^\omega$ defined by
\begin{equation}
y[t] = \lfloor(t+2)\theta\rfloor - \lfloor(t+1)\theta\rfloor
\end{equation}

Characteristic words are a fundamental object in combinatorics on words (see \citealt{Allouche2003}, Chapter 9). They are the prototypical example of sequences that are \emph{not} morphic in general (a characteristic word is morphic if and only if $\theta$ is a quadratic irrational), yet still have logarithmic $\operatorname{SLC}$.

\begin{proposition}
\label{prp:sturmian_slc}
For any $\theta \in (0, 1)$, the corresponding characteristic word $y$ and any $n \in \mathbb{N}$,
\begin{equation}
\operatorname{SLC}(y[:n]) \leq 24\log n + 8
\end{equation}
\end{proposition}

Notice that this bound is uniform w.r.t.\ $\theta$, and hence any efficient predictor for SLC has a uniform polylog mistake bound over this entire class of sequences.

The proof relies on the theory of continued fractions and standard words, which we briefly recall.

\subsubsection{Characteristic Blocks}

Every irrational $\theta \in (0, 1)$ has a unique continued fraction expansion $\theta = [0; a_1, a_2, a_3, \dots]$ with partial quotients $a_j \geq 1$. The \emph{characteristic blocks} $s_0, s_1, s_2, \dots \in \{0, 1\}^*$ are defined by $s_0 = 0$, $s_1 = 0^{a_1 - 1} 1$, and the recurrence
\begin{equation}
s_{j+1} = s_j^{a_{j+1}} s_{j-1}
\label{eq:std_word_rec}
\end{equation}

We will denote $q_j \coloneqq |s_j|$. Obviously, we have $q_0 = 1$, $q_1 = a_1$ and
\begin{equation}
q_{j+1} = a_{j+1} q_j + q_{j-1}
\end{equation}

The key property we need is that $y[:q_j] = s_j$ for $j \geq 1$ (see \citealt{Allouche2003}, Theorem 9.1.10).

We also have the obvious lower bound:
\begin{equation}
q_{j+1} = a_{j+1} q_j + q_{j-1} \geq a_{j+1} q_j
\label{eq:qj_lower}
\end{equation}

\subsubsection{Proof of Proposition~\ref{prp:sturmian_slc}}

\begin{proof}
First, we reduce to the case of irrational $\theta$. Suppose that $\theta$ is rational, and fix $n$. Since the floor function is right-continuous, we can choose an irrational $\theta' \in (\theta, 1)$ s.t.\ $\lfloor m\theta' \rfloor = \lfloor m\theta \rfloor$ for all $m \in \{1, \dots, n+1\}$. The first $n$ symbols of a characteristic word only depend on these values, and hence the characteristic word with slope $\theta'$ has the same prefix of length $n$ as $y$. Therefore, the claim for $\theta'$ implies the claim for $\theta$. From now on, we assume that $\theta$ is irrational.

We construct a sequence of symbols $A_0, A_1, A_2, \dots$, each of which is either a terminal or a nonterminal, together with sets of rules $P_0 \subseteq P_1 \subseteq P_2 \subseteq \dots$, s.t.\ $\operatorname{val}(A_j) = s_j$, where the values are defined as in Definition~\ref{def:slp} using the rules in $P_j$. Each $P_{j+1}$ extends $P_j$ by rules for fresh nonterminals, and we denote by $|P_j|$ the total length of the rules in $P_j$. Because of this cumulative construction, the rules defining $A_{j-1}$ are already present in $P_j$, so that $A_{j-1}$ and $A_j$ can be referenced freely when building $A_{j+1}$.

We use the fact that for any symbol $A$ (a terminal or a nonterminal) producing a word $w$ and any integer $r \geq 1$, a symbol producing $w^r$ can be built from $A$ by repeated squaring, introducing at most $4\lfloor\log r\rfloor$ fresh edges. Indeed, writing $r$ in binary as $b_\ell \dots b_0$ with $\ell = \lfloor \log r \rfloor$ and $b_\ell = 1$, the binary method computes
\begin{equation}
w^2, w^4, \dots, w^{2^\ell}
\end{equation}
using $\ell$ squaring rules, then accumulates the product over the remaining set bits using at most $\ell$ concatenation rules (one per set bit among $b_{\ell-1}, \dots, b_0$). Each rule is binary (2 edges), giving at most $4\ell = 4\lfloor\log r\rfloor$ fresh edges in total. For $r = 1$, no fresh edges are needed, since $A$ itself produces $w$.

\paragraph{Construction and size bound.} $A_0$ is the terminal $0$ and $P_0 = \emptyset$, so that $|P_0| = 0$. For $A_1$: if $a_1 = 1$ then $s_1 = 1$, $A_1$ is the terminal $1$ and $|P_1| = 0$; otherwise, we build $0^{a_1 - 1}$ by repeated squaring and concatenate the terminal $1$, giving $|P_1| \leq 4\lfloor\log(a_1 - 1)\rfloor + 2 \leq 4\log a_1 + 2$.

For $j \geq 1$, given $A_j$ and $A_{j-1}$, we build $A_{j+1}$ producing $s_{j+1} = s_j^{a_{j+1}} s_{j-1}$ as follows. We raise $A_j$ to the power $a_{j+1}$ by repeated squaring ($\leq 4\lfloor\log a_{j+1}\rfloor$ fresh edges), then concatenate with $A_{j-1}$ via one fresh binary rule ($2$ edges). Since the rules defining $A_{j-1}$ already reside inside $P_j$, the only new edges are from the squaring chain and the final concatenation, giving
\begin{equation}
|P_{j+1}| \leq |P_j| + 4\lfloor\log a_{j+1}\rfloor + 2
\label{eq:slp_simplified}
\end{equation}

It follows that for all $j \geq 1$:
\begin{equation}
|P_j| \leq 4\sum_{i=1}^{j} \log a_i + 2j
\label{eq:slp_sum}
\end{equation}

We bound each term. By Equation~\eqref{eq:qj_lower}, $a_i \leq q_i / q_{i-1}$, so the sum telescopes:
\begin{equation}
\sum_{i=1}^{j} \log a_i \leq \sum_{i=1}^{j} (\log q_i - \log q_{i-1}) = \log q_j
\label{eq:log_telescope}
\end{equation}
For the linear term, $q_{j+1} \geq q_j + q_{j-1}$ (since $a_{j+1} \geq 1$), which gives $q_{j+1} \geq 2q_{j-1}$. Iterating, $q_j \geq 2^{\lfloor j/2 \rfloor}$ for all $j \geq 0$ (since $q_0 = 1$ and $q_1 \geq 1$). Hence
\begin{equation}
j \leq 2\log q_j + 2
\label{eq:j_bound}
\end{equation}
Substituting Equation~\eqref{eq:log_telescope} and Equation~\eqref{eq:j_bound} into Equation~\eqref{eq:slp_sum}:
\begin{equation}
|P_j| \leq 4\log q_j + 2(2\log q_j + 2) = 8\log q_j + 4
\label{eq:Pj_final}
\end{equation}

\paragraph{Handling arbitrary prefixes.} We now show that $\operatorname{SLC}(y[:n]) \leq 24\log n + 8$ for every $n \in \mathbb{N}$. If $n \leq 1$, this holds since $\operatorname{SLC}(y[:n]) \leq |y[:n]| \leq 1$, so we assume $n \geq 2$. If $n < q_1$, then $y[:n] = 0^n$, since $y[:q_1] = s_1 = 0^{a_1 - 1} 1$. Raising the terminal $0$ to the power $n$ by repeated squaring yields a nonterminal producing $0^n$, whose rules have total length at most $4\lfloor\log n\rfloor$ and form a relaxed SLP (Definition~\ref{def:relaxed_slp}). Hence, $\operatorname{SLC}(y[:n]) \leq 4\log n$ by Lemma~\ref{lem:slp_normalize}.

Finally, assume $n \geq q_1$, and let $j \geq 1$ be the largest index with $q_j \leq n$, so that $n < q_{j+1}$. We first observe that every proper prefix of $s_{j-1}$ is a prefix of $s_j$: for $j \geq 2$, this holds since $s_j = s_{j-1}^{a_j} s_{j-2}$ and $a_j \geq 1$, and for $j = 1$, it holds since the only proper prefix of $s_0 = 0$ is the empty word. Since $y[:q_{j+1}] = s_{j+1} = s_j^{a_{j+1}} s_{j-1}$ and $n < q_{j+1}$, the word $y[:n]$ is a prefix of $s_j^{a_{j+1}} u$ for some proper prefix $u$ of $s_{j-1}$. Therefore, $y[:n] \sqsubseteq s_j^{a_{j+1}+1}$.

Now let $r = \lceil n / q_j \rceil$, so that $rq_j \geq n$ and $y[:n] = s_j^{a_{j+1}+1}[:n] \sqsubseteq s_j^r$. We have $r \leq n$, since $q_j \geq 1$. Raising $A_j$ to the power $r$ by repeated squaring yields a symbol $B$ with $\operatorname{val}(B) = s_j^r$, using at most $4\lfloor\log r\rfloor$ fresh edges. Since $|s_j^r| \geq n \geq 2$, $B$ is a nonterminal. Therefore, the rules in $P_j$ together with the fresh rules form a relaxed SLP with root $B$, value $s_j^r$ and size at most $|P_j| + 4\lfloor\log r\rfloor$. By Lemma~\ref{lem:slp_normalize}, Equation~\eqref{eq:Pj_final} and the bounds $\log q_j \leq \log n$ and $\lfloor\log r\rfloor \leq \log n$:
\begin{equation}
\operatorname{SLC}(s_j^r) \leq 8\log n + 4 + 4\log n = 12\log n + 4
\end{equation}
Finally, Proposition~\ref{prp:slc_mon} gives
\begin{equation}
\operatorname{SLC}(y[:n]) \leq 2 \cdot \operatorname{SLC}(s_j^r) \leq 24\log n + 8
\end{equation}
\end{proof}

\section{Worked Examples of HDP}
\label{apx:hdp_examples}

In this appendix, we illustrate the update rule (Algorithm~\ref{alg:hdp_update}) and the prediction rule (Algorithm~\ref{alg:hdp_predict}) of HDP by running them on short words. Throughout, $k = 2$ and $\Sigma = \{a, b\}$, where $a$ and $b$ have the indices $0$ and $1$ in $D^{(0)} = \Sigma$, respectively. As in Appendix~\ref{apx:hdp}, we identify these indices with the symbols themselves. Hence, the entries of $D^{(1)}$ are pairs of symbols, whereas for $j \geq 2$ the entries of $D^{(j)}$ are pairs of indices of $D^{(j-1)}$. We denote the empty word (e.g.\ an empty buffer) by $\lambda$. For readability, we also indicate the word of length $2^j$ that every entry $D^{(j)}[q]$ represents (denoted $\mathrm{str}_j(q)$ in Appendix~\ref{apx:hdp}), although HDP only stores the pairs. Finally, ties in the plurality vote (line~9 of Algorithm~\ref{alg:hdp_predict}) are broken in favor of $a$.

\subsection{The Update Rule}

\begin{example}
\label{ex:hdp_update}
Consider the Thue-Morse sequence of Example~\ref{ex:thue_morse}, with the symbols $0$ and $1$ renamed to $a$ and $b$, respectively, to distinguish them from dictionary indices. Its first $16$ symbols are
\begin{equation}
x = abbabaab\,baababba
\label{eq:hdp_ex_tm}
\end{equation}
Table~\ref{tab:hdp_update} traces the updates of HDP on $x$, and Figure~\ref{fig:hdp_blocks} illustrates the resulting dictionaries.
\end{example}

\begin{table}[tb]
\floatconts{tab:hdp_update}
{\caption{The updates of HDP on the word $x$ of Example~\ref{ex:hdp_update}. In round $t$, the column $D^{(j)}$ shows the tuple that is formed from the full buffer $u^{(j-1)}$ (line~8 of Algorithm~\ref{alg:hdp_update}) and looked up in $D^{(j)}$, followed by the index that it receives (line~11 or line~14). An asterisk marks a tuple that is not found in $D^{(j)}$ and is therefore appended to it (line~13). The remaining columns show the buffers after the update, where an empty cell stands for the empty buffer $\lambda$.}}
{\small
\begin{tabular}{@{}rccccccccccc@{}}
\toprule
& & \multicolumn{4}{c}{Tuples looked up} & & \multicolumn{5}{c}{Buffers after the update} \\
\cmidrule(lr){3-6} \cmidrule(l){8-12}
$t$ & $x[t]$ & $D^{(1)}$ & $D^{(2)}$ & $D^{(3)}$ & $D^{(4)}$ & & $u^{(0)}$ & $u^{(1)}$ & $u^{(2)}$ & $u^{(3)}$ & $u^{(4)}$ \\
\midrule
0 & $a$ &  &  &  &  & & $a$ &  &  &  &  \\
1 & $b$ & $(a,b) \mapsto 0^{\ast}$ &  &  &  & &  & $0$ &  &  &  \\
2 & $b$ &  &  &  &  & & $b$ & $0$ &  &  &  \\
3 & $a$ & $(b,a) \mapsto 1^{\ast}$ & $(0,1) \mapsto 0^{\ast}$ &  &  & &  &  & $0$ &  &  \\
4 & $b$ &  &  &  &  & & $b$ &  & $0$ &  &  \\
5 & $a$ & $(b,a) \mapsto 1$ &  &  &  & &  & $1$ & $0$ &  &  \\
6 & $a$ &  &  &  &  & & $a$ & $1$ & $0$ &  &  \\
7 & $b$ & $(a,b) \mapsto 0$ & $(1,0) \mapsto 1^{\ast}$ & $(0,1) \mapsto 0^{\ast}$ &  & &  &  &  & $0$ &  \\
8 & $b$ &  &  &  &  & & $b$ &  &  & $0$ &  \\
9 & $a$ & $(b,a) \mapsto 1$ &  &  &  & &  & $1$ &  & $0$ &  \\
10 & $a$ &  &  &  &  & & $a$ & $1$ &  & $0$ &  \\
11 & $b$ & $(a,b) \mapsto 0$ & $(1,0) \mapsto 1$ &  &  & &  &  & $1$ & $0$ &  \\
12 & $a$ &  &  &  &  & & $a$ &  & $1$ & $0$ &  \\
13 & $b$ & $(a,b) \mapsto 0$ &  &  &  & &  & $0$ & $1$ & $0$ &  \\
14 & $b$ &  &  &  &  & & $b$ & $0$ & $1$ & $0$ &  \\
15 & $a$ & $(b,a) \mapsto 1$ & $(0,1) \mapsto 0$ & $(1,0) \mapsto 1^{\ast}$ & $(0,1) \mapsto 0^{\ast}$ & &  &  &  &  & $0$ \\
\bottomrule
\end{tabular}}
\end{table}

In round $0$, line~4 of Algorithm~\ref{alg:hdp_update} appends $a$ to $u^{(0)}$, and since $|u^{(0)}| = 1 < k$, the update returns on line~6. In round $1$, line~4 appends $b$ to $u^{(0)}$. Now, $u^{(0)} = ab$ has $k$ entries, i.e.\ it holds the complete block $x[0:2]$, so lines~8--9 set $\mathrm{tuple} = (a, b)$ and empty the buffer. Since $D^{(1)}$ is empty, lines~13--14 append $(a, b)$ to $D^{(1)}$, where it receives the index $0$. In the next iteration of the loop, this index is appended to $u^{(1)}$, and the update returns, since $|u^{(1)}| = 1 < k$. Thus, $u^{(1)} = 0$ records that the current aligned block of size $4$ begins with the block represented by $D^{(1)}[0]$, namely $ab$.

In round $7$, the update propagates up to level $3$. The symbol $b$ completes the block $x[6:8] = ab$, which is already represented by $D^{(1)}[0]$ (lines~10--11), so the index $0$ is appended to $u^{(1)}$. The resulting buffer $u^{(1)} = 1\,0$ represents the complete block $x[4:8] = baab$. The tuple $(1, 0)$ is not in $D^{(2)}$, so it is appended to $D^{(2)}$ with the index $1$, which is in turn appended to $u^{(2)}$. The resulting buffer $u^{(2)} = 0\,1$ represents the complete block $x[0:8]$. The tuple $(0, 1)$ is appended to the empty dictionary $D^{(3)}$ with the index $0$, which is appended to $u^{(3)}$, and the update returns. Notice that the tuple $(0, 1)$ is also stored in $D^{(2)}$, where it represents a different word, namely $abba$ rather than $abbabaab$: every level has its own dictionary.

More generally, the loop of Algorithm~\ref{alg:hdp_update} reaches iteration $j$ in round $t$ if and only if $2^j$ divides $t + 1$, i.e.\ if and only if position $t$ is the last position of an aligned block of size $2^j$ (see the proof of Lemma~\ref{lem:hdp_state}). Hence, an update resembles an increment of a binary counter, and the completion of a block of size $2^{j+1}$ corresponds to a carry from digit $j$ to digit $j + 1$. Indeed, after round $t$, the buffer $u^{(j)}$ is non-empty if and only if the $j$-th binary digit of $t + 1$ is $1$, where the digits are numbered from the least significant one, starting from $0$ (Lemma~\ref{lem:hdp_state}). For example, after round $10$, the buffers $u^{(0)}$, $u^{(1)}$ and $u^{(3)}$ are non-empty whereas $u^{(2)}$ is empty, and $11$ is $1011$ in binary. For general $k$, the number of indices in $u^{(j)}$ is the $j$-th digit of $t + 1$ in base $k$.

After round $15$, all the buffers are empty except $u^{(4)} = 0$, and the dictionaries are
\begin{equation}
\begin{aligned}
D^{(1)} &= \big( (a, b),\, (b, a) \big) & &\text{representing } ab \text{ and } ba \\
D^{(2)} &= \big( (0, 1),\, (1, 0) \big) & &\text{representing } abba \text{ and } baab \\
D^{(3)} &= \big( (0, 1),\, (1, 0) \big) & &\text{representing } abbabaab \text{ and } baababba \\
D^{(4)} &= \big( (0, 1) \big) & &\text{representing } abbabaabbaababba
\end{aligned}
\label{eq:hdp_ex_dicts}
\end{equation}
Figure~\ref{fig:hdp_blocks} shows how these entries correspond to the aligned blocks of $x$. Every dictionary has at most $2$ entries. This is not a coincidence: by Lemma~\ref{lem:block}, applied to an automaton with two states that generates the Thue-Morse sequence (see Example~\ref{ex:thue_morse}), the sequence has at most $2$ distinct complete aligned blocks of every size $2^j$ (namely, $u_j$ and $\overline{u}_j$, in the notation of Example~\ref{ex:thue_morse}). Hence, after $n$ rounds, the state of HDP consists of at most $\log n$ non-empty dictionaries with at most $2$ entries each, together with the buffers, in agreement with the compression bound of Theorem~\ref{thm:hdp_bounds}.

\begin{figure}[htbp]
\floatconts{fig:hdp_blocks}
{\caption{The aligned blocks of the word $x$ of Example~\ref{ex:hdp_update}, labeled by their indices in the dictionaries of HDP. The bottom row shows the positions $t$. For $1 \leq j \leq 4$, the row of level $j$ is divided into the aligned blocks of size $2^j$, and every block is labeled by the index $q$ s.t.\ $D^{(j)}[q]$ represents it. This is the index that Algorithm~\ref{alg:hdp_update} appends to $u^{(j)}$ when the block is completed. The entry $D^{(j)}[q]$ itself is the pair of labels of the two halves of the block, in the row of level $j - 1$ (where the indices of $D^{(0)}$ are the symbols). For example, the block $x[4:8] = baab$ is labeled by $1$ at level $2$, and its halves are labeled by $1$ and $0$ at level $1$. Accordingly, $D^{(2)}[1] = (1, 0)$.}}
{\renewcommand{\arraystretch}{1.3}%
\begin{tabular}{@{}l@{\quad}|*{16}{@{}c@{}|}}
\cline{2-17}
level $4$ & \multicolumn{16}{c|}{$0$} \\ \cline{2-17}
level $3$ & \multicolumn{8}{c|}{$0$} & \multicolumn{8}{c|}{$1$} \\ \cline{2-17}
level $2$ & \multicolumn{4}{c|}{$0$} & \multicolumn{4}{c|}{$1$} & \multicolumn{4}{c|}{$1$} & \multicolumn{4}{c|}{$0$} \\ \cline{2-17}
level $1$ & \multicolumn{2}{c|}{$0$} & \multicolumn{2}{c|}{$1$} & \multicolumn{2}{c|}{$1$} & \multicolumn{2}{c|}{$0$} & \multicolumn{2}{c|}{$1$} & \multicolumn{2}{c|}{$0$} & \multicolumn{2}{c|}{$0$} & \multicolumn{2}{c|}{$1$} \\ \cline{2-17}
level $0$ & \makebox[1.6em]{$a$} & \makebox[1.6em]{$b$} & \makebox[1.6em]{$b$} & \makebox[1.6em]{$a$} & \makebox[1.6em]{$b$} & \makebox[1.6em]{$a$} & \makebox[1.6em]{$a$} & \makebox[1.6em]{$b$} & \makebox[1.6em]{$b$} & \makebox[1.6em]{$a$} & \makebox[1.6em]{$a$} & \makebox[1.6em]{$b$} & \makebox[1.6em]{$a$} & \makebox[1.6em]{$b$} & \makebox[1.6em]{$b$} & \makebox[1.6em]{$a$} \\ \cline{2-17}
\multicolumn{1}{@{}l@{\quad}}{$t$} & \multicolumn{1}{c}{\scriptsize $0$} & \multicolumn{1}{c}{\scriptsize $1$} & \multicolumn{1}{c}{\scriptsize $2$} & \multicolumn{1}{c}{\scriptsize $3$} & \multicolumn{1}{c}{\scriptsize $4$} & \multicolumn{1}{c}{\scriptsize $5$} & \multicolumn{1}{c}{\scriptsize $6$} & \multicolumn{1}{c}{\scriptsize $7$} & \multicolumn{1}{c}{\scriptsize $8$} & \multicolumn{1}{c}{\scriptsize $9$} & \multicolumn{1}{c}{\scriptsize $10$} & \multicolumn{1}{c}{\scriptsize $11$} & \multicolumn{1}{c}{\scriptsize $12$} & \multicolumn{1}{c}{\scriptsize $13$} & \multicolumn{1}{c}{\scriptsize $14$} & \multicolumn{1}{c}{\scriptsize $15$}
\end{tabular}}
\end{figure}

\subsection{The Prediction Rule}

Recall that in round $t$, HDP computes its prediction $\hat{x}_t$ from the state $s_t$, i.e.\ from the state after the update in round $t - 1$. For every $j$, we call the aligned block of size $2^j$ that contains position $t$ the \emph{current block} of size $2^j$. Its symbols that were observed before round $t$ form the word $x[\lfloor t \rfloor_j : t]$, in the notation of Appendix~\ref{apx:hdp}. We call the words represented by the entries of $D^{(j)}$ the \emph{known blocks} of size $2^j$. By Lemma~\ref{lem:hdp_state}, for $j \geq 1$, these are the complete aligned blocks of size $2^j$ that occurred before round $t$. A version $(q, c)$ of level $j$ (called $(q, b)$ in Algorithm~\ref{alg:hdp_predict}, but here $b$ is a symbol) stands for the known block represented by $D^{(j)}[q]$, which votes for the symbol $c$.

\begin{example}
\label{ex:hdp_predict}
We continue Example~\ref{ex:hdp_update}. Table~\ref{tab:hdp_predict} shows the predictions of HDP in the first $16$ rounds on the word $x$ of equation~\eqref{eq:hdp_ex_tm}.
\end{example}

\begin{table}[tb]
\floatconts{tab:hdp_predict}
{\caption{The predictions of HDP in the first $16$ rounds on the word $x$ of Example~\ref{ex:hdp_update}. For $j \in \{1, 2, 3\}$, the column $V_j$ shows the set $V$ at the beginning of iteration $j$ of the loop in Algorithm~\ref{alg:hdp_predict}, where an empty cell means that the loop does not reach iteration $j$. (At the beginning of iteration $0$, $V = \{(a, a), (b, b)\}$ in every round.) The algorithm returns in iteration $J_t$, and the column ``Votes'' shows how many of the versions in $V$ vote for $a$ and for $b$ in this iteration. Wrong predictions are marked by $\times$.}}
{\small
\begin{tabular}{@{}rcccccccc@{}}
\toprule
$t$ & $V_1$ & $V_2$ & $V_3$ & $J_t$ & Votes ($a : b$) & $\hat{x}_t$ & $x[t]$ & \\
\midrule
0 &  &  &  & 0 & $1 : 1$ & $a$ & $a$ &  \\
1 &  &  &  & 0 & $1 : 1$ & $a$ & $b$ & $\times$ \\
2 & $\{(0, a)\}$ &  &  & 1 & $1 : 0$ & $a$ & $b$ & $\times$ \\
3 &  &  &  & 0 & $1 : 1$ & $a$ & $a$ &  \\
4 & $\{(0, a), (1, b)\}$ & $\{(0, a)\}$ &  & 2 & $1 : 0$ & $a$ & $b$ & $\times$ \\
5 & $\{(1, a)\}$ &  &  & 1 & $1 : 0$ & $a$ & $a$ &  \\
6 & $\{(0, a), (1, b)\}$ &  &  & 1 & $1 : 1$ & $a$ & $a$ &  \\
7 & $\{(0, b)\}$ &  &  & 1 & $0 : 1$ & $b$ & $b$ &  \\
8 & $\{(0, a), (1, b)\}$ & $\{(0, a), (1, b)\}$ & $\{(0, a)\}$ & 3 & $1 : 0$ & $a$ & $b$ & $\times$ \\
9 & $\{(1, a)\}$ & $\{(1, a)\}$ &  & 2 & $1 : 0$ & $a$ & $a$ &  \\
10 & $\{(0, a), (1, b)\}$ & $\{(1, a)\}$ &  & 2 & $1 : 0$ & $a$ & $a$ &  \\
11 & $\{(0, b)\}$ & $\{(1, b)\}$ &  & 2 & $0 : 1$ & $b$ & $b$ &  \\
12 & $\{(0, a), (1, b)\}$ & $\{(0, a), (1, b)\}$ &  & 2 & $1 : 1$ & $a$ & $a$ &  \\
13 & $\{(0, b)\}$ & $\{(0, b)\}$ &  & 2 & $0 : 1$ & $b$ & $b$ &  \\
14 & $\{(0, a), (1, b)\}$ & $\{(0, b)\}$ &  & 2 & $0 : 1$ & $b$ & $b$ &  \\
15 & $\{(1, a)\}$ & $\{(0, a)\}$ &  & 2 & $1 : 0$ & $a$ & $a$ &  \\
\bottomrule
\end{tabular}}
\end{table}

Let us trace Algorithm~\ref{alg:hdp_predict} in round $t = 14$. By Table~\ref{tab:hdp_update}, the state $s_{14}$ consists of the dictionaries $D^{(1)}$ and $D^{(2)}$ of equation~\eqref{eq:hdp_ex_dicts} and $D^{(3)} = \big( (0, 1) \big)$, and of the buffers $u^{(0)} = \lambda$, $u^{(1)} = 0$, $u^{(2)} = 1$ and $u^{(3)} = 0$.
\begin{itemize}
\item Line~2 sets $V = \{(a, a), (b, b)\}$. In iteration $j = 0$, since $u^{(0)} = \lambda$, line~6 selects the entries of $D^{(1)}$ whose first component is $a$ or $b$, i.e.\ both $D^{(1)}[0] = (a, b)$ and $D^{(1)}[1] = (b, a)$. Hence, $V = \{(0, a), (1, b)\}$ at the beginning of iteration $1$. Indeed, no symbol of the current block of size $2$, namely $x[14:16]$, has been observed yet, so both known blocks of size $2$, $ab$ and $ba$, are consistent with it, and they vote for their first symbols.
\item In iteration $j = 1$, the buffer $u^{(1)} = 0$ records that the current block of size $4$, namely $x[12:16]$, begins with the block $ab$ represented by $D^{(1)}[0]$. Hence, line~6 selects the entries of $D^{(2)}$ of the form $(0, q)$ s.t.\ $(q, c) \in V$ for some $c$. The only such entry is $D^{(2)}[0] = (0, 1)$, whose second component comes from the version $(1, b)$. Hence, $V = \{(0, b)\}$ at the beginning of iteration $2$: the only known block of size $4$ that begins with $ab$ is $abba$, and it votes for its symbol at the offset of position $14$ in $x[12:16]$, namely $b$.
\item In iteration $j = 2$, the buffer $u^{(2)} = 1$ records that the current block of size $8$, namely $x[8:16]$, begins with the block $baab$ represented by $D^{(2)}[1]$. No entry of $D^{(3)}$ is of the form $(1, q)$, and hence $V_{\mathrm{next}} = \emptyset$ and line~9 returns the plurality vote over $V = \{(0, b)\}$, i.e.\ $\hat{x}_{14} = b$.
\end{itemize}
This prediction is correct. Notice that a vote at level $1$ would have been a tie, which is resolved by the context of the current block of size $4$, stored in $u^{(1)}$.

In round $t = 8$, a new block of size $8$ begins, so that $u^{(0)} = u^{(1)} = u^{(2)} = \lambda$ and $u^{(3)} = 0$. In iterations $0$, $1$ and $2$, since the buffers are empty, line~6 selects every entry of the next dictionary whose first component comes from a version in $V$. This yields $V = \{(0, a), (1, b)\}$ at levels $1$ and $2$, and $V = \{(0, a)\}$ at level $3$, since the only entry of $D^{(3)}$ represents $abbabaab$. In iteration $3$, $D^{(4)}$ is empty, so the algorithm returns $\hat{x}_8 = a$. In other words, HDP predicts that the new block of size $8$ repeats the only block of size $8$ seen so far. This prediction is wrong: the new block $x[8:16] = baababba$ is the negation of $x[0:8]$. The mistakes in rounds $2$ and $4$ have the same cause, at levels $1$ and $2$ respectively. The mistake in round $1$ is due to the tie-breaking rule: at that point, no block of size $2$ has been completed yet, so the vote takes place at level $0$ and is a tie. After round $8$, the votes take place at level $2$, where both blocks of size $4$ are already known, and all the predictions in rounds $9$ to $15$ are correct.

In terms of the analysis in Appendix~\ref{apx:hdp}, the versions in $V$ at the beginning of iteration $j$ correspond to the version space $\mathcal{W}_j(t)$, i.e.\ to the known blocks of size $2^j$ that are consistent with the observed part $x[\lfloor t \rfloor_j : t]$ of the current block of size $2^j$ (Lemma~\ref{lem:hdp_predict}). For example, $\mathcal{W}_1(14) = \{ab, ba\}$ and $\mathcal{W}_2(14) = \{abba\}$, whereas $\mathcal{W}_3(14) = \emptyset$, since the only known block of size $8$ in round $14$ is $abbabaab$, which does not begin with $x[8:14] = baabab$. Hence $J_{14} = 2$, i.e.\ the vote takes place at level $2$. Similarly, the mistake in round $8$ is made by the vote at level $J_8 = 3$, and it eliminates the only version of this level: $\mathcal{W}_3(8) = \{abbabaab\}$, whereas $\mathcal{W}_3(9) = \emptyset$. This is the mechanism behind the mistake bound: as long as the current block of size $2^j$ is not completed, the version space of level $j$ can only shrink, and every mistake made by a vote at level $j$ shrinks it at least by half (see the proof of Lemma~\ref{lem:hdp_factor}).

\begin{example}
\label{ex:hdp_plurality}
In Example~\ref{ex:hdp_predict}, every vote is either unanimous or a tie, since the Thue-Morse sequence has only two distinct complete aligned blocks of every size, and they differ at every position. To illustrate a vote among more versions, consider the prediction of HDP in round $t = 12$ on a word $x$ that begins with
\begin{equation}
x[:12] = aaab\,baab\,baaa
\label{eq:hdp_ex_plur}
\end{equation}
\end{example}

The complete aligned blocks of size $2$ of $x[:12]$ are $aa$, $ab$, $ba$, $ab$, $ba$ and $aa$, those of size $4$ are $aaab$, $baab$ and $baaa$, and there is a single one of size $8$, namely $aaabbaab$. Accordingly, running Algorithm~\ref{alg:hdp_update} on $x[:12]$ produces the dictionaries
\begin{equation}
\begin{aligned}
D^{(1)} &= \big( (a, a),\, (a, b),\, (b, a) \big) & &\text{representing } aa,\ ab \text{ and } ba \\
D^{(2)} &= \big( (0, 1),\, (2, 1),\, (2, 0) \big) & &\text{representing } aaab,\ baab \text{ and } baaa \\
D^{(3)} &= \big( (0, 1) \big) & &\text{representing } aaabbaab
\end{aligned}
\end{equation}
and the buffers $u^{(0)} = u^{(1)} = \lambda$, $u^{(2)} = 2$ and $u^{(3)} = 0$. That is, the current block of size $8$, namely $x[8:16]$, begins with $baaa$, and the current block of size $16$ begins with $aaabbaab$. Algorithm~\ref{alg:hdp_predict} proceeds as follows.
\begin{itemize}
\item In iteration $j = 0$, since $u^{(0)} = \lambda$, line~6 selects all the entries of $D^{(1)}$, and hence $V = \{(0, a), (1, a), (2, b)\}$ at the beginning of iteration $1$. A vote at this level would predict $a$, by $2$ votes (of $aa$ and $ab$) against $1$ (of $ba$).
\item In iteration $j = 1$, since $u^{(1)} = \lambda$, line~6 selects the entries of $D^{(2)}$ whose first component is $0$, $1$ or $2$. The entry $D^{(2)}[0] = (0, 1)$ extends the version $(0, a)$, and the entries $D^{(2)}[1] = (2, 1)$ and $D^{(2)}[2] = (2, 0)$ both extend the version $(2, b)$, whereas the version $(1, a)$ has no extension, since no known block of size $4$ begins with $ab$. Hence, $V = \{(0, a), (1, b), (2, b)\}$ at the beginning of iteration $2$, and now $b$ has $2$ votes (of $baab$ and $baaa$) against $1$ (of $aaab$).
\item In iteration $j = 2$, since $u^{(2)} = 2$, line~6 looks for entries of $D^{(3)}$ of the form $(2, q)$. There are none, since the only known block of size $8$ does not begin with $baaa$. Hence, $V_{\mathrm{next}} = \emptyset$, and line~9 returns $\hat{x}_{12} = b$.
\end{itemize}
Thus, moving from level $1$ to level $2$ removed one version and split another one into two, which reversed the outcome of the vote. This is because the votes at level $j$ are cast by distinct known blocks of size $2^j$: the block $ba$ begins two distinct known blocks of size $4$, whereas $ab$ begins none. In the notation of Appendix~\ref{apx:hdp}, $\mathcal{W}_1(12) = \{aa, ab, ba\}$, $\mathcal{W}_2(12) = \{aaab, baab, baaa\}$ and $\mathcal{W}_3(12) = \emptyset$, so that $J_{12} = 2$. Notice that a version of level $j$ can have several extensions only if $|u^{(j)}| < k - 1$ (i.e.\ only if $u^{(j)} = \lambda$ for $k = 2$), since otherwise $u^{(j)} q$ is a complete $k$-tuple, and the entries of $D^{(j+1)}$ are distinct.

\section{Examples of Straight-Line Programs}
\label{apx:slp_examples}

In this appendix, we illustrate Definition~\ref{def:slp} with a few examples. Recall that an SLP is a type of context-free grammar whose language consists of a single word. Accordingly, we call the elements of $Q$ \emph{nonterminals}, the elements of $\Sigma$ \emph{terminals}, $\delta(q)$ the \emph{rule} of $q$ and $q_0$ the \emph{root}.

\begin{example}
\label{ex:slp_abracadabra}
Let $\Sigma = \{a, b, c, d, r\}$ and $P = (Q, q_0, \delta)$, where $Q = \{q_0, q_1\}$ and
\begin{equation}
\delta(q_0) = q_1\, c\, a\, d\, q_1, \qquad \delta(q_1) = abra
\end{equation}
Then, $\operatorname{val}_P(q_1) = abra$ and
\begin{equation}
\operatorname{val}(P) = \operatorname{val}_P(q_0) = abra \cdot c \cdot a \cdot d \cdot abra = abracadabra
\end{equation}
\end{example}

\begin{figure}[htbp]
\floatconts{fig:slp_dag}
{\caption{The SLP $P$ of Example~\ref{ex:slp_abracadabra} as a directed acyclic graph. There is an edge for every position in every rule, labeled by this position. For example, the two edges from $q_0$ to $q_1$ correspond to the positions $0$ and $4$ of $\delta(q_0) = q_1\,c\,a\,d\,q_1$, and the two edges from $q_1$ to $a$ correspond to the positions $0$ and $3$ of $\delta(q_1) = abra$.}}
{\begin{tikzpicture}[>=stealth, x=1.9cm, y=1.45cm, position/.style={font=\footnotesize, fill=white, inner sep=1pt}]
\node (q0) at (2.5, 2) {$q_0$};
\node (q1) at (1, 1) {$q_1$};
\node (b) at (0, 0) {$b$};
\node (r) at (1, 0) {$r$};
\node (a) at (2, 0) {$a$};
\node (c) at (3, 0) {$c$};
\node (d) at (4, 0) {$d$};
\draw[->] (q0) to[bend right=22] node[position] {$0$} (q1);
\draw[->] (q0) to[bend left=22] node[position] {$4$} (q1);
\draw[->] (q0) to node[position] {$1$} (c);
\draw[->] (q0) to node[position] {$2$} (a);
\draw[->] (q0) to node[position] {$3$} (d);
\draw[->] (q1) to[bend right=22] node[position] {$0$} (a);
\draw[->] (q1) to[bend left=22] node[position] {$3$} (a);
\draw[->] (q1) to node[position] {$1$} (b);
\draw[->] (q1) to node[position] {$2$} (r);
\end{tikzpicture}}
\end{figure}

All three requirements of Definition~\ref{def:slp} hold: the rule of $q_1$ has length $4$, $\delta$ is acyclic since the rule of $q_1$ consists of terminals, and $q_0$ is the only nonterminal that doesn't appear in any rule, since $q_1$ appears in the rule of $q_0$. Figure~\ref{fig:slp_dag} shows $P$ as a directed acyclic graph. The size of $P$ is its number of edges, $|P| = |\delta(q_0)| + |\delta(q_1)| = 5 + 4 = 9$. This is less than $|abracadabra| = 11$, which is the size of the SLP with the single rule $\delta(q_0) = abracadabra$, because $P$ contains the word $abra$ only once, in the rule of $q_1$, and refers to it twice, in the rule of $q_0$. In particular, $\operatorname{SLC}(abracadabra) \leq 9$.

We consider modifications of $P$ to illustrate what the requirements of Definition~\ref{def:slp} rule out. If the rule of $q_1$ were $ab\,q_1$, then $\delta$ would not be acyclic, and $\operatorname{val}_P(q_1)$ would not be defined, since no finite word $w$ satisfies $w = ab\,w$. If we added to $P$ a nonterminal $q_2$ with $\delta(q_2) = ab$, then $q_2$ would not appear in any rule, violating the third requirement. Such a nonterminal would not affect the value, but would increase the size. More generally, nonterminals that don't appear in any rule, as well as unary rules (i.e.\ rules of length $1$) of nonterminals other than the root, can be eliminated without changing the value or increasing the size (Lemma~\ref{lem:slp_normalize}). Hence, allowing them would not change $\operatorname{SLC}$. In contrast, the rule of the root may have length $0$ or $1$, since otherwise the empty word and the words of length $1$ would not be the values of any SLP. The following two examples illustrate this.

\begin{example}
\label{ex:slp_empty}
Let $\lambda$ denote the empty word, and let $P = (Q, q_0, \delta)$, where $Q = \{q_0\}$ and $\delta(q_0) = \lambda$. Then, $\operatorname{val}(P) = \lambda$, since $\operatorname{val}_P(q_0)$ is the concatenation of the values of the entries of $\delta(q_0)$, and there are none. Moreover, $|P| = 0$, and hence $\operatorname{SLC}(\lambda) = 0$.
\end{example}

\begin{example}
\label{ex:slp_letter}
Let $a \in \Sigma$, and let $P = (Q, q_0, \delta)$, where $Q = \{q_0\}$ and $\delta(q_0) = a$. Then, $\operatorname{val}(P) = a$ and $|P| = 1$. Therefore, $\operatorname{SLC}(a) = 1$: an SLP of size $0$ has only empty rules, so its value is the empty word.
\end{example}

\begin{example}
\label{ex:slp_fibonacci}
The recurrence defining the words $u_j$ of Example~\ref{ex:fibonacci} (where $\Sigma = \{0, 1\}$) describes the following SLP. Fix $j \geq 2$, and let $P = (Q, A_j, \delta)$, where $Q = \{A_1, A_2, \dots, A_j\}$ and
\begin{equation}
\delta(A_1) = 01, \qquad \delta(A_2) = A_1\, 0, \qquad \delta(A_i) = A_{i-1} A_{i-2} \quad \text{for } 3 \leq i \leq j
\end{equation}
\end{example}

All rules have length $2$, the rule of every $A_i$ contains only terminals and nonterminals with smaller indices (and hence $\delta$ is acyclic), and $A_j$ is the only nonterminal that doesn't appear in any rule, since every $A_i$ with $i < j$ appears in the rule of $A_{i+1}$. Hence, $P$ is an SLP. Moreover, since $u_0 = 0$, $u_1 = 01$ and $u_i = u_{i-1} u_{i-2}$ for $i \geq 2$, induction on $i$ shows that $\operatorname{val}_P(A_i) = u_i$ for all $1 \leq i \leq j$. Therefore, $\operatorname{val}(P) = u_j$ and $|P| = 2j$, so that $\operatorname{SLC}(u_j) \leq 2j$.

On the other hand, the lengths $|u_0|, |u_1|, |u_2|, \dots$ are the Fibonacci numbers $1, 2, 3, 5, 8, \dots$, which grow exponentially. For example, $|u_{10}| = 144$, whereas $|P| = 20$ for $j = 10$. Thus, the value of an SLP can be exponentially longer than its size. This is possible because a nonterminal can appear in several rules (or several times in the same rule), and each of its occurrences contributes a copy of its value. In this SLP, every $A_i$ with $i \leq j - 2$ appears in the rules of both $A_{i+1}$ and $A_{i+2}$.

\end{document}